\documentclass[aps,twocolumn,pra,superscriptaddress,amsmath,amssymb]{revtex4-1} 
\usepackage{dcolumn}
\usepackage{bm}
\usepackage{amsmath}
\usepackage{txfonts}
\usepackage[T1]{fontenc}
\usepackage{xspace}
\usepackage{ulem}
\usepackage{comment}
\newcommand{\bras}[1]{\langle#1\rvert}
\newcommand{\kets}[1]{\lvert#1\rangle}

\newcommand{\mean}[1]{\left<#1\right>}
\newcommand{\means}[1]{\langle#1\rangle}

\ifx\pdfoutput\undefined
\usepackage[dvipdfmx]{graphicx}
\usepackage[dvipdfmx]{hyperref}
\usepackage[dvipdfmx]{color}
\usepackage[dvipdfmx]{xcolor}
\else
\usepackage{graphicx}
\usepackage{hyperref}
\usepackage{color}
\usepackage{xcolor}
\fi
\begin{document}
\let\emph\textit

\title{
Scattering phenomena for spin transport in Kitaev spin liquid
}
\author{Joji Nasu}
\affiliation{
  Department of Physics, Tohoku University, Sendai 980-8578, Japan
}

\author{Yuta Murakami}
\affiliation{
  Center for Emergent Matter Science, RIKEN, Wako, Saitama 351-0198, Japan
}
\author{Akihisa Koga}
\affiliation{
  Department of Physics, Tokyo Institute of Technology,
  Meguro, Tokyo 152-8551, Japan
}

\date{\today}
\begin{abstract}
  The Kitaev model exhibits a canonical quantum spin liquid as a ground state and hosts two fractional quasiparticles, itinerant Majorana fermion and localized flux excitation.
  The former can carry heat and spin modulations in the quantum spin liquid, but the role of the latter remains unknown for the transport phenomena.
  Here, we focus on spin transport in the presence of excited fluxes and report that they yield strong interference in the propagation of the Majorana fermions, which feel gauge-like potential emergent around the fluxes.
  We examine the transient spin dynamics triggered by a pulsed magnetic field at an edge.
  In the absence of excited fluxes, the magnetic-field pulse creates the plane wave of the Majorana fermions, which flows in the quantum spin liquid.
  Although this wave does not accompany the change of local spin moments in bulk, it induces local moments at the side opposite to the edge under the magnetic-field pulse.
  We observe the spatial modulation of induced spin moments when fluxes are excited in the bulk region.
  This behavior is more striking than the case of lattice defects.
  Moreover, we find that, although the amplitude of the spatial change is almost independent of the distance between lattice defects, it is strongly enhanced by increasing the distance for the case of excited fluxes.
  The difference is understood from the influence on the itinerant Majorana fermions; the lattice defects change the system locally, but flux excitations alter all the transfer integrals on the string connecting them.
  The present results will provide another route to observing intrinsic flux excitations distinguished from extrinsic effects such as lattice defects.
\end{abstract}
\maketitle


\section{Introduction}

Spin transport phenomena have recently been investigated not only in metallic materials but also for insulating magnets~\cite{Maekawa2013rev,adachi2013theory}.
In itinerant electron systems, since an electron possesses the spin-1/2 degree of freedom, a spin-polarized electric current is always associated with the spin current.
On the other hand, when electric currents with different spins flow in the opposite directions, the spin current can occur without electric current, which leads to intriguing phenomena such as the spin Hall effect and its inverse one~\cite{Hirsch1999,Brataas2002,Watson2003,murakami2003dissipationless}.
As electrical conduction is not essential for the spin current generation, spin transport has been studied in Mott insulators where the spin degree of freedom is active.
Among them, magnetically ordered systems have been intensively investigated for spin transport where collective modes from a magnetic order, magnons, carry spin excitations~\cite{uchida2010spin,uchida2010,Xiao2010,Adachi2011,Zhang_magnon2012,Zhang_magnon_full2012,Rezende2014}.
In addition to ferromagnetic and antiferromagnetic insulators, Mott insulators without magnetic orders have been examined as a playground for spin transport when itinerant elementary excitations are present~\cite{qiu2018spin,naka2019spin,hirobe2017one,Nasu_spinseebeck2021}.
For instance, the spin Seebeck effect has been observed in the quasi-one-dimensional copper oxide, where spinons, quasiparticles in quantum spin liquids (QSLs), are regarded as carriers of spin excitations.

Until now, theoretical and experimental works have proposed many QSL candidates.
Among them, the Kitaev QSL has attracted considerable attention as a canonical example for QSLs~\cite{RevModPhys.87.1,Trebst2017pre,Hermanns2018rev,Knolle2019rev,takagi2019rev,Motome2020rev}. 
This QSL state appears in the spin-1/2 quantum spin model on a honeycomb lattice, where the magnetic interactions are bond-dependent Ising-type~\cite{Kitaev2006}.
The model, referred to as the Kitaev model, hosts Majorana quasiparticles as elementary excitations from the QSL ground state.
Although the Kitaev model has such peculiar magnetic interactions, this model is realized in a certain kind of Mott insulators whose magnetism originates from magnetic ions with strong spin-orbit coupling~\cite{PhysRevLett.102.017205,Winter2016,Winter_2017rev,Jang_AFKitaev2019,Stavropoulos2019}.
As the candidate materials, iridium oxides and ruthenium compounds are intensively investigated~\cite{PhysRevLett.105.027204,PhysRevB.82.064412,PhysRevLett.108.127203,PhysRevB.88.035107,PhysRevLett.110.097204,1367-2630-16-1-013056,PhysRevLett.113.107201,PhysRevB.90.041112,PhysRevB.91.094422,Johnson2015,PhysRevB.91.144420,yadav2016kitaev,PhysRevB.93.155143,Koitzsch2016}.
Moreover, other candidate materials such as cobalt oxides have also been proposed recently, and hence, the range of compounds with Kitaev-like nature is expanding~\cite{Liu2018Pseudospin,sano2018,Lefran2016,Bera2017,Zhong_Field-induced2018,wildes2017magnetic,Yan_Magnetic2019,zhong2020weak,Yuan2020}.

One of the most attractive features of the Kitaev model is the presence of the fractional quasiparticles, Majorana fermion and flux excitation, the latter of which is referred to also as vison or vortex.
The signatures of these fractional quasiparticles manifest themselves in the measurements for the specific heat and excitation spectra, such as neutron and Raman scatterings in the candidate materials~\cite{PhysRevB.92.115122,PhysRevLett.112.207203,PhysRevB.92.115127,PhysRevLett.113.187201,PhysRevLett.114.147201,winter2017breakdown,Song2016,Nasu2016nphys,Halasz2016,yoshitake2016,tanaka2022thermodynamic}.
Interestingly, applying weak magnetic fields causes the topologically nontrivial band structure of the Majorana fermions~\cite{Nasu2017,Cookmeyer2018,kasahara2018majorana,Hentrich2019,hickey2019emergence,Gordon2019,Gao2019_thermal,Lee2020_Magnetic,yokoi2021half,Chern2021_Sign,Zhang2021,Koyama2021,czajka2021oscillations,hwang2022identification}.
Moreover, as a counterpart of a Majorana edge mode, each flux excitation traps a Majorana zero mode, which behaves as a non-Abelian anyon~\cite{Kitaev2006}.
This anyonic quasiparticle can be an operation element in topological quantum computing, which should be implemented by braiding with other anyons.
To establish the braiding operations, one desires the creation, observation, and manipulation of the flux excitations.
In these viewpoints, several theoretical proposals have been made very recently~\cite{Feldmeier2020,Konig2020,Pereira2020,Udagawa2021,Klocke2021,Wei2021,Jang2021,Yue-Liu2021pre}.
However, experimental work has not achieved capturing flux states thus far.
The difficulty is based on the charge neutrality of the quasiparticles.

One can also expect that the quasiparticles in the Kitaev QSL propagate accompanied by the spin excitations because these originate from quantum spin.
In particular, spin transport attributed to the itinerant Majorana fermions in the Kitaev QSL has been discussed theoretically~\cite{Yao_Lee2011,Carvalho2018,Aftergood2020,Minakawa2020,Taguchi2021,Koga_majorana2021,Taguchi_Thermally2022,Takikawa2021pre}.
The Majorana fermion system in the QSL exhibits gapless linear dispersions, implying a long-ranged spin transport.
On the other hand, less is known about the role of the other fractional quasiparticles, the flux excitations, in the spin transport~\cite{Minakawa2020}.
The mobility of the flux excitations is much smaller than that of the Majorana quasiparticles under a weak magnetic field.
However, excited fluxes should behave as scatterers for the flow of the Majorana quasiparticles, which contributes to the spin transport in the Kitaev QSL.
On the other hand, lattice defects may disturb the propagation of the Majorana quasiparticles, but it remains unclear whether such an extrinsic effect is different from the impact of the intrinsic flux excitations for the Majorana-fermion flow.
This issue should be essential to detect the fluxes and control their positions by discriminating from extrinsic lattice defects.

\begin{figure}[t]
  \begin{center}
  \includegraphics[width=\columnwidth,clip]{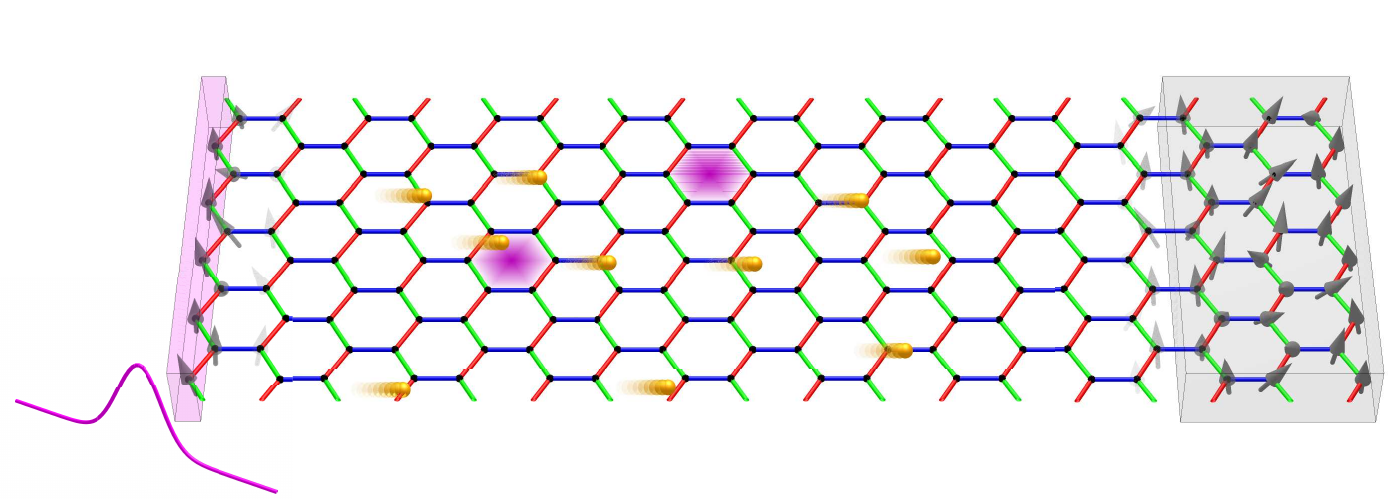}
  \caption{
    Schematic picture of spin transport carried by fractional Majorana quasiparticles (orange spheres), triggered by a time-dependent magnetic field in the magenta region at the left side.
    The flow of Majorana quasiparticles induces local spin moments in the gray region with a weak static magnetic filed.
    Excited fluxes are shown as purple hexagons.
  }
  \label{fig:intro}
  \end{center}
  \end{figure}

In this paper, we investigate the spin transport ascribed to the fractional Majorana quasiparticles in the presence of flux excitations.
We introduce the Kitaev model on a honeycomb lattice with two edges at which the time-dependent and static magnetic fields are applied, respectively (see Fig.~\ref{fig:intro}).
We consider a Gaussian magnetic-field pulse as the time-dependent field and examine the time evolution using a time-dependent Majorana mean-field (MF) theory, which we develop in the present work.
We calculate the time evolutions of the induced local spin moments and spin correlations in the presence of two fluxes or two lattice defects.
In their absence, the pulsed magnetic field yields the wavepacket of spin excitations at one of the edges, propagating to the other edges.
Unlike conventional magnon flows, local spin moments never appear in bulk during the propagation.
Instead, the modulation of local spin correlations accompanies the flow of the wavepacket in the Kitaev QSL.
When the wavepacket reaches the other edge, it induces the local spin moment with the delay determined by the velocity of the itinerant Majorana fermions.
We find that the spatial distributions of the spin correlations averaged along the wavefront remain largely intact both against the flux excitations and lattice defects.
However, spatial distributions along the wavefront after the fluxes scattering the wavepacket are distinctly different from the case with lattice defects.
The spatial variation of the scattered wave by the fluxes is more significant than the defects.
In particular, the variation is strongly enhanced by increasing the distance between the two fluxes while it is almost unchanged for the defect case.
The result suggests that the scattering phenomenon by the flux excitations is dominated by the change of gauge-like fields emerging on the string connecting between the fluxes, which is similar to the Aharonov-Bohm effect.
We also discuss the application of the present results to the observation of the flux excitations.

This paper is organized as follows. In the next section, we first introduce the Kitaev model and its Majorana fermion representation.
The local conserved quantities intrinsic in this model and the gauge structure related to the present work are also presented in Sec.~\ref{sec:model-flux}.
Moreover, we show another Majorana-fermion representation for the Kitaev model based on the Jordan-Wigner transformation to address the time-dependent magnetic field without the gauge redundancy in Sec.~\ref{sec:time-dep-hamil}.
In Sec.~\ref{sec:method}, we detail the numerical method to solve the time-dependent Hamiltonian in the Majorana representation.
We introduce the time-dependent MF theory for the Majorana fermion system based on the von Neumann equation for the single particle density matrix, which is significantly efficient for inhomogeneous cases like the system addressed in the current study.
The results are given in Sec.~\ref{sec:result}.
In Sec.~\ref{sec:sv-spin}, we present the calculation results for the time evolution of the spatial change of spin correlations and local spin moment caused by injecting a pulsed magnetic field at the edge of the system with flux excitations or lattice defects.
The pulse-intensity dependence is given in Sec.~\ref{sec:pulse-intensity-dep}.
We also examine the effect of scatterer configurations.
The dependence on the distance between the fluxes or defects are shown in Sec.~\ref{sec:distance-dep}.
In Sec.~\ref{sec:discussion}, we discuss the difference between fluxes and defects as the scatterers of the spin transport and applications to detections of the flux excitations in the Kitaev spin liquid.
Finally, Sec.~\ref{sec:summary} is devoted to the summary.

\section{Model}\label{sec:model}


In this study, we address the time evolution of magnetic excitations in the Kitaev model.
First, we introduce the Kitaev model and its local conserved quantity, flux, based on the formalism given originally by Kitaev.
Then, we explain the setup to clarify the spin transport via itinerant Majorana fermions in the presence of flux excitations.
This is formulated by the Jordan-Wigner transformation, different from Kitaev's original formalism.

\subsection{Kitaev model and its flux excitation}\label{sec:model-flux}

\begin{figure*}[t]
\begin{center}
\includegraphics[width=2\columnwidth,clip]{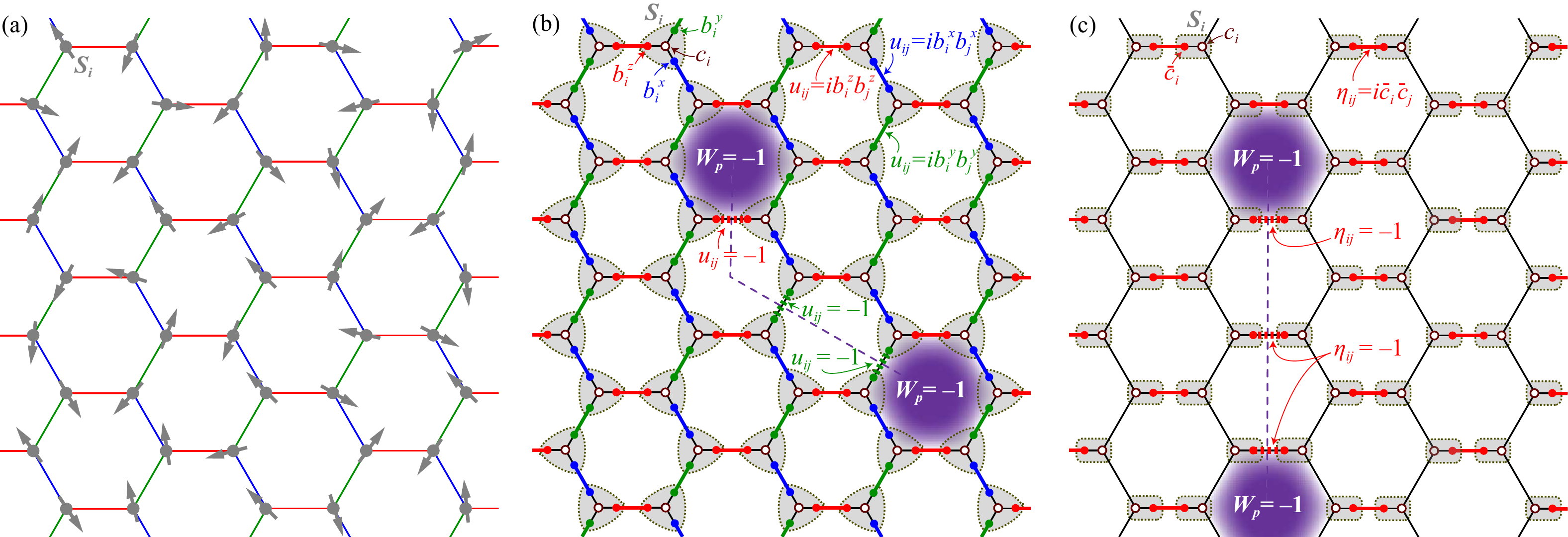}
\caption{
(a) Schematic figure of the Kitaev model on a honeycomb lattice.
On each lattice site, an $S=1/2$ spin is located.
Blue, green, and red lines represent the NN $x$, $y$ and $z$ bonds, respectively.
(b) Four-Majorana representation for the Kitaev model.
Each spin $\bm{S}_i$ is written by the Majorana fermions $\{c_i,b_i^x,b_i^y,b_i^z\}$ surrounded by a gray rounded triangle.
The $Z_2$ gauge field $u_{ij}=ib_i^\gamma b_j^\gamma$ on the $\gamma$ bond is also shown by a colored line.
The bold solid (dotted) lines stand for $u_{ij}=+1$ ($-1$).
Two fluxes with $W_p=-1$ are excited in the hexagons at the both ends of the purple dashed line crossing the bonds with $u_{ij}=-1$.
(c) Two-Majorana representation for the Kitaev model.
Each spin $\bm{S}_i$ is written by the Majorana fermions $\{c_i,\bar{c}_i\}$ surrounded by a gray rounded rectangle.
The $Z_2$ variables $\eta_{ij}$ are defined on the $z$ bonds.
The two excited fluxes denoted by the purple hexagons are located at the ends of the vertical dashed line crossing the dotted red bonds representing $\eta_{ij}=-1$.
}
\label{fig:majorana}
\end{center}
\end{figure*}

The Kitaev model is given by
\begin{align}
  {\cal H}_K=-J\sum_{\gamma=x,y,z}\sum_{\means{ij}_\gamma}S_i^\gamma S_j^\gamma,\label{eq:Kitaev}
\end{align}
where $S_i^\gamma$ is the $\gamma$ component of an $S=1/2$ spin defined on a honeycomb lattice, and $J$ is the exchange constant.
The nearest neighbor (NN) bonds of the honeycomb lattice are decomposed into three types, denoted by $\means{ij}_x$, $\means{ij}_y$, and $\means{ij}_z$, on which the $\gamma$ component of spins interact as Ising type [see Fig.~\ref{fig:majorana}(a)].
Kitaev showed that this model can be solved exactly by introducing four Majorana fermions, $\{c_i,b^x_i,b^y_i,b^z_i\}$, at each site~\cite{Kitaev2006}.
Using the Majorana fermions, an $S=1/2$ spin operator is represented by $S_i^\gamma=\frac{1}{2}ib_i^\gamma c_i$, and the Hamiltonian is written as
\begin{align}
  {\cal H}_K=\frac{J}{4}\sum_{\means{ij}_\gamma}u_{ij}i c_i c_j,\label{eq:Kitaev_gauge}
\end{align}
where $u_{ij}=i b_i^\gamma b_j^\gamma$ on the $\gamma$ bond, as shown in Fig.~\ref{fig:majorana}(b).
Note that $u_{ij}^2=1$, and all $u_{ij}$ commute with ${\cal H}_K$.
Thus, $u_{ij}$ is a $Z_2$ local conserved quantity defined on each bond.
In the Majorana representation, there is a redundancy for $u_{ij}$, and it cannot be written by the spin operator.
Therefore, $u_{ij}$ is regarded as the $Z_2$ gauge field.

The local flux is another $Z_2$ local covered quantity, which is attributed to the gauge field.
This is defined by a six product of spins on hexagon plaquette $p$ as
\begin{align}
  W_p=2^6 \prod_{i\in p} S_i^{\gamma_i},\label{eq:wp}
\end{align}
where $\gamma_i$ is the component determined by the type of bonds not belonging to the hexagon at site $i$.
This is written by using the six product of $u_{ij}$ on the edges of the hexagon $p$ as
\begin{align}
  W_p=\prod_{\means{ij}\in p} u_{ij},\label{eq:wp_gauge}
\end{align}
which is gauge-invariant.
Here, we introduce the $Z_2$ gauge potential $A_{ij}(=0,\pi)$ on each bond of the honeycomb lattice, which is defined such that $u_{ij}=\exp\left[iA_{ij}\right]$.
Correspondingly, $W_p$ is written as $W_p=\exp\left[i\Phi_p\right]$
with $\Phi_p=\sum_{\means{ij}\in p}A_{ij}$.
This equation recalls the relation between the flux and vector potential in electromagnetism, and therefore, $W_p=-1$, corresponding to $\Phi_p$ taking $\pi$, is termed ``flux''.
When the fluxes with $W_p=-1$ are present in the Kitaev QSL, the flux configuration is realized by flipping the $Z_2$ gauge fields $u_{ij}$ to $-1$ on the string connecting the excited fluxes, which is schematically shown in Fig.~\ref{fig:majorana}(b).
The changed gauge fields modulate the transfer integrals of the itinerant Majorana fermions $\{c_i\}$ in Eq.~\eqref{eq:Kitaev_gauge}.
These suggest that flux excitations not only affect the local structure of the Majorana fermions in the vicinity of the excited fluxes but also have an effect throughout the itinerant Majorana fermion system, which is considered as an analogy of the Aharonov-Bohm effect.
In particular, we expect that interference between the Majorana wavepackets propagating across the string connecting the excited fluxes and in the region without flipped gauge fields.
The observation of the interference effect could be a hallmark of excited fluxes, which are applicable to topological quantum computation.

\subsection{Time-dependent Hamiltonian with pulsed magnetic field}
\label{sec:time-dep-hamil}

\begin{figure}[t]
  \begin{center}
  \includegraphics[width=\columnwidth,clip]{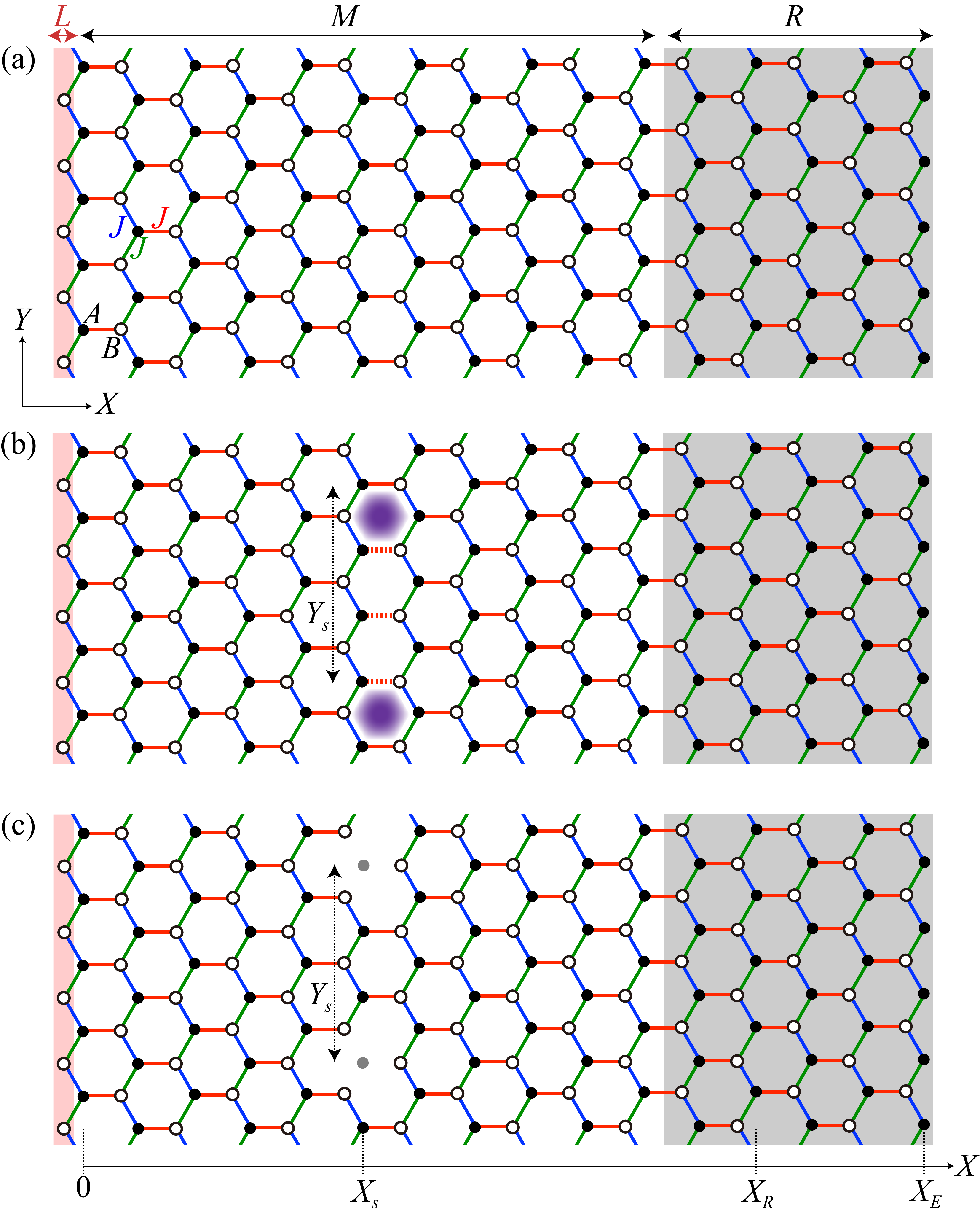}
  \caption{
    (a) Honeycomb lattice cluster with $2N_X N_Y$ sites addressed in the present calculations, where a periodic boundary condition is imposed along the zigzag chain consisting of $x$ and $y$ bonds, parallel to the $Y$ axis.
    The filled and open circles stand for the sites belonging to the $A$ and $B$ sublattices, respectively.
    The lattice sites are discussed in three parts: $L$, $M$, and $R$.
    The region $L$ includes the leftmost sites on the $B$ sublattice.
    The sites on the $N_{R}$ zigzag chains from the right side are defined to belong to the region $R$.
    The region $M$ is composed of the other sites.
    The origin of the $X$ coordinate is taken to the position of the leftmost sites in the $A$ sublattice.
    The $X$ coordinates of the rightmost site and the site belonging to the second chain from the left of the region $R$ are defined as $X_E$ and $X_R$, respectively.
    (b) The Kitaev model with two excited fluxes highlighted as purple with the distance $Y_s$ along $Y$ direction between their center.
    Flipped $Z_2$ variables $\eta_{ij}=-1$ are denoted by the dotted lines on $z$ bonds.
    The $X$ coordinate of the bottom left vertex of the hexagons with excited fluxes is defined as $X_s$.
    (a) Honeycomb lattice with two lattice defects shown by the gray circles in the sublattice $A$ with the distance $Y_s$ and their $X$ coordinate is $X_s$.
    The $X$ coordinates, $X_s$, $X_R$, and $X_E$, are given by $X_s=\frac{3}{2}\frac{N_X}{2}$, $X_R=\frac{3}{2}(N_X-N_R+1)$, and $X_E=\frac{3}{2}(N_X-1)$.
    The numerical calculations are performed for $N_X=N_Y=60$ and $N_{R}=10$.
  }
  \label{fig:lattice}
  \end{center}
\end{figure}

Since the Majorana fermion is electrically neutral, its transport phenomena is difficult to be observed directly.
Here, we focus on the spin transport in the Kitaev model.
By following the previous study~\cite{Minakawa2020}, we examine the propagation of the Majorana fermions excited by a pulsed magnetic field at an edge and observe the induced spin moment at the opposite edge.
To consider this setup, we impose the periodic boundary condition along the $Y$ direction, namely, the zigzag chain consisting of the $x$ and $y$ bonds including $2N_Y$ sites, as shown in Fig.~\ref{fig:lattice}(a).
We impose the open boundary condition such that two zigzag edges appear, where the number of the zigzag chains is $N_X$ and the total number of sites is $N=2N_X N_Y$ in the cluster.
Hereafter, we assume that the length of an NN bond on the honeycomb lattice is unity.

While Kitaev originally introduced four Majorana fermions at each site as explained before, we here use another approach to solve the Kitaev model.
This model given in Eq.~\eqref{eq:Kitaev} can be solved exactly by applying the Jordan-Wigner transformation and introducing two-types of Majorana fermions, $c_i$ and $\bar{c}_i$, at each site, as shown in Fig.~\ref{fig:majorana}(c)~\cite{PhysRevB.76.193101,PhysRevLett.98.087204,1751-8121-41-7-075001,PhysRevLett.113.197205}.
Since the honeycomb lattice is bipartite, this is divided into two sublattices, $A$ and $B$ in Fig.~\ref{fig:lattice}(a).
The spins on these sublattices are represented by 
\begin{align}
  S_i^x=\frac{1}{2}c_i\tau_i,\quad S_i^y=-\frac{1}{2}\bar{c}_i\tau_i,\quad S_i^z=\frac{1}{2}ic_i\bar{c}_i
  \quad {\rm for}\ i\in A,\\
  S_i^x=\frac{1}{2}\bar{c}_i\tau_i,\quad S_i^y=-\frac{1}{2}c_i\tau_i,\quad S_i^z=\frac{1}{2}i\bar{c}_i c_i
  \quad {\rm for}\ i\in B,
\end{align}
where $\tau_i$ is the string operator defined on each chain consisting of $x$ and $y$ bonds along the $Y$ direction, which is given by
\begin{align}
  \tau_i=\prod_{j<i}\left(-2S_j^z\right).
\end{align}
Here, we assume that the site numbering is made for the $-Y$ direction and neglect an additional effect originating from the periodic boundary condition in the Jordan-Wigner transformation, which appears on the bonds across the boundary.
Using this transformation, Eq.~\eqref{eq:Kitaev} is rewritten as
\begin{align}
  {\cal H}_K=-\frac{J}{4}\sum_{\gamma=x,y}\sum_{[ij]_\gamma}ic_i c_j
  -\frac{J}{4}\sum_{[ij]_z}ic_i c_j i\bar{c}_i \bar{c}_j,\label{eq:Majorana_H}
\end{align}
where $[ij]_\gamma$ denotes the ordered NN pair with $i$ ($j$) belonging to $A$ ($B$) sublattice.
In this model, $\eta_{ij}=i\bar{c}_i \bar{c}_j$ on each $z$ bond commutes with ${\cal H}_K$, and therefore, this is regarded as a classical number taking $\pm 1$.
Then, the Hamiltonian is mapped onto a free Majorana fermion model.
The local conserved quantity $W_p$ in Eq.~\eqref{eq:wp} is written as the product of two $\eta_{ij}$ on the hexagonal plaquette $p$,
\begin{align}
  W_p=\prod_{\means{ij}\in p} \eta_{ij},
\end{align}
which is similar to Eq.~\eqref{eq:wp_gauge}.
Note that, in the approach using Jordan-Wigner transformation, there is no macroscopic redundancy in $\eta_{ij}$, which are defined only on the $z$ bonds.
This allows us to proceed calculations without a local gauge fixing or projection onto the physical space, which is essential in the Kitaev's original representation.
Therefore, we use the Jordan-Wigner approach in this study.
However, this has a drawback.
Figure~\ref{fig:majorana}(c) is a schematic picture for the Jordan-Wigner transformation with two Majorana fermions introduced at each site.
As shown in this figure, local conserved quantities are absent on $x$ and $y$ bonds, which avoids redundancy.
On the other hand, in this representation, $\eta_{ij}$ are defined only on $z$ bonds, and we assume the periodic boundary condition along the $Y$ direction.
These conditions restrict flux excitations with $W_p=-1$ to pair creations along the zigzag direction, namely the $Y$ axis, by flipping $\eta_{ij}$ to $-1$ on the straight line connecting between the two fluxes.
Thus, in the present study, we consider the flux configuration shown in Fig.~\ref{fig:lattice}(b) with the distance $Y_s$.

To examine the effect triggered by a spin pumping at a edge, we introduce magnetic fields around edges.
Here, we consider the time-dependent magnetic field $h(t)$ in the region $L$ and the static field $h_s$ along the $S^z$ direction in the region $R$, which includes $N_R$ zigzag chains, as shown in Fig.~\ref{fig:lattice}(a).
The total Hamiltonian is written as
\begin{align}
  {\cal H}(t)&={\cal H}_K -h(t)\sum_{i\in L}S_i^z - h_s\sum_{i\in R}S_i^z\nonumber\\
  &={\cal H}_K - \frac{h(t)}{2}\sum_{i\in L} i\bar{c}_i c_i
  - \frac{h_s}{2}\left[\sum_{i\in A\ {\rm in}\ R} ic_i \bar{c}_i +\sum_{i\in B\ {\rm in}\ R} i\bar{c}_i c_i \right].
\end{align}
Note that the introduction of the magnetic field violates the exact solvability in the Kitaev model because of a mixing between two Majorana fermions $c$ and $\bar{c}$ originating from the magnetic field, and $W_p$ and $\eta_{ij}$ are no longer conserved quantities.
Nevertheless, as the mixing is local, these quantities are conserved in the region $M$ shown in Fig.~\ref{fig:lattice}.
Therefore, in this region, an excited flux $W_p=-1$ is well defined even in the presence of $h(t)$ and $h_s$.

We here consider the transient spin dynamics in the presence of excited fluxes at $X_s$ for the $X$ axis with the distance $Y_s$, as shown in Fig.~\ref{fig:lattice}(b).
To reveal characteristics intrinsic in fluxes as scatterers for spin propagation, we also examine the case with lattice defects instead of fluxes, as shown in Fig.~\ref{fig:lattice}(c).
Defects are introduced as vacancies of spins in the lattice, which are implemented by setting the exchange constants on the bonds connecting each defect to zero.
In the present study, we only focus on the situation that two defects are located on the $A$ sublattice with distance $Y_s$.

\section{Method}\label{sec:method}

Since the second term in Eq.~\eqref{eq:Majorana_H} describes the interaction between Majorana fermions, the Hamiltonian under the magnetic field cannot be solved exactly.
To analyze the Hamiltonian, we apply the MF approximation to the interaction terms as
\begin{align}
  ic_i c_j i\bar{c}_i \bar{c}_j \simeq 
  \means{ic_i c_j} i\bar{c}_i \bar{c}_j +ic_i c_j \means{i\bar{c}_i \bar{c}_j}-\means{ic_i c_j} \means{i\bar{c}_i \bar{c}_j}\nonumber\\
  -\means{ic_i \bar{c}_i} i c_j \bar{c}_j-ic_i \bar{c}_i \means{i c_j \bar{c}_j}
  +\means{ic_i \bar{c}_i} \means{i c_j \bar{c}_j}\nonumber\\
  -\means{ic_i\bar{c}_j} i\bar{c}_i c_j -ic_i\bar{c}_j \means{i\bar{c}_i c_j}
  +\means{ic_i\bar{c}_j} \means{i\bar{c}_i c_j }.\label{eq:mf_decoupling}
\end{align}
Using the MF decouplings, the original Hamiltonian ${\cal H}$ is approximately written as a bilinear form of Majorana fermions, which is symbolically given by 
\begin{align}
  {\cal H}_{\rm MF}=\frac{i}{4}\sum_{kk'}\gamma_k \mathbf{A}_{kk'}\gamma_{k'}+C,\label{eq:MFmat}
\end{align}
where $\mathbf{A}$ is a $2N\times 2N$ skew-symmetric matrix, $C$ is the constant originating from the remnants of the MF decouplings in Eq.~\eqref{eq:mf_decoupling}, and $\{\gamma_k\}=\{c_1,c_2, \cdots, c_N, \bar{c}_1,\bar{c}_2, \cdots, \bar{c}_N\}$ are Majorana fermions, which satisfy the anticommutation relation $\{\gamma_k,\gamma_{k'}\}=2\delta_{kk'}$.
When $h(t)=0$, the Hamiltonian is not time-dependent, and the MFs $\bras{\Psi_0}\gamma_k\gamma_{k'}\kets{\Psi_0}$ can be determined self-consistently, where $\kets{\Psi_0}$ is the ground state of ${\cal H}_{\rm MF}$.
This procedure is achieved by diagonalizing the matrix $\mathbf{A}$ in Eq.~\eqref{eq:MFmat} as 
\begin{align}
  U^\dagger i\mathbf{A} U=\Lambda\equiv \{\varepsilon_1,\varepsilon_2,\cdots,\varepsilon_N\},
\end{align}
where $\varepsilon_\lambda$ is the positive eigenvalues of $\mathbf{A}$.
Note that $\mathbf{A}$ also possesses negative eigenvalues with opposite sign, $-\varepsilon_\lambda$, because $\mathbf{A}$ is skew-symmetric.
$U$ is a $2N\times N$ matrix satisfying $U^\dagger U=1$.
The diagonal form of the MF Hamiltonian is given by
\begin{align}
  {\cal H}_{\rm MF}=\sum_{\lambda=1}^N\varepsilon_\lambda\left(f_{\lambda}^\dagger f_{\lambda}-\frac{1}{2}\right)+C,
\end{align}
where $\gamma_k=\sqrt{2}\sum_{\lambda=1}^N\left(U_{k\lambda}f_\lambda + U_{k\lambda}^\dagger f_\lambda^\dagger\right)$.
Here, we introduce the density matrix $\rho$ as
\begin{align}
  \rho_{kk'}^0=\frac{i}{4}\left(\bras{\Psi_0}\gamma_{k'}\gamma_k\kets{\Psi_0}-\delta_{kk'}\right).
\end{align}
Note that $\rho_{kk'}^0$ is a real skew-symmetric matrix, whose coefficient is introduced such that the MF energy is simply written as $\bras{\Psi_0}{\cal H}_{\rm MF}\kets{\Psi_0}={\rm Tr} [\rho \mathbf{A}]+C$.
Using the the transform matrix $U$, the density matrix is given by
\begin{align}
  \rho_{kk'}^0=\frac{1}{2}{\rm Im}\left(UU^\dagger\right)_{kk'}.
\end{align}
Since $U$ depends on $\rho_{kk'}$, which is a set of the MFs, and thus, the above is nothing but the self-consistent equation.

In the presence of the time-dependent field $h(t)$, the density matrix is also time dependent.
In this case, the density matrix $\rho(t)$ is defined as
\begin{align}
  \rho_{kk'}(t)=\frac{i}{4}\left(\bras{\Psi(t)}\gamma_{k'}\gamma_k\kets{\Psi(t)}-\delta_{kk'}\right),
\end{align}
where $\kets{\Psi(t)}$ is time-dependent wave function, which is determined by the following Schr\"odinger equation:
\begin{align}
  i\frac{\partial}{\partial t}\kets{\Psi(t)}={\cal H}_{\rm MF}(t)\kets{\Psi(t)},\label{eq:schrodinger}
\end{align}
where we assume that $\hbar$ is unity.
This Schr\"odinger equation gives the following von Neumann equation:
\begin{align}
  i\frac{\partial \rho_{kk'}(t)}{\partial t}=\sum_{k''}\left(i\mathbf{A}_{kk''}\rho_{k''k'}-\rho_{kk''} i\mathbf{A}_{k''k'}\right)
  =[i\mathbf{A},\rho(t)]_{kk'}.\label{eq:Neumann}
\end{align}
Since the Hamiltonian matrix $i\mathbf{A}$ is a function of $t$ and $\rho$, the time-evolution of $\rho$ can be evaluated by the Runge-Kutta method.

Another way to compute the time evolution of MFs is based on the Schr\"odinger equation given in Eq.~\eqref{eq:schrodinger}.
In this method, we need to perform the diagonalization of $i\mathbf{A}$ at every time step.
On the other hand, the present method based on the von Neumann equation, which is originally introduced for complex fermion systems~\cite{volkov1973collisionless,Tsuji_theoryof2015,Murakami_collective2020}, demands only the matrix product of $\rho$ and $i\mathbf{A}$, as shown in Eq~\eqref{eq:Neumann}.
Moreover, the calculation cost can be reduced using the property of $i\mathbf{A}$ being usually a sparse matrix.
This approach is highly efficient for the case with larger clusters such as the system addressed in the present study.

\section{Result}\label{sec:result}

\begin{figure*}[t]
  \begin{center}
  \includegraphics[width=2\columnwidth,clip]{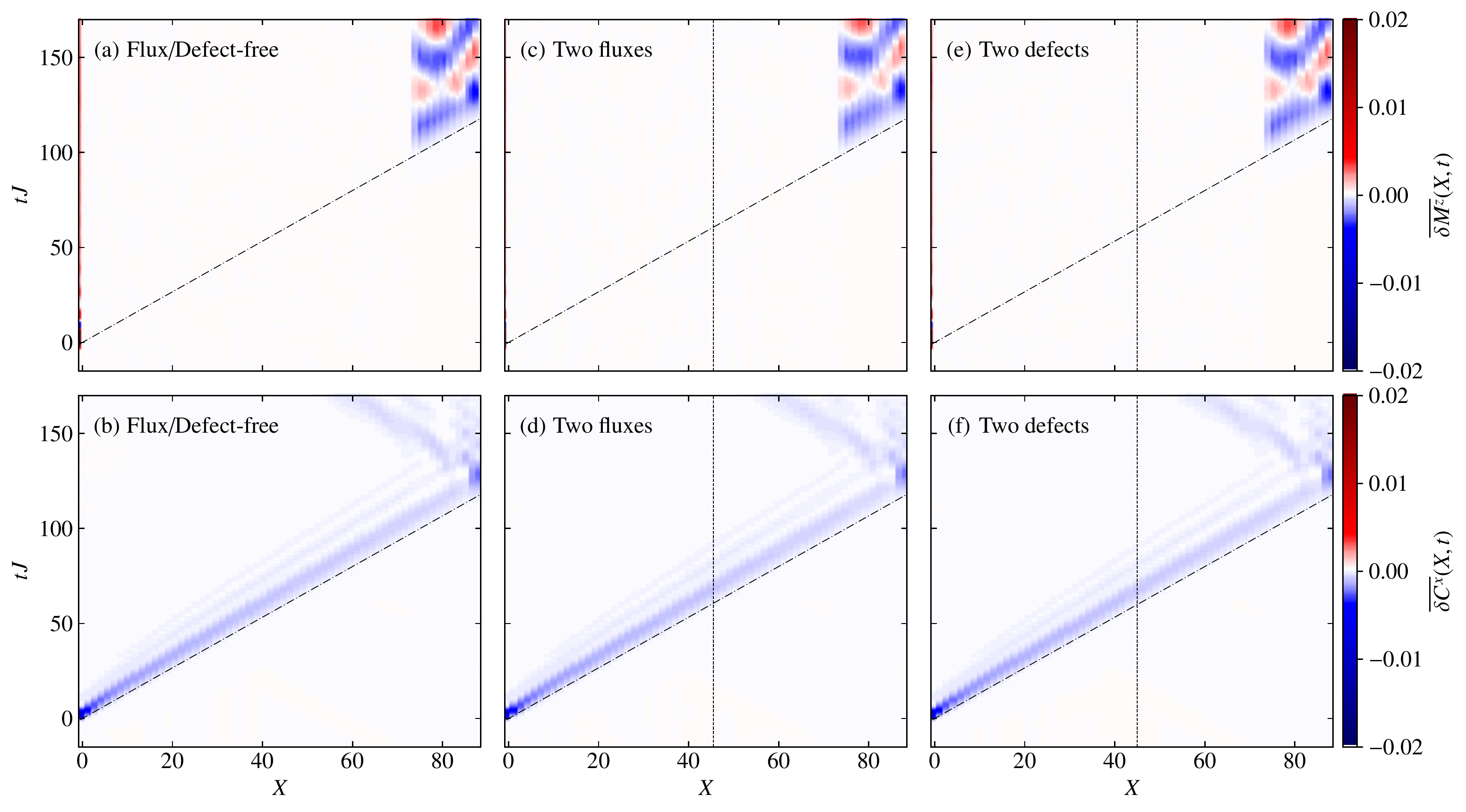}
  \caption{
    Spatiotemporal maps of the local spin moment $\overline{\delta M^z} (X,t)$ and spin correlations $\overline{\delta C^x} (X,t)$ averaged along the $Y$ direction for (a),(b) the absence of fluxes and lattice defects, (c),(d) the presence of the two fluxes, and (e),(f) the presence of the two lattice defects.
    The dashed-dotted lines represents $t=X/v$ with $v=3/4$ being the velocity of the linear dispersion of the itinerant Majorana fermions. 
    The vertical dotted lines in (c)--(f) indicate $X_s$, which is the position of fluxes or defects.
    The pulse intensity is set to $A=1$, and the positions of fluxes or defects are given by $(X_s,Y_s)=(45,30\sqrt{3})$.
    }
\label{fig:ave}
\end{center}
\end{figure*}

In this section, we show the results for the transient dynamics triggered by the time dependent magnetic field, which we choose as a Gaussian form given by
\begin{align}
  h(t)=\frac{A}{\sqrt{2\pi}\sigma}\exp\left[-\frac{t^2}{2\sigma^2}\right],
\end{align}
where $A$ and $\sigma$ are the amplitude and width of the pulsed magnetic field, respectively.
From now on, we set $\sigma=2/J$.

In our calculations, we address three situations: (i) the flux-free and defect-free case [Fig.~\ref{fig:lattice}(a)], (ii) the case with two fluxes are excited in the initial state [Fig.~\ref{fig:lattice}(b)], (iii) the flux-free case with two defects [Fig.~\ref{fig:lattice}(c)] on the cluster with $N_X=N_Y=60$.
As mentioned before, $W_p$ is conserved quantity in the region $M$, we can selectively excite $W_p$ to $-1$ by flipping $\eta_{ij}$ into $-1$ on the line connecting two fluxes in the initial state.
The corresponding $\eta$ are shown by the dotted lines in Fig.~\ref{fig:lattice}(b).
In the case (iii), we introduce two lattice defects in the sublattice $A$.
Note that previous studies pointed out the possibility that a vacancy traps an excited flux with a significantly small trapping energy~\cite{Willans2010,Willans2011}.
In the present study, we do not consider this effect, and the flux-free state is only assumed in the existence of lattice defects to compare the effects of fluxes and defects separately on the scattering process of the itinerant Majorana fermions.

The time-dependent simulation starts from $t_0 =-20/J$ using an initial state obtained by the equilibrium MF approximation, where $\rho^0$ are determined self-consistently in the absence of $h(t)$.
Namely, $\rho(t_0)=\rho^0$.
The time evolution is performed by discretizing time in Eq.~\eqref{eq:Neumann} and using the fourth-order Runge-Kutta method.
In the simulation, the time step is chosen to $\Delta t=0.1/J$.

\subsection{Spatiotemporal variation of spin states}
\label{sec:sv-spin}

\begin{figure*}[t]
  \begin{center}
  \includegraphics[width=2\columnwidth,clip]{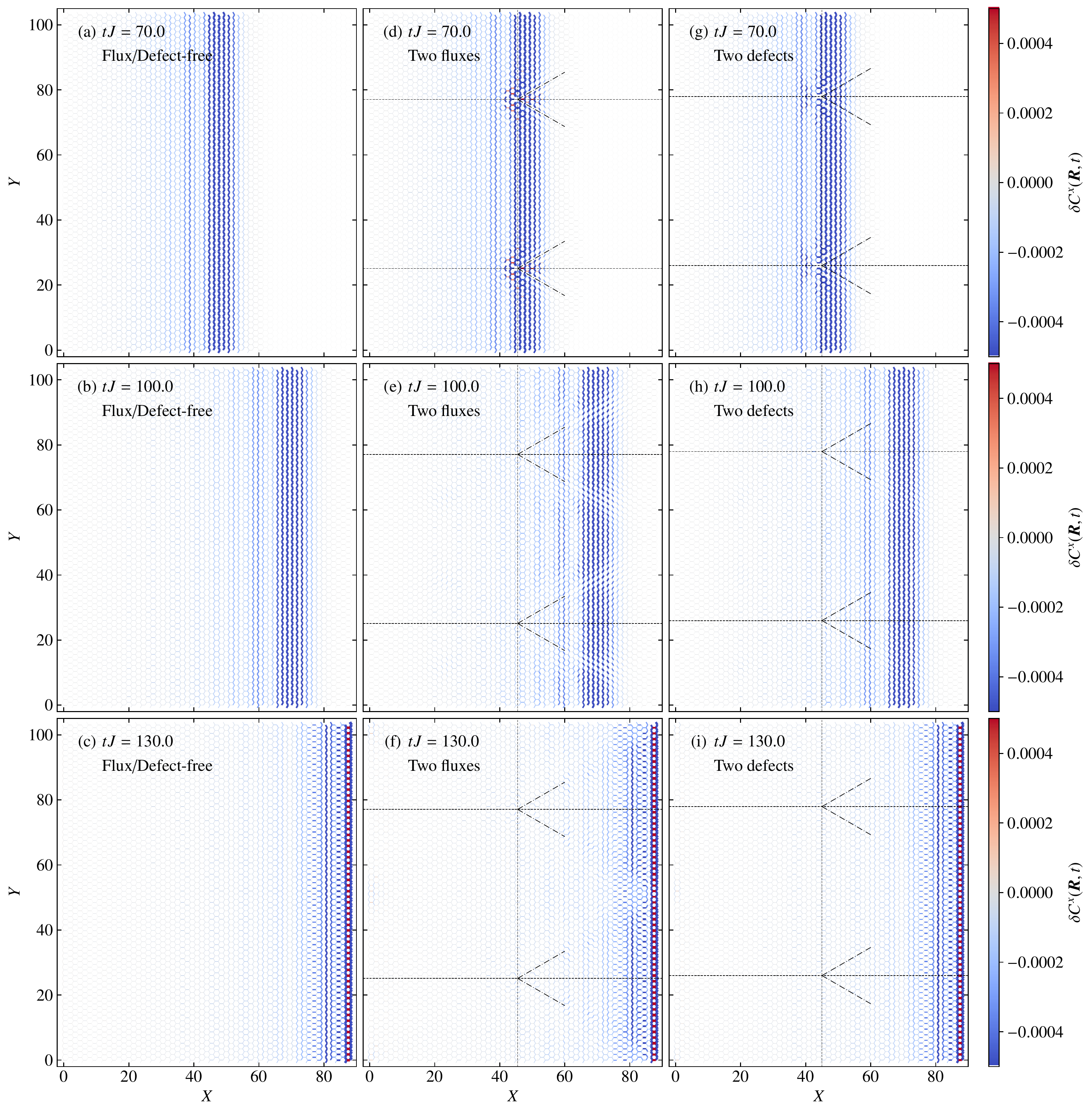}
  \caption{
    Spatial map for the change of the spin correlation on the $x$ bonds from the initial time $t=t_0$ at several $t$ in (a)--(c) the absence of fluxes and lattice defects, (d)--(f) the presence of the two fluxes, and (g)--(i) the presence of the two lattice defects.
    In (d)--(i), the fluxes or lattice defects are located at the crossing points of the dotted lines.
    The dashed-dotted lines in (d)--(i) indicate the $\pm 30$-degree directions from the scatterers.
    The pulse intensity is set to $A=1$, and the positions of fluxes or defects are given by $(X_s,Y_s)=(45,30\sqrt{3})$.
  }
\label{fig:map_bond}
\end{center}
\end{figure*}

We calculate the time evolutions of the local spin moment defined by
\begin{align}
  M^z(\bm{R}_i,t)=\means{S_i^z}(t)
\end{align}
with $\bm{R}_i=(X_i,Y_i)$ being the site position and NN spin correlation
\begin{align}
  C^\gamma(\bm{R}_{ij},t)=\means{S_i^\gamma S_j^\gamma}(t)
\end{align}
on the $\gamma$ bond $\means{ij}_\gamma$ with $\bm{R}_{ij}=(\bm{R}_i+\bm{R}_j)/2$.
First, to see the overall structure, we show the spatiotemporal map of the local moment and spin correlation averaged along the $Y$ direction on the plane of $X$ and $t$.
We evaluate the changes from the initial value at $t=t_0$, which are introduced by
\begin{align}
  \overline{\delta M^z} (X_i,t) = \frac{1}{N_Y}\sum_{Y_i}\delta M^z(\bm{R}_i,t),
\end{align}
and
\begin{align}
  \overline{\delta C^x} (X_i,t) = \frac{1}{N_Y}\sum_{Y_i}\delta C^x(\bm{R}_i,t),
\end{align}
where $\delta {\cal O}(t) = {\cal O}(t)-{\cal O}(t_0)$.

Figure~\ref{fig:ave} shows the spatiotemporal map of these quantities, where the amplitude of the pulsed magnetic field at the left edge is assumed to be $A=1$.
Figure~\ref{fig:ave}(a) shows the time evolution of the local spin moment $\overline{\delta M^z} (X,t)$ in the absence of excited fluxes and lattice defects.
Note that, in this case, no spatial variation appears along the $Y$ direction.
After the magnetic pulse is introduced at $t=0$, the local moments begin to change in the region $R$ with some time delay but they do not change in the bulk region $M$.
Figure~\ref{fig:ave}(b) shows the time evolution of $\overline{\delta C^x} (X,t)$.
The magnetic pulse causes the negative change of it, which propagates to the right side.
We note that $S_i^x S_j^x=\frac{1}{4}c_i c_j$ on the $x$ bond, indicating that its change is regarded as a wavepacket of the itinerant Majorana fermion.
Thus, the velocity of the propagation is determined by the low-energy dispersion of the Majorana fermion system, as discussed in Ref.~\cite{Minakawa2020}.
In the present study, we choose the length of the NN bond on the honeycomb lattice to be unity, and the velocity determined from the linear dispersion of the fermionic band is $v=3J/4$.
The dashed-dotted lines in Fig~\ref{fig:ave} represent the line of $x=vt$.
The results clearly indicate that the Majorana wavepacket created by the pulsed magnetic field at the left edge travels to the other side and yields the change of the spin moments at the right edge~\cite{Minakawa2020}.
As pointed out in Ref.~\cite{Minakawa2020}, the propagation should be long-ranged because of the gapless excitation of the Majorana fermions.
In other words, the long-ranged spin correlation is present between the edges although the short-ranged spin correlation is only permitted in the bulk due to the existence of the local conserved quantities.

In the case without excited fluxes and lattice defects, the Majorana wavepacket created at the left edge is uniform along the $Y$ direction.
We expect that inhomogeneity of the wavepacket along the $Y$ direction should appear in the presence of excited fluxes or defects.
We show the spatiotemporal maps for $\overline{\delta M^z} (X,t)$ and $\overline{\delta C^x} (X,t)$ in Figs~\ref{fig:ave}(c) and \ref{fig:ave}(d) [Figs~\ref{fig:ave}(e) and \ref{fig:ave}(f)] for the case with two fluxes (two defects), whose $X$ position, $X_s$, is indicated by the vertical dashed line.
In the present calculations, $(X_s,Y_s)=\frac{1}{2}(\frac{3N_x}{2},\sqrt{3}N_Y)=(45,30\sqrt{3})$ both for the cases with two fluxes and two defects.
Contrary to the expectation, these quantities remain almost unchanged.
This result suggests that the presence of the fluxes and defects does not strongly affect the averaged value.

To reveal the inhomogeneous nature due to the fluxes and defects, we show the spatial distribution of $\delta C^x (\bm{R},t)$ in Fig.~\ref{fig:map_bond}.
Figures~\ref{fig:map_bond}(a)--\ref{fig:map_bond}(c) present the snapshots at $tJ=70$, $100$, and $130$, respectively, in the absence of fluxes and defects.
The plane wavepacket moves forward to the right side without change along the $Y$ direction.
In the presence of the fluxes or defects, scattering phenomena should be observed.
Figures~\ref{fig:map_bond}(d)--\ref{fig:map_bond}(f) show the time evolution of the snapshot for $\delta C^x (\bm{R},t)$ in the presence of the two fluxes, which are located at the crossing points between the dashed lines [see also Fig.~\ref{fig:lattice}(b)].
As shown in Fig.~\ref{fig:map_bond}(d), the wavepacket is scattered by the fluxes, and the scattered waves appear to propagate to the $\pm 30$-degree directions from the fluxes, as shown in Fig.~\ref{fig:map_bond}(e).
These angles correspond to the directions of the zigzag chains on the honeycomb lattice (see Fig.~\ref{fig:lattice}).
The inhomogeneity along the $Y$ direction survives even at the time where the wavepacket reaches the right edge [Fig.~\ref{fig:map_bond}(f)].
We find that $\delta C^x (\bm{R},t)$ around the middle point of the flux positions is strongly suppressed due to the superposition of the scattered waves from two fluxes as shown in Fig.~\ref{fig:map_bond}(f).

\begin{figure*}[t]
  \begin{center}
  \includegraphics[width=2\columnwidth,clip]{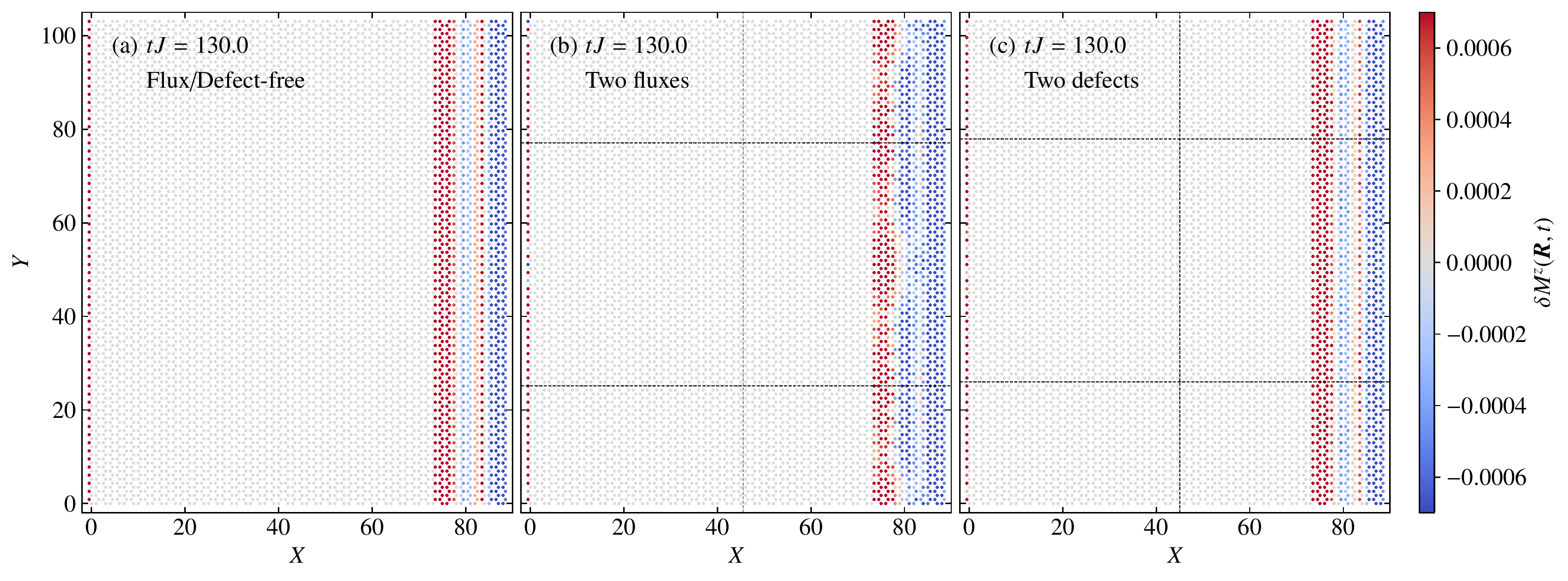}
  \caption{
    Spatial map for the change of the local spin moments from the initial time $t=t_0$ at several $t$ in (a) the absence of fluxes and lattice defects, (b) the presence of the two fluxes, and (c) the presence of the two lattice defects.
    In (b) and (c), the fluxes or lattice defects are located at the crossing points of the dotted lines.
    The pulse intensity is set to $A=1$, and the positions of fluxes or defects are given by $(X_s,Y_s)=(45,30\sqrt{3})$.
  }
\label{fig:map_mag}
\end{center}
\end{figure*}

To clarify whether this scattering phenomenon is intrinsic to the excited fluxes, we examine the case with lattice defects.
The time evolution of $\delta C^x (\bm{R},t)$ in the presence of two defects is shown in Figs.~\ref{fig:map_bond}(g)--\ref{fig:map_bond}(i).
The crossing points of the vertical and horizontal dashed lines indicate the positions of the lattice defects, which are the same as those of the two fluxes [see also Figs.~\ref{fig:lattice}(b) and \ref{fig:lattice}(c)].
Figure~\ref{fig:map_bond}(g) presents the spatial map of $\delta C^x (\bm{R},t)$ just after the scattering of the wavepacket by the defects.
The spatial distribution is similar to that in the case with the two fluxes shown in Fig.~\ref{fig:map_bond}(d).
Nonetheless, a distinct difference between the spatial distributions for the cases with defects and fluxes appears after a sufficient time has elapsed from the scattering.
Figure~\ref{fig:map_bond}(h) shows $\delta C^x (\bm{R},t)$ at $tJ=100$ in the presence of the two lattice defects.
As shown in this figure, the spatial homogeneity along the $Y$ direction appears to be almost recovered at this time, compared to the case with two fluxes shown in Fig.~\ref{fig:map_bond}(e).
The homogeneity of the wavefront is retained when the wavepacket reaches the right edge.
These results suggest that the scattering processes by the fluxes and defects are clearly different; the plane wavepacket generated at the left edge is strongly affected by excited fluxes in comparison with the case with defects.
This difference is understood as follows:
In the case with lattice defects, the transfer integrals of the Majorana fermions are modulated only around the defects.
On the other hand, for the flux case, the transfer integrals changes not only in the vicinity of the excited fluxes but also on the string connecting the excited fluxes, as explained in Sec.~\ref{sec:model-flux}.
Thus, nonlocal modulations for the transfer integrals are yielded by the flux excitations, which leads to a substantial spatial change of the Majorana wavepacket shown in Figs.~\ref{fig:map_bond}(e) and \ref{fig:map_bond}(f).
The nonlocal change due to the flux excitations is interpreted as a similar phenomenon to the Aharonov-Bohm effect, which is also discussed in Sec.~\ref{sec:discussion}.
We find that the phenomenon for the Majorana fermions can be observed by the spin transport in the Kitaev spin liquids with flux excitations.

Since the direction observation of $\delta C^x (\bm{R},t)$ is a difficult issue in experiments, we calculate the spatial dependence of the local spin moments.
Figure~\ref{fig:map_mag} shows the spatial distributions of $\delta M^z (\bm{R},t)$ at $tJ=130$, corresponding to the time when the wavepacket arrives at the right edge.
In the case without excited fluxes and defects, the local spin moments are uniformly induced along the $Y$ direction at the right edge, as shown in Fig.~\ref{fig:map_mag}(a).
On the other hand, for the case with the two-flux excitations, we observe an inhomogeneity for the induced local moments [Fig.~\ref{fig:map_mag}(b)].
In particular, a significant change appears around the middle $Y$ point of the flux positions, which arises due to the suppression of $\delta C^x (\bm{R},t)$ observed in Fig.~\ref{fig:map_bond}(f), as mentioned before.
In contrast to the flux case, the spatial modulation along the $Y$ direction is significantly weak for the case with two lattice defects [Fig.~\ref{fig:map_mag}(c)].
The results imply that the difference between the flux and defect scatterings can be deduced from a spatial change of the induced spin moments. 

\begin{figure*}[t]
  \begin{center}
  \includegraphics[width=2\columnwidth,clip]{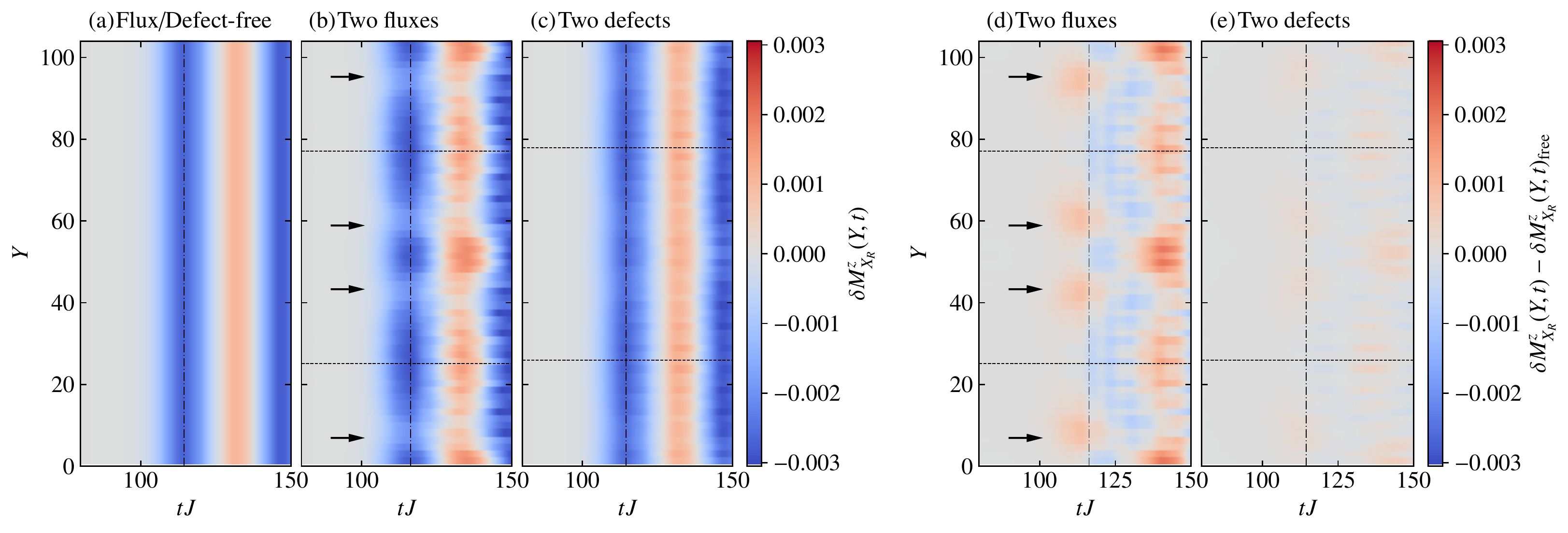}
  \caption{
   (a)--(c) Spatiotemporal map of the induced spin moments at $X=X_R$ in the region $R$ in (a) the absence of fluxes and lattice defects, (b) the presence of the two fluxes, and (c) the presence of the two lattice defects.
  Plots in the presence of two-fluxes and two-defects as the difference from the flux/defect-free case are also given in (d) and (e), respectively.
The horizontal dotted lines represent the $Y$ coordinates of the flux/defect positions.
The vertical dashed-dotted line indicates the time $t=t_m$ with $\left|\means{\delta M^z_{X_R}}_Y(t)\right|$ taking the first peak.
    The pulse intensity is set to $A=1$, and the positions of fluxes or defects are given by $(X_s,Y_s)=(45,30\sqrt{3})$.
    The arrows in (b) and (d) indicate $Y=Y_s\pm \Delta Y$, where $\Delta Y=(X_R-X_s)/\sqrt{3}$.
  }
\label{fig:tdep_map}
\end{center}
\end{figure*}

To see the time evolution of the induced spin moments in detail, we evaluate $\delta M^z (\bm{R},t)$ at $X_R=\frac{3}{2}(N_X-N_R+1)$, corresponding to the left side of the region $R$ (see Fig.~\ref{fig:lattice}).
Figures~\ref{fig:tdep_map}(a)--\ref{fig:tdep_map}(c) show the spatiotemporal map of $\delta M^z (\bm{R},t)$, which is given by
\begin{align}
  \delta M^z_{X_R}(Y,t)=\delta M^z (\bm{R}=(X_R,Y),t).\label{eq:MzXR}
\end{align}
Figure~\ref{fig:tdep_map}(a) shows $\delta M^z_{X_R}(Y,t)$ for the case without excited fluxes and lattice defects.
The change of it appears after $tJ\sim 100$, which corresponds to $X_R/v$.
It is $Y$-independent because of the absence of scatters for the transport of the Majorana fermions.
This feature disappears by the introduction of flux excitations.
The case with the two fluxes is shown in Fig.~\ref{fig:tdep_map}(b), where the horizontal dashed lines represent the $Y$ positions of the fluxes.
As shown in this figure, suppressions of $\delta M^z$ are found in four regions [see the arrows in Fig.~\ref{fig:tdep_map}(b)].
The reductions in the upper two regions originate from the upper excited flux because the middle point of these $Y$ positions coincides with the flux position, and those in the lower ones from the flux with the lower position.
Note that the upper two dips are $\sim 40$ apart from each other in the $Y$ direction.
This is understood as follows: The two scattered waves yielded from each flux propagate to the $\pm 30$-degree directions from the $X$ axis, as discussed before.
We estimate the distance between the two waves at $X_R$ as $2\Delta Y=\frac{2}{\sqrt{3}}(X_R-X_s)\sim 36$ in the present cluster, which is close to the distance of the two dips.
Note that the arrows in Fig.~\ref{fig:tdep_map}(b) indicate $Y=Y_s\pm \Delta Y$, which coincide with the dips.
We expect that these dips become deeper when the distance between fluxes is $Y_s\sim 36$.
The distance dependence will be discussed in Sec.~\ref{sec:distance-dep} in detail.
On the other hand, for the case with the two lattice defects, the induced spin moments at $X=X_R$ appears to be almost $Y$ independent while small structure whose scale is comparable to the lattice length is observed as shown in Fig.~\ref{fig:tdep_map}(c).
To see the difference between the flux and defect more clearly, we present the data from which the values of the flux/defect-free case are subtracted.
As shown in Figs.~\ref{fig:tdep_map}(d) and \ref{fig:tdep_map}(e), we find a lager spatial change in the case with flux excitations.
The difference of maximum and minimum values in Fig.~\ref{fig:tdep_map}(d) is about four times larger than that in Fig.~\ref{fig:tdep_map}(e).
We note that the first peaks indicated by the vertical dashed-dotted lines in Fig.~\ref{fig:tdep_map} is located at the almost same time.

\subsection{Pulse-intensity dependence}
\label{sec:pulse-intensity-dep}

\begin{figure*}[t]
  \begin{center}
  \includegraphics[width=2\columnwidth,clip]{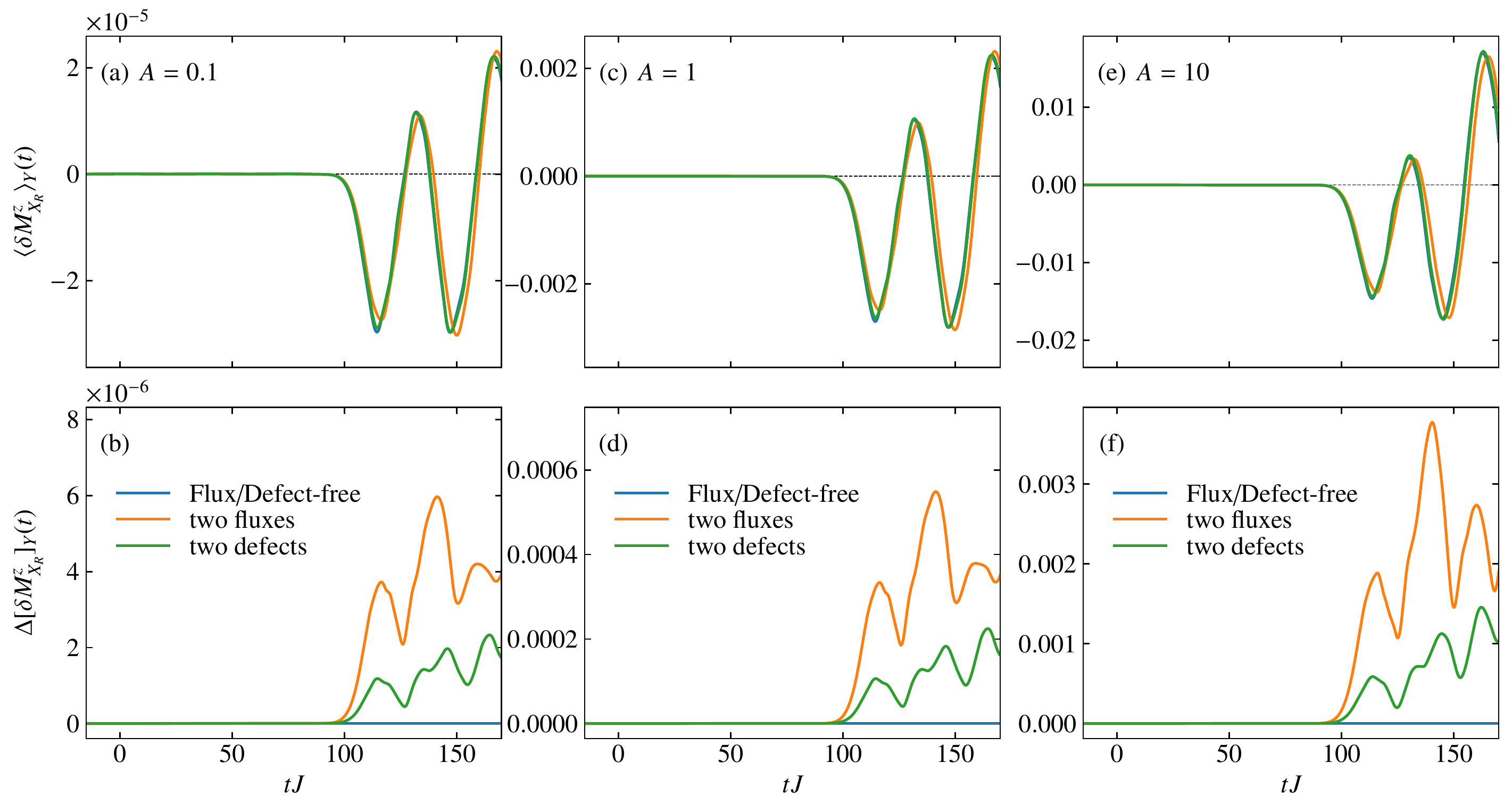}
  \caption{
    (a),(b) Time evolution of the spatial average and fluctuation of the induced spin moment at $X=X_R$ along the $Y$ direction for the pulse intensity $A=0.1$.
    The corresponding plots for (c),(d) $A=1$ and (e),(f) $A=10$.
  }
\label{fig:tdep_sd}
\end{center}
\end{figure*}

So far, we have discussed the results by fixing the shape of the magnetic field pulse with the amplitude $A=1$.
In this section, we examine the dependence of the pulse intensity $A$.
Figure~\ref{fig:tdep_sd} shows the time evolutions of the spatial distributions along the $Y$ direction for the induced spin moments at $X=X_R$ at several values of $A$.
Figures~\ref{fig:tdep_sd}(c) and \ref{fig:tdep_sd}(d) present the average and spatial variance of $\delta M^z_{X_R}(Y,t)$ along the $Y$ axis at $A=1$, which are defined as
\begin{align}
  \means{\delta M^z_{X_R}}_Y(t)= \frac{1}{N_Y}\sum_{Y_i}\delta M^z_{X_R}(Y_i,t)
\end{align}
and
\begin{align}
  \Delta[\delta M^z_{X_R}]_Y(t)= \sqrt{\mean{\left(\delta M^z_{X_R}-\means{\delta M^z_{X_R}}_Y\right)^2}_Y},
  \label{eq:Mz_sd}
\end{align}
respectively.
We find that the time dependence of $\means{\delta M^z_{X_R}}_Y(t)$ is almost unchanged against the introductions of fluxes and defects; it starts to oscillate at $tJ\sim 100$.
The spatial fluctuation $\Delta[\delta M^z_{X_R}]_Y(t)$ also increases at this time but it depends on the existence of the fluxes or defects.
In the absence of fluxes and defects, the spatial fluctuation trivially vanishes.
Figure~\ref{fig:tdep_sd}(d) indicates that the spatial fluctuation for the two-flux case is stronger than that for the two-defect case, which is also implied from Fig.~\ref{fig:tdep_map}.
For the dependence of the pulse intensity, We find that the overall structure of $\means{\delta M^z_{X_R}}_Y
(t)$ and $\Delta[\delta M^z_{X_R}]_Y(t)$ is unrelated to the amplitude $A$ while the absolute values depend on it.
Note that the previous study pointed out that the amplitude of the induced moment in the $R$ region is proportional to $A^2$~\cite{Minakawa2020}.
This appears to be observed in $A\lesssim 1$ but not to be satisfied for $A\sim 10$, which is interpreted as the case beyond the small $A$ regime.

\subsection{Distance dependence between fluxes/defects}
\label{sec:distance-dep}

\begin{figure*}[t]
  \begin{center}
  \includegraphics[width=2\columnwidth,clip]{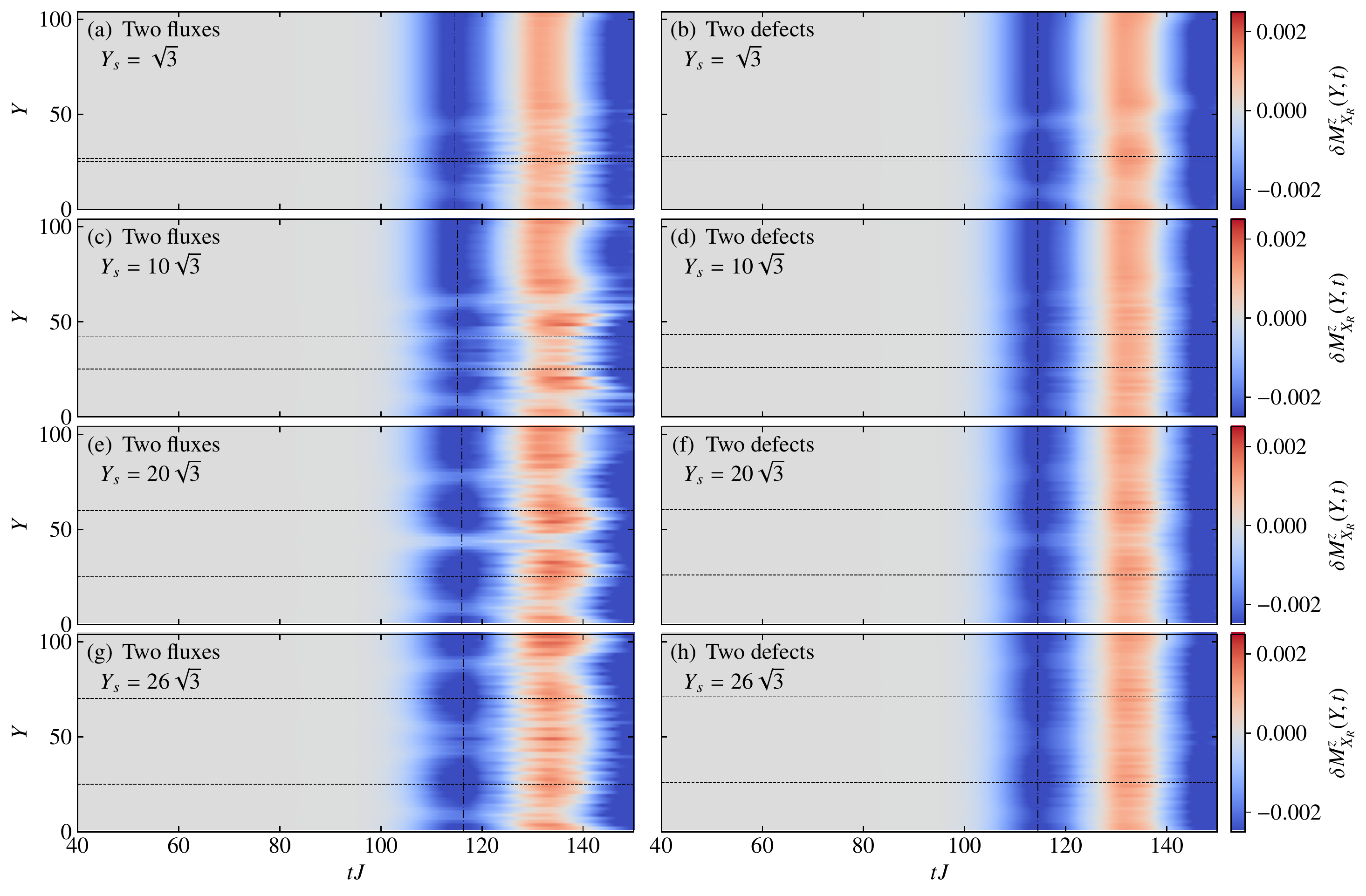}
  \caption{
    (a),(b) Spatiotemporal map of the induced spin moments at $X=X_R$ in the region $R$ in the presence of (a) the two fluxes and (b) the two lattice defects with the distance being $Y_s=\sqrt{3}$.
    The horizontal dashed lines represent the $Y$ coordinates of the flux/defect positions.
    The vertical dashed-dotted line indicates the time $t=t_m$ with $\left|\means{\delta M^z_{X_R}}_Y(t)\right|$ taking the first peak.
    The figures in [(c), (d)], [(e), (f)], and [(g), (h)] are the corresponding plots for $Y_s=10\sqrt{3}$, $20\sqrt{3}$, and $26\sqrt{3}$, respectively.
  }
\label{fig:tdep_map_dist}
\end{center}
\end{figure*}

Finally, we examine how the positions of the fluxes and defects affect the spin transport.
Figure~\ref{fig:tdep_map_dist} shows the counter plot of the induced magnetic moment $\delta M^z_{X_R}(Y,t)$ at $X=X_R$ given in Eq.~\eqref{eq:MzXR} for the several distances $Y_s$ between fluxes and defects.
As shown in Fig.~\ref{fig:tdep_map_dist}(a), in the case with two fluxes being adjacent, the modulation of $\delta M^z_{X_R}(Y,t)$ along the $Y$ direction is small in comparison with the case shown in Fig.~\ref{fig:tdep_map}(b).
Moreover, the overall structure appears to be similar to that for the defect case presented in Fig.~\ref{fig:tdep_map_dist}(b).
This is in contrast to the case shown in Fig.~\ref{fig:tdep_map}, suggesting that the distance $Y_s$ is crucial for the difference between the roles of flux and defect. 
In both cases, there are two weak dips in the first peak of $\delta M^z_{X_R}(Y,t)$ at $tJ\sim 115$, which is indicated by the vertical dashed-dotted line.
The distance between the dips are estimated as $\sim 40$, which is determined by $2\Delta Y=\frac{2}{\sqrt{3}}(X_R-X_s)$, as discussed in Sec.~\ref{sec:sv-spin}.

While the scattering phenomena of the spin transport by the fluxes and defects are similar to each other in the case with $Y_s=\sqrt{3}$, we find distinctly different behavior in the larger $Y_s$ cases.
Figure~\ref{fig:tdep_map_dist}(c) shows the time evolution of $\delta M^z_{X_R}(Y,t)$ in the presence of the two fluxes whose distance is $Y_s=10\sqrt{3}$.
As shown in this figure, four dips are observed in the first peak along the $Y$ direction.
On the other hand, for the defect case, the $Y$ dependence is hardly observed as shown in Figure~\ref{fig:tdep_map_dist}(d).
This distinctive difference is also observed in the cases with larger $Y_s$.
Figure~\ref{fig:tdep_map_dist}(d) shows the result for $Y_s=20\sqrt{3}$.
Since the value of $Y_s$ is close to $2\Delta Y=\frac{2}{\sqrt{3}}(X_R-X_s)$, we observe the strong suppression in the middle region of the two fluxes.
On the other hand, in the presence of the two defects, the modulation along the $Y$ direction is significantly weak in comparison with the flux case as presented in Fig.~\ref{fig:tdep_map_dist}(f).
The apparent difference is present even for the further increase of $Y_s$ as shown in Figs.~\ref{fig:tdep_map_dist}(g) and \ref{fig:tdep_map_dist}(h).

\begin{figure}[t]
  \begin{center}
  \includegraphics[width=\columnwidth,clip]{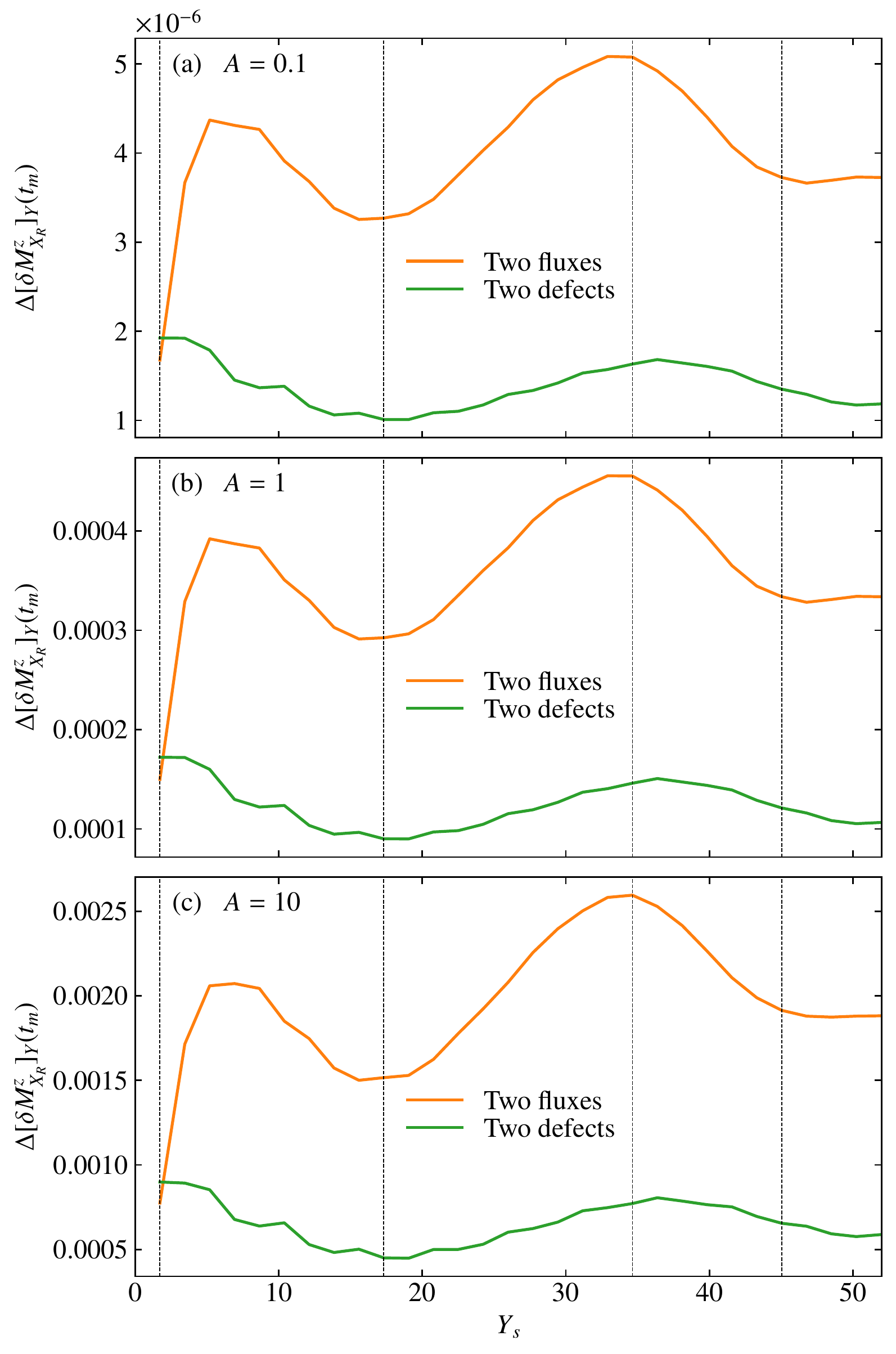}
  \caption{
    $Y_s$ dependence of the spatial variance for the spin moments along the $Y$ direction at $X=X_R$ and $t=t_m$ at (a) $A=0.1$, (b) $A=1$, and (c) $A=10$.
    The vertical dashed lines represent the values of $Y_s$ shown in Fig.~\ref{fig:tdep_map_dist}.
  }
\label{fig:dist_dep}
\end{center}
\end{figure}

To clarify the difference between the roles of the fluxes and defects as scatterers, we examine the $Y_s$ dependence of the spatial modulation of $\delta M^z_{X_R}(Y,t)$ along the $Y$ direction at $t=t_m$, where
 $t_m$ is the time of the first-peak time of $\means{\delta M^z_{X_R}}_Y(t)$, which is indicated by the vertical dashed-dotted lines in Fig.~\ref{fig:tdep_map_dist}.
 Figure~\ref{fig:dist_dep} shows the $Y_s$ dependence of $\Delta[\delta M^z_{X_R}]_Y(t_m)$ given in Eq.~\eqref{eq:Mz_sd} for several pulse intensities.
The overall structure of $\Delta[\delta M^z_{X_R}]_Y(t_m)$ as a function of $Y_s$ is the same except for its absolute value.
Hereafter, we focus on the case with $A=1$, which is shown in Fig.~\ref{fig:dist_dep}(b).
The values of $\Delta[\delta M^z_{X_R}]_Y(t_m)$ at $Y_s=\sqrt{3}$ for the two cases are close to each other.
With increasing $Y_s$, $\Delta[\delta M^z_{X_R}]_Y(t_m)$ for the two-flux case abruptly increases and shows nonmonotonic $Y_s$ dependence.
In particular, this quantity takes a peak around $Y_s\sim 35\sim 20\sqrt{3}$, which is reflected by the superposition of the scattered waves by the two fluxes discussed in Sec.~\ref{sec:sv-spin}.
On the other hand, $\Delta[\delta M^z_{X_R}]_Y(t_m)$ in the presence of two defects decreases with increasing $Y_s$ and does not exhibit a large change in contrast to the two-flux case.

\section{Discussion}
\label{sec:discussion}

Here, we discuss the origin of the difference between the two types of scatterers for the spin transport carried by the Majorana fermions.
In the presence of lattice defects, a defect modulates only its vicinity and little changes the Majorana fermion systems effectively.
If the number of defects is unchanged, the scattering phenomenon is expected to be almost independent of the defect configurations, which results in the weak defect-distance dependence shown in Fig.~\ref{fig:dist_dep}.
On the other hand, flux excitations affect the Majorana fermion system differently.
As mentioned in Sec.~\ref{sec:model-flux}, flux excitations modulate the transfer integrals of the itinerant Majorana fermions $\{c_i\}$ via flipped local variables $\eta_{ij}$ on the strings connecting the excited fluxes in the Jordan-Wigner transformation.
When two excited fluxes share their edges and are adjacent to each other, only one local variable $\eta_{ij}$ on the shared edge is flipped.
As the result, an influence upon the spin transport in the presence of adjacent fluxes is similar to that with lattice defects, as shown in Figs.~\ref{fig:tdep_map_dist}(a), \ref{fig:tdep_map_dist}(b), and Fig.~\ref{fig:dist_dep}(a).
However, the configuration of excited fluxes brings about a significant change to the spin transport compared with the system with defects, which strongly suggests that the hidden local variables $\eta_{ij}$ play a crucial role in the scattering phenomena of the spin transport.
Although the number of the excited fluxes is fixed to two, the number of the flipped local variables increases proportionally to the distance of excited fluxes.
This result indicates the nonlocal contribution to the Majorana fermions not only for the vicinity of the excited fluxes.
The nonlocal modulations of the transfer integrals cause large spatial dependence of the induced spin moments shown in Fig.~\ref{fig:tdep_map_dist}, which occurs due to the interference between the Majorana wavepackets propagating in the uniform $\eta$ region and through the bonds with flipped $\eta$.
This phenomenon should be interpreted as the Majorana version of the Aharonov-Bohm effect.

The present results suggest that spin transport is advantageous for detecting flux excitations in distinction from lattice defects, whereas the generation and control of fluxes have been recently proposed~\cite{Jang2021,Yue-Liu2021pre}.
For observing excited fluxes, scanning tunneling microscope (STM) or atomic force microscope (AFM) measurements are also promising as local measurements~\cite{Feldmeier2020,Konig2020,Pereira2020,Udagawa2021}.
Moreover, recent studies also suggest that the presence of flux excitations is detected using quantum interferometry with appropriate microfabrication~\cite{Klocke2021,Wei2021}.
Our present results pave another path for observing excited fluxes, which become non-Abelian anyons by applying magnetic fields.
The present situation can be reproduced in real materials by considering a setup with a Kitaev candidate material sandwiched between ferromagnetic materials.
The spin pumping in one of the ferromagnets generates a spin excitation at the edge of the Kitaev candidate material, and the resultant change of spin moments at the other edge could be measured by the induced magnetic moment in the ferromagnet connected with this side.
In this case, we need to observe spatially-resolved transient dynamics of the induced spin moment.
This might be achieved by magneto-optical measurements.
Time-resolved techniques using the Faraday or Kerr effect have been developed with picosecond-order resolutions~\cite{Beaurepaire1996,Hiebert1997,Kise2000,Guidoni2002,Ogasawara2005,kimel2005ultrafast,Cinchetti2006}.
Moreover, spatial imaging of time-resolved measurements has also been attempted recently~\cite{Takahashi2021}.
Exploiting these techniques, induced spin moments could be observed in the ferromagnet connected to a Kitaev candidate material, such as iridium oxides and $\alpha$-RuCl$_3$.
Note that the resultant spatial modulation of spin moments at an edge spreads over the length scale ten times larger than the lattice constant, as shown in Fig.~\ref{fig:tdep_map}(d).
This result suggests that extremely high spatial resolution is not required to observe the induced moment in the present scheme.
Moreover, the gapless Majorana fermions carry the spin modulation, and thereby the spin transport is long-ranged, which may facilitate the observation of the spin transport.

In the present calculations, we only consider the case with two fluxes or two defects and do not the case where both fluxes and defects exist simultaneously to discuss each effect separately.
Indeed, a lattice defect potentially traps a flux excitation with extremely small binding energy, as mentioned before~\cite{Willans2010,Willans2011}.
It might be interesting to consider the coexistence of fluxes and defects in the system for the spin transport of the Kitaev quantum spin liquid and the introductions of more than two fluxes and defects.
In this work, we introduce lattice defects on one of the two sublattices as vacancies for simplicity.
Comparing the effects of flux excitations with other local lattice deformations such as the introduction of vacancies on both sublattices, bond disorders, and dislocations are also intriguing issues. 
These remain future works.

\section{Summary}
\label{sec:summary}

In summary, we have investigated spin transport carried by fractional Majorana quasiparticles in the Kitaev quantum spin liquid with flux excitations, which are other fractional excitations intrinsic in this state.
To clarify the spin transport, we have considered the simple setup where the time-dependent magnetic field is applied at the left edge of the cluster to mimic spin injection by spin pumping, and we have examined induced spin moments at the other edge.
We have calculated the time evolution of the change of spin correlations and local spin moments by employing the time-dependent mean-field theory that we develop for Majorana fermion systems.
In the absence of flux excitations, the wavepacket of the Majorana fermions created by the magnetic-field pulse propagates to the other side with the velocity determined from the slope of the linear dispersion of the Majorana fermion system.
When the wavepacket accompanied by the change of spin correlations reaches the other edge, it induces the local spin moments.
In the presence of the flux excitations, the wavepacket is scattered by them, and induced local spin moments exhibit spatial inhomogeneity.
We have found that the change of spin correlations in bulk and the inhomogeneity of spin moments at the edge are more significant than that in the case where the excited fluxes are replaced with lattice defects.
Furthermore, we have also clarified that the inhomogeneity strongly depends on the distance between two excited fluxes compared with the case with lattice defects.
From the difference, we deduce that the gauge-like hidden degree of freedom plays an essential
role in the scattering phenomena for the flux case, which recalls the Aharonov-Bohm effect.
This is in contrast to the case with defects, where the scattering happens solely near a defect.
We have also discussed the possibility of applications to observe fluxes excitations in the Kitaev spin liquids, which is important for realizing topological quantum computation.

\begin{acknowledgments}
The authors thank Y.~Motome for fruitful discussions.
Parts of the numerical calculations were performed in the supercomputing
systems in ISSP, the University of Tokyo.
This work was supported by Grant-in-Aid for Scientific Research from
JSPS, KAKENHI Grant Nos. JP19K03742 (J.N.), JP20H00122 (J.N.), JP20K14412 (Y.M.), JP21H05017 (Y.M.),  JP17K05536 (A.K.), JP19H05821 (A.K.), JP21H01025 (A.K.), JP22K03525 (A.K.), JST PRESTO Grant No. JPMJPR19L5 (J.N.), and JST CREST Grant No. JPMJCR1901 (Y.M.).
\end{acknowledgments}

\bibliography{refs}

\begin{thebibliography}{110}%
\makeatletter
\providecommand \@ifxundefined [1]{%
 \@ifx{#1\undefined}
}%
\providecommand \@ifnum [1]{%
 \ifnum #1\expandafter \@firstoftwo
 \else \expandafter \@secondoftwo
 \fi
}%
\providecommand \@ifx [1]{%
 \ifx #1\expandafter \@firstoftwo
 \else \expandafter \@secondoftwo
 \fi
}%
\providecommand \natexlab [1]{#1}%
\providecommand \enquote  [1]{``#1''}%
\providecommand \bibnamefont  [1]{#1}%
\providecommand \bibfnamefont [1]{#1}%
\providecommand \citenamefont [1]{#1}%
\providecommand \href@noop [0]{\@secondoftwo}%
\providecommand \href [0]{\begingroup \@sanitize@url \@href}%
\providecommand \@href[1]{\@@startlink{#1}\@@href}%
\providecommand \@@href[1]{\endgroup#1\@@endlink}%
\providecommand \@sanitize@url [0]{\catcode `\\12\catcode `\$12\catcode
  `\&12\catcode `\#12\catcode `\^12\catcode `\_12\catcode `\%12\relax}%
\providecommand \@@startlink[1]{}%
\providecommand \@@endlink[0]{}%
\providecommand \url  [0]{\begingroup\@sanitize@url \@url }%
\providecommand \@url [1]{\endgroup\@href {#1}{\urlprefix }}%
\providecommand \urlprefix  [0]{URL }%
\providecommand \Eprint [0]{\href }%
\providecommand \doibase [0]{http://dx.doi.org/}%
\providecommand \selectlanguage [0]{\@gobble}%
\providecommand \bibinfo  [0]{\@secondoftwo}%
\providecommand \bibfield  [0]{\@secondoftwo}%
\providecommand \translation [1]{[#1]}%
\providecommand \BibitemOpen [0]{}%
\providecommand \bibitemStop [0]{}%
\providecommand \bibitemNoStop [0]{.\EOS\space}%
\providecommand \EOS [0]{\spacefactor3000\relax}%
\providecommand \BibitemShut  [1]{\csname bibitem#1\endcsname}%
\let\auto@bib@innerbib\@empty
\bibitem [{\citenamefont {Maekawa}\ \emph {et~al.}(2013)\citenamefont
  {Maekawa}, \citenamefont {Adachi}, \citenamefont {Uchida}, \citenamefont
  {Ieda},\ and\ \citenamefont {Saitoh}}]{Maekawa2013rev}%
  \BibitemOpen
  \bibfield  {author} {\bibinfo {author} {\bibfnamefont {S.}~\bibnamefont
  {Maekawa}}, \bibinfo {author} {\bibfnamefont {H.}~\bibnamefont {Adachi}},
  \bibinfo {author} {\bibfnamefont {K.-i.}\ \bibnamefont {Uchida}}, \bibinfo
  {author} {\bibfnamefont {J.}~\bibnamefont {Ieda}}, \ and\ \bibinfo {author}
  {\bibfnamefont {E.}~\bibnamefont {Saitoh}},\ }\href {\doibase
  10.7566/JPSJ.82.102002} {\bibfield  {journal} {\bibinfo  {journal} {J. Phys.
  Soc. Jpn.}\ }\textbf {\bibinfo {volume} {82}},\ \bibinfo {pages} {102002}
  (\bibinfo {year} {2013})}\BibitemShut {NoStop}%
\bibitem [{\citenamefont {Adachi}\ \emph {et~al.}(2013)\citenamefont {Adachi},
  \citenamefont {Uchida}, \citenamefont {Saitoh},\ and\ \citenamefont
  {Maekawa}}]{adachi2013theory}%
  \BibitemOpen
  \bibfield  {author} {\bibinfo {author} {\bibfnamefont {H.}~\bibnamefont
  {Adachi}}, \bibinfo {author} {\bibfnamefont {K.-i.}\ \bibnamefont {Uchida}},
  \bibinfo {author} {\bibfnamefont {E.}~\bibnamefont {Saitoh}}, \ and\ \bibinfo
  {author} {\bibfnamefont {S.}~\bibnamefont {Maekawa}},\ }\href {\doibase
  10.1088/0034-4885/76/3/036501} {\bibfield  {journal} {\bibinfo  {journal}
  {Rep. Prog. Phys.}\ }\textbf {\bibinfo {volume} {76}},\ \bibinfo {pages}
  {036501} (\bibinfo {year} {2013})}\BibitemShut {NoStop}%
\bibitem [{\citenamefont {Hirsch}(1999)}]{Hirsch1999}%
  \BibitemOpen
  \bibfield  {author} {\bibinfo {author} {\bibfnamefont {J.~E.}\ \bibnamefont
  {Hirsch}},\ }\href {\doibase 10.1103/PhysRevLett.83.1834} {\bibfield
  {journal} {\bibinfo  {journal} {Phys. Rev. Lett.}\ }\textbf {\bibinfo
  {volume} {83}},\ \bibinfo {pages} {1834} (\bibinfo {year}
  {1999})}\BibitemShut {NoStop}%
\bibitem [{\citenamefont {Brataas}\ \emph {et~al.}(2002)\citenamefont
  {Brataas}, \citenamefont {Tserkovnyak}, \citenamefont {Bauer},\ and\
  \citenamefont {Halperin}}]{Brataas2002}%
  \BibitemOpen
  \bibfield  {author} {\bibinfo {author} {\bibfnamefont {A.}~\bibnamefont
  {Brataas}}, \bibinfo {author} {\bibfnamefont {Y.}~\bibnamefont
  {Tserkovnyak}}, \bibinfo {author} {\bibfnamefont {G.~E.~W.}\ \bibnamefont
  {Bauer}}, \ and\ \bibinfo {author} {\bibfnamefont {B.~I.}\ \bibnamefont
  {Halperin}},\ }\href {\doibase 10.1103/PhysRevB.66.060404} {\bibfield
  {journal} {\bibinfo  {journal} {Phys. Rev. B}\ }\textbf {\bibinfo {volume}
  {66}},\ \bibinfo {pages} {060404} (\bibinfo {year} {2002})}\BibitemShut
  {NoStop}%
\bibitem [{\citenamefont {Watson}\ \emph {et~al.}(2003)\citenamefont {Watson},
  \citenamefont {Potok}, \citenamefont {Marcus},\ and\ \citenamefont
  {Umansky}}]{Watson2003}%
  \BibitemOpen
  \bibfield  {author} {\bibinfo {author} {\bibfnamefont {S.~K.}\ \bibnamefont
  {Watson}}, \bibinfo {author} {\bibfnamefont {R.~M.}\ \bibnamefont {Potok}},
  \bibinfo {author} {\bibfnamefont {C.~M.}\ \bibnamefont {Marcus}}, \ and\
  \bibinfo {author} {\bibfnamefont {V.}~\bibnamefont {Umansky}},\ }\href
  {\doibase 10.1103/PhysRevLett.91.258301} {\bibfield  {journal} {\bibinfo
  {journal} {Phys. Rev. Lett.}\ }\textbf {\bibinfo {volume} {91}},\ \bibinfo
  {pages} {258301} (\bibinfo {year} {2003})}\BibitemShut {NoStop}%
\bibitem [{\citenamefont {Murakami}\ \emph {et~al.}(2003)\citenamefont
  {Murakami}, \citenamefont {Nagaosa},\ and\ \citenamefont
  {Zhang}}]{murakami2003dissipationless}%
  \BibitemOpen
  \bibfield  {author} {\bibinfo {author} {\bibfnamefont {S.}~\bibnamefont
  {Murakami}}, \bibinfo {author} {\bibfnamefont {N.}~\bibnamefont {Nagaosa}}, \
  and\ \bibinfo {author} {\bibfnamefont {S.-C.}\ \bibnamefont {Zhang}},\ }\href
  {\doibase 10.1126/science.1087128} {\bibfield  {journal} {\bibinfo  {journal}
  {Science}\ }\textbf {\bibinfo {volume} {301}},\ \bibinfo {pages} {1348}
  (\bibinfo {year} {2003})}\BibitemShut {NoStop}%
\bibitem [{\citenamefont {Uchida}\ \emph
  {et~al.}(2010{\natexlab{a}})\citenamefont {Uchida}, \citenamefont {Xiao},
  \citenamefont {Adachi}, \citenamefont {Ohe}, \citenamefont {Takahashi},
  \citenamefont {Ieda}, \citenamefont {Ota}, \citenamefont {Kajiwara},
  \citenamefont {Umezawa}, \citenamefont {Kawai} \emph
  {et~al.}}]{uchida2010spin}%
  \BibitemOpen
  \bibfield  {author} {\bibinfo {author} {\bibfnamefont {K.-i.}\ \bibnamefont
  {Uchida}}, \bibinfo {author} {\bibfnamefont {J.}~\bibnamefont {Xiao}},
  \bibinfo {author} {\bibfnamefont {H.}~\bibnamefont {Adachi}}, \bibinfo
  {author} {\bibfnamefont {J.-i.}\ \bibnamefont {Ohe}}, \bibinfo {author}
  {\bibfnamefont {S.}~\bibnamefont {Takahashi}}, \bibinfo {author}
  {\bibfnamefont {J.}~\bibnamefont {Ieda}}, \bibinfo {author} {\bibfnamefont
  {T.}~\bibnamefont {Ota}}, \bibinfo {author} {\bibfnamefont {Y.}~\bibnamefont
  {Kajiwara}}, \bibinfo {author} {\bibfnamefont {H.}~\bibnamefont {Umezawa}},
  \bibinfo {author} {\bibfnamefont {H.}~\bibnamefont {Kawai}},  \emph
  {et~al.},\ }\href@noop {} {\bibfield  {journal} {\bibinfo  {journal} {Nat.
  Mater.}\ }\textbf {\bibinfo {volume} {9}},\ \bibinfo {pages} {894} (\bibinfo
  {year} {2010}{\natexlab{a}})}\BibitemShut {NoStop}%
\bibitem [{\citenamefont {Uchida}\ \emph
  {et~al.}(2010{\natexlab{b}})\citenamefont {Uchida}, \citenamefont {Adachi},
  \citenamefont {Ota}, \citenamefont {Nakayama}, \citenamefont {Maekawa},\ and\
  \citenamefont {Saitoh}}]{uchida2010}%
  \BibitemOpen
  \bibfield  {author} {\bibinfo {author} {\bibfnamefont {K.-i.}\ \bibnamefont
  {Uchida}}, \bibinfo {author} {\bibfnamefont {H.}~\bibnamefont {Adachi}},
  \bibinfo {author} {\bibfnamefont {T.}~\bibnamefont {Ota}}, \bibinfo {author}
  {\bibfnamefont {H.}~\bibnamefont {Nakayama}}, \bibinfo {author}
  {\bibfnamefont {S.}~\bibnamefont {Maekawa}}, \ and\ \bibinfo {author}
  {\bibfnamefont {E.}~\bibnamefont {Saitoh}},\ }\href@noop {} {\bibfield
  {journal} {\bibinfo  {journal} {Appl. Phys. Lett.}\ }\textbf {\bibinfo
  {volume} {97}},\ \bibinfo {pages} {172505} (\bibinfo {year}
  {2010}{\natexlab{b}})}\BibitemShut {NoStop}%
\bibitem [{\citenamefont {Xiao}\ \emph {et~al.}(2010)\citenamefont {Xiao},
  \citenamefont {Bauer}, \citenamefont {Uchida}, \citenamefont {Saitoh},\ and\
  \citenamefont {Maekawa}}]{Xiao2010}%
  \BibitemOpen
  \bibfield  {author} {\bibinfo {author} {\bibfnamefont {J.}~\bibnamefont
  {Xiao}}, \bibinfo {author} {\bibfnamefont {G.~E.~W.}\ \bibnamefont {Bauer}},
  \bibinfo {author} {\bibfnamefont {K.-c.}\ \bibnamefont {Uchida}}, \bibinfo
  {author} {\bibfnamefont {E.}~\bibnamefont {Saitoh}}, \ and\ \bibinfo {author}
  {\bibfnamefont {S.}~\bibnamefont {Maekawa}},\ }\href {\doibase
  10.1103/PhysRevB.81.214418} {\bibfield  {journal} {\bibinfo  {journal} {Phys.
  Rev. B}\ }\textbf {\bibinfo {volume} {81}},\ \bibinfo {pages} {214418}
  (\bibinfo {year} {2010})}\BibitemShut {NoStop}%
\bibitem [{\citenamefont {Adachi}\ \emph {et~al.}(2011)\citenamefont {Adachi},
  \citenamefont {Ohe}, \citenamefont {Takahashi},\ and\ \citenamefont
  {Maekawa}}]{Adachi2011}%
  \BibitemOpen
  \bibfield  {author} {\bibinfo {author} {\bibfnamefont {H.}~\bibnamefont
  {Adachi}}, \bibinfo {author} {\bibfnamefont {J.-i.}\ \bibnamefont {Ohe}},
  \bibinfo {author} {\bibfnamefont {S.}~\bibnamefont {Takahashi}}, \ and\
  \bibinfo {author} {\bibfnamefont {S.}~\bibnamefont {Maekawa}},\ }\href
  {\doibase 10.1103/PhysRevB.83.094410} {\bibfield  {journal} {\bibinfo
  {journal} {Phys. Rev. B}\ }\textbf {\bibinfo {volume} {83}},\ \bibinfo
  {pages} {094410} (\bibinfo {year} {2011})}\BibitemShut {NoStop}%
\bibitem [{\citenamefont {Zhang}\ and\ \citenamefont
  {Zhang}(2012{\natexlab{a}})}]{Zhang_magnon2012}%
  \BibitemOpen
  \bibfield  {author} {\bibinfo {author} {\bibfnamefont {S.~S.-L.}\
  \bibnamefont {Zhang}}\ and\ \bibinfo {author} {\bibfnamefont
  {S.}~\bibnamefont {Zhang}},\ }\href {\doibase 10.1103/PhysRevLett.109.096603}
  {\bibfield  {journal} {\bibinfo  {journal} {Phys. Rev. Lett.}\ }\textbf
  {\bibinfo {volume} {109}},\ \bibinfo {pages} {096603} (\bibinfo {year}
  {2012}{\natexlab{a}})}\BibitemShut {NoStop}%
\bibitem [{\citenamefont {Zhang}\ and\ \citenamefont
  {Zhang}(2012{\natexlab{b}})}]{Zhang_magnon_full2012}%
  \BibitemOpen
  \bibfield  {author} {\bibinfo {author} {\bibfnamefont {S.~S.-L.}\
  \bibnamefont {Zhang}}\ and\ \bibinfo {author} {\bibfnamefont
  {S.}~\bibnamefont {Zhang}},\ }\href {\doibase 10.1103/PhysRevB.86.214424}
  {\bibfield  {journal} {\bibinfo  {journal} {Phys. Rev. B}\ }\textbf {\bibinfo
  {volume} {86}},\ \bibinfo {pages} {214424} (\bibinfo {year}
  {2012}{\natexlab{b}})}\BibitemShut {NoStop}%
\bibitem [{\citenamefont {Rezende}\ \emph {et~al.}(2014)\citenamefont
  {Rezende}, \citenamefont {Rodr\'{\i}guez-Su\'arez}, \citenamefont {Cunha},
  \citenamefont {Rodrigues}, \citenamefont {Machado}, \citenamefont
  {Fonseca~Guerra}, \citenamefont {Lopez~Ortiz},\ and\ \citenamefont
  {Azevedo}}]{Rezende2014}%
  \BibitemOpen
  \bibfield  {author} {\bibinfo {author} {\bibfnamefont {S.~M.}\ \bibnamefont
  {Rezende}}, \bibinfo {author} {\bibfnamefont {R.~L.}\ \bibnamefont
  {Rodr\'{\i}guez-Su\'arez}}, \bibinfo {author} {\bibfnamefont {R.~O.}\
  \bibnamefont {Cunha}}, \bibinfo {author} {\bibfnamefont {A.~R.}\ \bibnamefont
  {Rodrigues}}, \bibinfo {author} {\bibfnamefont {F.~L.~A.}\ \bibnamefont
  {Machado}}, \bibinfo {author} {\bibfnamefont {G.~A.}\ \bibnamefont
  {Fonseca~Guerra}}, \bibinfo {author} {\bibfnamefont {J.~C.}\ \bibnamefont
  {Lopez~Ortiz}}, \ and\ \bibinfo {author} {\bibfnamefont {A.}~\bibnamefont
  {Azevedo}},\ }\href {\doibase 10.1103/PhysRevB.89.014416} {\bibfield
  {journal} {\bibinfo  {journal} {Phys. Rev. B}\ }\textbf {\bibinfo {volume}
  {89}},\ \bibinfo {pages} {014416} (\bibinfo {year} {2014})}\BibitemShut
  {NoStop}%
\bibitem [{\citenamefont {Qiu}\ \emph {et~al.}(2018)\citenamefont {Qiu},
  \citenamefont {Hou}, \citenamefont {Barker}, \citenamefont {Yamamoto},
  \citenamefont {Gomonay},\ and\ \citenamefont {Saitoh}}]{qiu2018spin}%
  \BibitemOpen
  \bibfield  {author} {\bibinfo {author} {\bibfnamefont {Z.}~\bibnamefont
  {Qiu}}, \bibinfo {author} {\bibfnamefont {D.}~\bibnamefont {Hou}}, \bibinfo
  {author} {\bibfnamefont {J.}~\bibnamefont {Barker}}, \bibinfo {author}
  {\bibfnamefont {K.}~\bibnamefont {Yamamoto}}, \bibinfo {author}
  {\bibfnamefont {O.}~\bibnamefont {Gomonay}}, \ and\ \bibinfo {author}
  {\bibfnamefont {E.}~\bibnamefont {Saitoh}},\ }\href {\doibase
  10.1038/s41563-018-0087-4} {\bibfield  {journal} {\bibinfo  {journal} {Nature
  materials}\ }\textbf {\bibinfo {volume} {17}},\ \bibinfo {pages} {577}
  (\bibinfo {year} {2018})}\BibitemShut {NoStop}%
\bibitem [{\citenamefont {Naka}\ \emph {et~al.}(2019)\citenamefont {Naka},
  \citenamefont {Hayami}, \citenamefont {Kusunose}, \citenamefont {Yanagi},
  \citenamefont {Motome},\ and\ \citenamefont {Seo}}]{naka2019spin}%
  \BibitemOpen
  \bibfield  {author} {\bibinfo {author} {\bibfnamefont {M.}~\bibnamefont
  {Naka}}, \bibinfo {author} {\bibfnamefont {S.}~\bibnamefont {Hayami}},
  \bibinfo {author} {\bibfnamefont {H.}~\bibnamefont {Kusunose}}, \bibinfo
  {author} {\bibfnamefont {Y.}~\bibnamefont {Yanagi}}, \bibinfo {author}
  {\bibfnamefont {Y.}~\bibnamefont {Motome}}, \ and\ \bibinfo {author}
  {\bibfnamefont {H.}~\bibnamefont {Seo}},\ }\href {\doibase
  10.1038/s41467-019-12229-y} {\bibfield  {journal} {\bibinfo  {journal} {Nat.
  Commun.}\ }\textbf {\bibinfo {volume} {10}},\ \bibinfo {pages} {1} (\bibinfo
  {year} {2019})}\BibitemShut {NoStop}%
\bibitem [{\citenamefont {Hirobe}\ \emph {et~al.}(2017)\citenamefont {Hirobe},
  \citenamefont {Sato}, \citenamefont {Kawamata}, \citenamefont {Shiomi},
  \citenamefont {Uchida}, \citenamefont {Iguchi}, \citenamefont {Koike},
  \citenamefont {Maekawa},\ and\ \citenamefont {Saitoh}}]{hirobe2017one}%
  \BibitemOpen
  \bibfield  {author} {\bibinfo {author} {\bibfnamefont {D.}~\bibnamefont
  {Hirobe}}, \bibinfo {author} {\bibfnamefont {M.}~\bibnamefont {Sato}},
  \bibinfo {author} {\bibfnamefont {T.}~\bibnamefont {Kawamata}}, \bibinfo
  {author} {\bibfnamefont {Y.}~\bibnamefont {Shiomi}}, \bibinfo {author}
  {\bibfnamefont {K.-i.}\ \bibnamefont {Uchida}}, \bibinfo {author}
  {\bibfnamefont {R.}~\bibnamefont {Iguchi}}, \bibinfo {author} {\bibfnamefont
  {Y.}~\bibnamefont {Koike}}, \bibinfo {author} {\bibfnamefont
  {S.}~\bibnamefont {Maekawa}}, \ and\ \bibinfo {author} {\bibfnamefont
  {E.}~\bibnamefont {Saitoh}},\ }\href {\doibase 10.1038/nphys3895} {\bibfield
  {journal} {\bibinfo  {journal} {Nat. Phys.}\ }\textbf {\bibinfo {volume}
  {13}},\ \bibinfo {pages} {30} (\bibinfo {year} {2017})}\BibitemShut {NoStop}%
\bibitem [{\citenamefont {Nasu}\ and\ \citenamefont
  {Naka}(2021)}]{Nasu_spinseebeck2021}%
  \BibitemOpen
  \bibfield  {author} {\bibinfo {author} {\bibfnamefont {J.}~\bibnamefont
  {Nasu}}\ and\ \bibinfo {author} {\bibfnamefont {M.}~\bibnamefont {Naka}},\
  }\href {\doibase 10.1103/PhysRevB.103.L121104} {\bibfield  {journal}
  {\bibinfo  {journal} {Phys. Rev. B}\ }\textbf {\bibinfo {volume} {103}},\
  \bibinfo {pages} {L121104} (\bibinfo {year} {2021})}\BibitemShut {NoStop}%
\bibitem [{\citenamefont {Nussinov}\ and\ \citenamefont {van~den
  Brink}(2015)}]{RevModPhys.87.1}%
  \BibitemOpen
  \bibfield  {author} {\bibinfo {author} {\bibfnamefont {Z.}~\bibnamefont
  {Nussinov}}\ and\ \bibinfo {author} {\bibfnamefont {J.}~\bibnamefont {van~den
  Brink}},\ }\href {\doibase 10.1103/RevModPhys.87.1} {\bibfield  {journal}
  {\bibinfo  {journal} {Rev. Mod. Phys.}\ }\textbf {\bibinfo {volume} {87}},\
  \bibinfo {pages} {1} (\bibinfo {year} {2015})}\BibitemShut {NoStop}%
\bibitem [{\citenamefont {Trebst}(shed)}]{Trebst2017pre}%
  \BibitemOpen
  \bibfield  {author} {\bibinfo {author} {\bibfnamefont {S.}~\bibnamefont
  {Trebst}},\ }\href {http://arxiv.org/abs/1701.07056} {\bibfield  {journal}
  {\bibinfo  {journal} {preprint}\ ,\ \bibinfo {pages} {arXiv:1701.07056}}
  (\bibinfo {year} {unpublished})}\BibitemShut {NoStop}%
\bibitem [{\citenamefont {Hermanns}\ \emph {et~al.}(2018)\citenamefont
  {Hermanns}, \citenamefont {Kimchi},\ and\ \citenamefont
  {Knolle}}]{Hermanns2018rev}%
  \BibitemOpen
  \bibfield  {author} {\bibinfo {author} {\bibfnamefont {M.}~\bibnamefont
  {Hermanns}}, \bibinfo {author} {\bibfnamefont {I.}~\bibnamefont {Kimchi}}, \
  and\ \bibinfo {author} {\bibfnamefont {J.}~\bibnamefont {Knolle}},\ }\href
  {\doibase 10.1146/annurev-conmatphys-033117-053934} {\bibfield  {journal}
  {\bibinfo  {journal} {Annu. Rev. Condens. Matter Phys.}\ }\textbf {\bibinfo
  {volume} {9}},\ \bibinfo {pages} {17} (\bibinfo {year} {2018})}\BibitemShut
  {NoStop}%
\bibitem [{\citenamefont {Knolle}\ and\ \citenamefont
  {Moessner}(2019)}]{Knolle2019rev}%
  \BibitemOpen
  \bibfield  {author} {\bibinfo {author} {\bibfnamefont {J.}~\bibnamefont
  {Knolle}}\ and\ \bibinfo {author} {\bibfnamefont {R.}~\bibnamefont
  {Moessner}},\ }\href {\doibase 10.1146/annurev-conmatphys-031218-013401}
  {\bibfield  {journal} {\bibinfo  {journal} {Annu. Rev. Condens. Matter
  Phys.}\ }\textbf {\bibinfo {volume} {10}},\ \bibinfo {pages} {451} (\bibinfo
  {year} {2019})}\BibitemShut {NoStop}%
\bibitem [{\citenamefont {Takagi}\ \emph {et~al.}(2019)\citenamefont {Takagi},
  \citenamefont {Takayama}, \citenamefont {Jackeli}, \citenamefont
  {Khaliullin},\ and\ \citenamefont {Nagler}}]{takagi2019rev}%
  \BibitemOpen
  \bibfield  {author} {\bibinfo {author} {\bibfnamefont {H.}~\bibnamefont
  {Takagi}}, \bibinfo {author} {\bibfnamefont {T.}~\bibnamefont {Takayama}},
  \bibinfo {author} {\bibfnamefont {G.}~\bibnamefont {Jackeli}}, \bibinfo
  {author} {\bibfnamefont {G.}~\bibnamefont {Khaliullin}}, \ and\ \bibinfo
  {author} {\bibfnamefont {S.~E.}\ \bibnamefont {Nagler}},\ }\href {\doibase
  10.1038/s42254-019-0038-2} {\bibfield  {journal} {\bibinfo  {journal} {Nat.
  Rev. Phys.}\ ,\ \bibinfo {pages} {1}} (\bibinfo {year} {2019})}\BibitemShut
  {NoStop}%
\bibitem [{\citenamefont {Motome}\ and\ \citenamefont
  {Nasu}(2020)}]{Motome2020rev}%
  \BibitemOpen
  \bibfield  {author} {\bibinfo {author} {\bibfnamefont {Y.}~\bibnamefont
  {Motome}}\ and\ \bibinfo {author} {\bibfnamefont {J.}~\bibnamefont {Nasu}},\
  }\href {\doibase 10.7566/JPSJ.89.012002} {\bibfield  {journal} {\bibinfo
  {journal} {J. Phys. Soc. Jpn.}\ }\textbf {\bibinfo {volume} {89}},\ \bibinfo
  {pages} {012002} (\bibinfo {year} {2020})}\BibitemShut {NoStop}%
\bibitem [{\citenamefont {Kitaev}(2006)}]{Kitaev2006}%
  \BibitemOpen
  \bibfield  {author} {\bibinfo {author} {\bibfnamefont {A.}~\bibnamefont
  {Kitaev}},\ }\href {\doibase 10.1016/j.aop.2005.10.005} {\bibfield  {journal}
  {\bibinfo  {journal} {Ann. Phys. (N. Y.)}\ }\textbf {\bibinfo {volume}
  {321}},\ \bibinfo {pages} {2} (\bibinfo {year} {2006})}\BibitemShut {NoStop}%
\bibitem [{\citenamefont {Jackeli}\ and\ \citenamefont
  {Khaliullin}(2009)}]{PhysRevLett.102.017205}%
  \BibitemOpen
  \bibfield  {author} {\bibinfo {author} {\bibfnamefont {G.}~\bibnamefont
  {Jackeli}}\ and\ \bibinfo {author} {\bibfnamefont {G.}~\bibnamefont
  {Khaliullin}},\ }\href {\doibase 10.1103/PhysRevLett.102.017205} {\bibfield
  {journal} {\bibinfo  {journal} {Phys. Rev. Lett.}\ }\textbf {\bibinfo
  {volume} {102}},\ \bibinfo {pages} {017205} (\bibinfo {year}
  {2009})}\BibitemShut {NoStop}%
\bibitem [{\citenamefont {Winter}\ \emph {et~al.}(2016)\citenamefont {Winter},
  \citenamefont {Li}, \citenamefont {Jeschke},\ and\ \citenamefont
  {Valent\'{\i}}}]{Winter2016}%
  \BibitemOpen
  \bibfield  {author} {\bibinfo {author} {\bibfnamefont {S.~M.}\ \bibnamefont
  {Winter}}, \bibinfo {author} {\bibfnamefont {Y.}~\bibnamefont {Li}}, \bibinfo
  {author} {\bibfnamefont {H.~O.}\ \bibnamefont {Jeschke}}, \ and\ \bibinfo
  {author} {\bibfnamefont {R.}~\bibnamefont {Valent\'{\i}}},\ }\href {\doibase
  10.1103/PhysRevB.93.214431} {\bibfield  {journal} {\bibinfo  {journal} {Phys.
  Rev. B}\ }\textbf {\bibinfo {volume} {93}},\ \bibinfo {pages} {214431}
  (\bibinfo {year} {2016})}\BibitemShut {NoStop}%
\bibitem [{\citenamefont {Winter}\ \emph
  {et~al.}(2017{\natexlab{a}})\citenamefont {Winter}, \citenamefont {Tsirlin},
  \citenamefont {Daghofer}, \citenamefont {van~den Brink}, \citenamefont
  {Singh}, \citenamefont {Gegenwart},\ and\ \citenamefont
  {Valent{\'{\i}}}}]{Winter_2017rev}%
  \BibitemOpen
  \bibfield  {author} {\bibinfo {author} {\bibfnamefont {S.~M.}\ \bibnamefont
  {Winter}}, \bibinfo {author} {\bibfnamefont {A.~A.}\ \bibnamefont {Tsirlin}},
  \bibinfo {author} {\bibfnamefont {M.}~\bibnamefont {Daghofer}}, \bibinfo
  {author} {\bibfnamefont {J.}~\bibnamefont {van~den Brink}}, \bibinfo {author}
  {\bibfnamefont {Y.}~\bibnamefont {Singh}}, \bibinfo {author} {\bibfnamefont
  {P.}~\bibnamefont {Gegenwart}}, \ and\ \bibinfo {author} {\bibfnamefont
  {R.}~\bibnamefont {Valent{\'{\i}}}},\ }\href {\doibase
  10.1088/1361-648x/aa8cf5} {\bibfield  {journal} {\bibinfo  {journal} {J.
  Phys.: Condens. Matter}\ }\textbf {\bibinfo {volume} {29}},\ \bibinfo {pages}
  {493002} (\bibinfo {year} {2017}{\natexlab{a}})}\BibitemShut {NoStop}%
\bibitem [{\citenamefont {Jang}\ \emph {et~al.}(2019)\citenamefont {Jang},
  \citenamefont {Sano}, \citenamefont {Kato},\ and\ \citenamefont
  {Motome}}]{Jang_AFKitaev2019}%
  \BibitemOpen
  \bibfield  {author} {\bibinfo {author} {\bibfnamefont {S.-H.}\ \bibnamefont
  {Jang}}, \bibinfo {author} {\bibfnamefont {R.}~\bibnamefont {Sano}}, \bibinfo
  {author} {\bibfnamefont {Y.}~\bibnamefont {Kato}}, \ and\ \bibinfo {author}
  {\bibfnamefont {Y.}~\bibnamefont {Motome}},\ }\href {\doibase
  10.1103/PhysRevB.99.241106} {\bibfield  {journal} {\bibinfo  {journal} {Phys.
  Rev. B}\ }\textbf {\bibinfo {volume} {99}},\ \bibinfo {pages} {241106}
  (\bibinfo {year} {2019})}\BibitemShut {NoStop}%
\bibitem [{\citenamefont {Stavropoulos}\ \emph {et~al.}(2019)\citenamefont
  {Stavropoulos}, \citenamefont {Pereira},\ and\ \citenamefont
  {Kee}}]{Stavropoulos2019}%
  \BibitemOpen
  \bibfield  {author} {\bibinfo {author} {\bibfnamefont {P.~P.}\ \bibnamefont
  {Stavropoulos}}, \bibinfo {author} {\bibfnamefont {D.}~\bibnamefont
  {Pereira}}, \ and\ \bibinfo {author} {\bibfnamefont {H.-Y.}\ \bibnamefont
  {Kee}},\ }\href {\doibase 10.1103/PhysRevLett.123.037203} {\bibfield
  {journal} {\bibinfo  {journal} {Phys. Rev. Lett.}\ }\textbf {\bibinfo
  {volume} {123}},\ \bibinfo {pages} {037203} (\bibinfo {year}
  {2019})}\BibitemShut {NoStop}%
\bibitem [{\citenamefont {Chaloupka}\ \emph {et~al.}(2010)\citenamefont
  {Chaloupka}, \citenamefont {Jackeli},\ and\ \citenamefont
  {Khaliullin}}]{PhysRevLett.105.027204}%
  \BibitemOpen
  \bibfield  {author} {\bibinfo {author} {\bibfnamefont {J.}~\bibnamefont
  {Chaloupka}}, \bibinfo {author} {\bibfnamefont {G.}~\bibnamefont {Jackeli}},
  \ and\ \bibinfo {author} {\bibfnamefont {G.}~\bibnamefont {Khaliullin}},\
  }\href {\doibase 10.1103/PhysRevLett.105.027204} {\bibfield  {journal}
  {\bibinfo  {journal} {Phys. Rev. Lett.}\ }\textbf {\bibinfo {volume} {105}},\
  \bibinfo {pages} {027204} (\bibinfo {year} {2010})}\BibitemShut {NoStop}%
\bibitem [{\citenamefont {Singh}\ and\ \citenamefont
  {Gegenwart}(2010)}]{PhysRevB.82.064412}%
  \BibitemOpen
  \bibfield  {author} {\bibinfo {author} {\bibfnamefont {Y.}~\bibnamefont
  {Singh}}\ and\ \bibinfo {author} {\bibfnamefont {P.}~\bibnamefont
  {Gegenwart}},\ }\href {\doibase 10.1103/PhysRevB.82.064412} {\bibfield
  {journal} {\bibinfo  {journal} {Phys. Rev. B}\ }\textbf {\bibinfo {volume}
  {82}},\ \bibinfo {pages} {064412} (\bibinfo {year} {2010})}\BibitemShut
  {NoStop}%
\bibitem [{\citenamefont {Singh}\ \emph {et~al.}(2012)\citenamefont {Singh},
  \citenamefont {Manni}, \citenamefont {Reuther}, \citenamefont {Berlijn},
  \citenamefont {Thomale}, \citenamefont {Ku}, \citenamefont {Trebst},\ and\
  \citenamefont {Gegenwart}}]{PhysRevLett.108.127203}%
  \BibitemOpen
  \bibfield  {author} {\bibinfo {author} {\bibfnamefont {Y.}~\bibnamefont
  {Singh}}, \bibinfo {author} {\bibfnamefont {S.}~\bibnamefont {Manni}},
  \bibinfo {author} {\bibfnamefont {J.}~\bibnamefont {Reuther}}, \bibinfo
  {author} {\bibfnamefont {T.}~\bibnamefont {Berlijn}}, \bibinfo {author}
  {\bibfnamefont {R.}~\bibnamefont {Thomale}}, \bibinfo {author} {\bibfnamefont
  {W.}~\bibnamefont {Ku}}, \bibinfo {author} {\bibfnamefont {S.}~\bibnamefont
  {Trebst}}, \ and\ \bibinfo {author} {\bibfnamefont {P.}~\bibnamefont
  {Gegenwart}},\ }\href {\doibase 10.1103/PhysRevLett.108.127203} {\bibfield
  {journal} {\bibinfo  {journal} {Phys. Rev. Lett.}\ }\textbf {\bibinfo
  {volume} {108}},\ \bibinfo {pages} {127203} (\bibinfo {year}
  {2012})}\BibitemShut {NoStop}%
\bibitem [{\citenamefont {Foyevtsova}\ \emph {et~al.}(2013)\citenamefont
  {Foyevtsova}, \citenamefont {Jeschke}, \citenamefont {Mazin}, \citenamefont
  {Khomskii},\ and\ \citenamefont {Valent\'{\i}}}]{PhysRevB.88.035107}%
  \BibitemOpen
  \bibfield  {author} {\bibinfo {author} {\bibfnamefont {K.}~\bibnamefont
  {Foyevtsova}}, \bibinfo {author} {\bibfnamefont {H.~O.}\ \bibnamefont
  {Jeschke}}, \bibinfo {author} {\bibfnamefont {I.~I.}\ \bibnamefont {Mazin}},
  \bibinfo {author} {\bibfnamefont {D.~I.}\ \bibnamefont {Khomskii}}, \ and\
  \bibinfo {author} {\bibfnamefont {R.}~\bibnamefont {Valent\'{\i}}},\ }\href
  {\doibase 10.1103/PhysRevB.88.035107} {\bibfield  {journal} {\bibinfo
  {journal} {Phys. Rev. B}\ }\textbf {\bibinfo {volume} {88}},\ \bibinfo
  {pages} {035107} (\bibinfo {year} {2013})}\BibitemShut {NoStop}%
\bibitem [{\citenamefont {Chaloupka}\ \emph {et~al.}(2013)\citenamefont
  {Chaloupka}, \citenamefont {Jackeli},\ and\ \citenamefont
  {Khaliullin}}]{PhysRevLett.110.097204}%
  \BibitemOpen
  \bibfield  {author} {\bibinfo {author} {\bibfnamefont {J.}~\bibnamefont
  {Chaloupka}}, \bibinfo {author} {\bibfnamefont {G.}~\bibnamefont {Jackeli}},
  \ and\ \bibinfo {author} {\bibfnamefont {G.}~\bibnamefont {Khaliullin}},\
  }\href {\doibase 10.1103/PhysRevLett.110.097204} {\bibfield  {journal}
  {\bibinfo  {journal} {Phys. Rev. Lett.}\ }\textbf {\bibinfo {volume} {110}},\
  \bibinfo {pages} {097204} (\bibinfo {year} {2013})}\BibitemShut {NoStop}%
\bibitem [{\citenamefont {Katukuri}\ \emph {et~al.}(2014)\citenamefont
  {Katukuri}, \citenamefont {Nishimoto}, \citenamefont {Yushankhai},
  \citenamefont {Stoyanova}, \citenamefont {Kandpal}, \citenamefont {Choi},
  \citenamefont {Coldea}, \citenamefont {Rousochatzakis}, \citenamefont
  {Hozoi},\ and\ \citenamefont {van~den Brink}}]{1367-2630-16-1-013056}%
  \BibitemOpen
  \bibfield  {author} {\bibinfo {author} {\bibfnamefont {V.~M.}\ \bibnamefont
  {Katukuri}}, \bibinfo {author} {\bibfnamefont {S.}~\bibnamefont {Nishimoto}},
  \bibinfo {author} {\bibfnamefont {V.}~\bibnamefont {Yushankhai}}, \bibinfo
  {author} {\bibfnamefont {A.}~\bibnamefont {Stoyanova}}, \bibinfo {author}
  {\bibfnamefont {H.}~\bibnamefont {Kandpal}}, \bibinfo {author} {\bibfnamefont
  {S.}~\bibnamefont {Choi}}, \bibinfo {author} {\bibfnamefont {R.}~\bibnamefont
  {Coldea}}, \bibinfo {author} {\bibfnamefont {I.}~\bibnamefont
  {Rousochatzakis}}, \bibinfo {author} {\bibfnamefont {L.}~\bibnamefont
  {Hozoi}}, \ and\ \bibinfo {author} {\bibfnamefont {J.}~\bibnamefont {van~den
  Brink}},\ }\href {http://stacks.iop.org/1367-2630/16/i=1/a=013056} {\bibfield
   {journal} {\bibinfo  {journal} {New J. Phys.}\ }\textbf {\bibinfo {volume}
  {16}},\ \bibinfo {pages} {013056} (\bibinfo {year} {2014})}\BibitemShut
  {NoStop}%
\bibitem [{\citenamefont {Yamaji}\ \emph {et~al.}(2014)\citenamefont {Yamaji},
  \citenamefont {Nomura}, \citenamefont {Kurita}, \citenamefont {Arita},\ and\
  \citenamefont {Imada}}]{PhysRevLett.113.107201}%
  \BibitemOpen
  \bibfield  {author} {\bibinfo {author} {\bibfnamefont {Y.}~\bibnamefont
  {Yamaji}}, \bibinfo {author} {\bibfnamefont {Y.}~\bibnamefont {Nomura}},
  \bibinfo {author} {\bibfnamefont {M.}~\bibnamefont {Kurita}}, \bibinfo
  {author} {\bibfnamefont {R.}~\bibnamefont {Arita}}, \ and\ \bibinfo {author}
  {\bibfnamefont {M.}~\bibnamefont {Imada}},\ }\href {\doibase
  10.1103/PhysRevLett.113.107201} {\bibfield  {journal} {\bibinfo  {journal}
  {Phys. Rev. Lett.}\ }\textbf {\bibinfo {volume} {113}},\ \bibinfo {pages}
  {107201} (\bibinfo {year} {2014})}\BibitemShut {NoStop}%
\bibitem [{\citenamefont {Plumb}\ \emph {et~al.}(2014)\citenamefont {Plumb},
  \citenamefont {Clancy}, \citenamefont {Sandilands}, \citenamefont {Shankar},
  \citenamefont {Hu}, \citenamefont {Burch}, \citenamefont {Kee},\ and\
  \citenamefont {Kim}}]{PhysRevB.90.041112}%
  \BibitemOpen
  \bibfield  {author} {\bibinfo {author} {\bibfnamefont {K.~W.}\ \bibnamefont
  {Plumb}}, \bibinfo {author} {\bibfnamefont {J.~P.}\ \bibnamefont {Clancy}},
  \bibinfo {author} {\bibfnamefont {L.~J.}\ \bibnamefont {Sandilands}},
  \bibinfo {author} {\bibfnamefont {V.~V.}\ \bibnamefont {Shankar}}, \bibinfo
  {author} {\bibfnamefont {Y.~F.}\ \bibnamefont {Hu}}, \bibinfo {author}
  {\bibfnamefont {K.~S.}\ \bibnamefont {Burch}}, \bibinfo {author}
  {\bibfnamefont {H.-Y.}\ \bibnamefont {Kee}}, \ and\ \bibinfo {author}
  {\bibfnamefont {Y.-J.}\ \bibnamefont {Kim}},\ }\href {\doibase
  10.1103/PhysRevB.90.041112} {\bibfield  {journal} {\bibinfo  {journal} {Phys.
  Rev. B}\ }\textbf {\bibinfo {volume} {90}},\ \bibinfo {pages} {041112}
  (\bibinfo {year} {2014})}\BibitemShut {NoStop}%
\bibitem [{\citenamefont {Kubota}\ \emph {et~al.}(2015)\citenamefont {Kubota},
  \citenamefont {Tanaka}, \citenamefont {Ono}, \citenamefont {Narumi},\ and\
  \citenamefont {Kindo}}]{PhysRevB.91.094422}%
  \BibitemOpen
  \bibfield  {author} {\bibinfo {author} {\bibfnamefont {Y.}~\bibnamefont
  {Kubota}}, \bibinfo {author} {\bibfnamefont {H.}~\bibnamefont {Tanaka}},
  \bibinfo {author} {\bibfnamefont {T.}~\bibnamefont {Ono}}, \bibinfo {author}
  {\bibfnamefont {Y.}~\bibnamefont {Narumi}}, \ and\ \bibinfo {author}
  {\bibfnamefont {K.}~\bibnamefont {Kindo}},\ }\href {\doibase
  10.1103/PhysRevB.91.094422} {\bibfield  {journal} {\bibinfo  {journal} {Phys.
  Rev. B}\ }\textbf {\bibinfo {volume} {91}},\ \bibinfo {pages} {094422}
  (\bibinfo {year} {2015})}\BibitemShut {NoStop}%
\bibitem [{\citenamefont {Johnson}\ \emph {et~al.}(2015)\citenamefont
  {Johnson}, \citenamefont {Williams}, \citenamefont {Haghighirad},
  \citenamefont {Singleton}, \citenamefont {Zapf}, \citenamefont {Manuel},
  \citenamefont {Mazin}, \citenamefont {Li}, \citenamefont {Jeschke},
  \citenamefont {Valent\'{\i}},\ and\ \citenamefont {Coldea}}]{Johnson2015}%
  \BibitemOpen
  \bibfield  {author} {\bibinfo {author} {\bibfnamefont {R.~D.}\ \bibnamefont
  {Johnson}}, \bibinfo {author} {\bibfnamefont {S.~C.}\ \bibnamefont
  {Williams}}, \bibinfo {author} {\bibfnamefont {A.~A.}\ \bibnamefont
  {Haghighirad}}, \bibinfo {author} {\bibfnamefont {J.}~\bibnamefont
  {Singleton}}, \bibinfo {author} {\bibfnamefont {V.}~\bibnamefont {Zapf}},
  \bibinfo {author} {\bibfnamefont {P.}~\bibnamefont {Manuel}}, \bibinfo
  {author} {\bibfnamefont {I.~I.}\ \bibnamefont {Mazin}}, \bibinfo {author}
  {\bibfnamefont {Y.}~\bibnamefont {Li}}, \bibinfo {author} {\bibfnamefont
  {H.~O.}\ \bibnamefont {Jeschke}}, \bibinfo {author} {\bibfnamefont
  {R.}~\bibnamefont {Valent\'{\i}}}, \ and\ \bibinfo {author} {\bibfnamefont
  {R.}~\bibnamefont {Coldea}},\ }\href {\doibase 10.1103/PhysRevB.92.235119}
  {\bibfield  {journal} {\bibinfo  {journal} {Phys. Rev. B}\ }\textbf {\bibinfo
  {volume} {92}},\ \bibinfo {pages} {235119} (\bibinfo {year}
  {2015})}\BibitemShut {NoStop}%
\bibitem [{\citenamefont {Sears}\ \emph {et~al.}(2015)\citenamefont {Sears},
  \citenamefont {Songvilay}, \citenamefont {Plumb}, \citenamefont {Clancy},
  \citenamefont {Qiu}, \citenamefont {Zhao}, \citenamefont {Parshall},\ and\
  \citenamefont {Kim}}]{PhysRevB.91.144420}%
  \BibitemOpen
  \bibfield  {author} {\bibinfo {author} {\bibfnamefont {J.~A.}\ \bibnamefont
  {Sears}}, \bibinfo {author} {\bibfnamefont {M.}~\bibnamefont {Songvilay}},
  \bibinfo {author} {\bibfnamefont {K.~W.}\ \bibnamefont {Plumb}}, \bibinfo
  {author} {\bibfnamefont {J.~P.}\ \bibnamefont {Clancy}}, \bibinfo {author}
  {\bibfnamefont {Y.}~\bibnamefont {Qiu}}, \bibinfo {author} {\bibfnamefont
  {Y.}~\bibnamefont {Zhao}}, \bibinfo {author} {\bibfnamefont {D.}~\bibnamefont
  {Parshall}}, \ and\ \bibinfo {author} {\bibfnamefont {Y.-J.}\ \bibnamefont
  {Kim}},\ }\href {\doibase 10.1103/PhysRevB.91.144420} {\bibfield  {journal}
  {\bibinfo  {journal} {Phys. Rev. B}\ }\textbf {\bibinfo {volume} {91}},\
  \bibinfo {pages} {144420} (\bibinfo {year} {2015})}\BibitemShut {NoStop}%
\bibitem [{\citenamefont {Yadav}\ \emph {et~al.}(2016)\citenamefont {Yadav},
  \citenamefont {Bogdanov}, \citenamefont {Katukuri}, \citenamefont
  {Nishimoto}, \citenamefont {van~den Brink},\ and\ \citenamefont
  {Hozoi}}]{yadav2016kitaev}%
  \BibitemOpen
  \bibfield  {author} {\bibinfo {author} {\bibfnamefont {R.}~\bibnamefont
  {Yadav}}, \bibinfo {author} {\bibfnamefont {N.~A.}\ \bibnamefont {Bogdanov}},
  \bibinfo {author} {\bibfnamefont {V.~M.}\ \bibnamefont {Katukuri}}, \bibinfo
  {author} {\bibfnamefont {S.}~\bibnamefont {Nishimoto}}, \bibinfo {author}
  {\bibfnamefont {J.}~\bibnamefont {van~den Brink}}, \ and\ \bibinfo {author}
  {\bibfnamefont {L.}~\bibnamefont {Hozoi}},\ }\href {\doibase
  10.1038/srep37925} {\bibfield  {journal} {\bibinfo  {journal} {Scientific
  Reports}\ }\textbf {\bibinfo {volume} {6}},\ \bibinfo {pages} {37925}
  (\bibinfo {year} {2016})}\BibitemShut {NoStop}%
\bibitem [{\citenamefont {Kim}\ and\ \citenamefont
  {Kee}(2016)}]{PhysRevB.93.155143}%
  \BibitemOpen
  \bibfield  {author} {\bibinfo {author} {\bibfnamefont {H.-S.}\ \bibnamefont
  {Kim}}\ and\ \bibinfo {author} {\bibfnamefont {H.-Y.}\ \bibnamefont {Kee}},\
  }\href {\doibase 10.1103/PhysRevB.93.155143} {\bibfield  {journal} {\bibinfo
  {journal} {Phys. Rev. B}\ }\textbf {\bibinfo {volume} {93}},\ \bibinfo
  {pages} {155143} (\bibinfo {year} {2016})}\BibitemShut {NoStop}%
\bibitem [{\citenamefont {Koitzsch}\ \emph {et~al.}(2016)\citenamefont
  {Koitzsch}, \citenamefont {Habenicht}, \citenamefont {M\"uller},
  \citenamefont {Knupfer}, \citenamefont {B\"uchner}, \citenamefont {Kandpal},
  \citenamefont {van~den Brink}, \citenamefont {Nowak}, \citenamefont
  {Isaeva},\ and\ \citenamefont {Doert}}]{Koitzsch2016}%
  \BibitemOpen
  \bibfield  {author} {\bibinfo {author} {\bibfnamefont {A.}~\bibnamefont
  {Koitzsch}}, \bibinfo {author} {\bibfnamefont {C.}~\bibnamefont {Habenicht}},
  \bibinfo {author} {\bibfnamefont {E.}~\bibnamefont {M\"uller}}, \bibinfo
  {author} {\bibfnamefont {M.}~\bibnamefont {Knupfer}}, \bibinfo {author}
  {\bibfnamefont {B.}~\bibnamefont {B\"uchner}}, \bibinfo {author}
  {\bibfnamefont {H.~C.}\ \bibnamefont {Kandpal}}, \bibinfo {author}
  {\bibfnamefont {J.}~\bibnamefont {van~den Brink}}, \bibinfo {author}
  {\bibfnamefont {D.}~\bibnamefont {Nowak}}, \bibinfo {author} {\bibfnamefont
  {A.}~\bibnamefont {Isaeva}}, \ and\ \bibinfo {author} {\bibfnamefont
  {T.}~\bibnamefont {Doert}},\ }\href {\doibase 10.1103/PhysRevLett.117.126403}
  {\bibfield  {journal} {\bibinfo  {journal} {Phys. Rev. Lett.}\ }\textbf
  {\bibinfo {volume} {117}},\ \bibinfo {pages} {126403} (\bibinfo {year}
  {2016})}\BibitemShut {NoStop}%
\bibitem [{\citenamefont {Liu}\ and\ \citenamefont
  {Khaliullin}(2018)}]{Liu2018Pseudospin}%
  \BibitemOpen
  \bibfield  {author} {\bibinfo {author} {\bibfnamefont {H.}~\bibnamefont
  {Liu}}\ and\ \bibinfo {author} {\bibfnamefont {G.}~\bibnamefont
  {Khaliullin}},\ }\href {\doibase 10.1103/PhysRevB.97.014407} {\bibfield
  {journal} {\bibinfo  {journal} {Phys. Rev. B}\ }\textbf {\bibinfo {volume}
  {97}},\ \bibinfo {pages} {014407} (\bibinfo {year} {2018})}\BibitemShut
  {NoStop}%
\bibitem [{\citenamefont {Sano}\ \emph {et~al.}(2018)\citenamefont {Sano},
  \citenamefont {Kato},\ and\ \citenamefont {Motome}}]{sano2018}%
  \BibitemOpen
  \bibfield  {author} {\bibinfo {author} {\bibfnamefont {R.}~\bibnamefont
  {Sano}}, \bibinfo {author} {\bibfnamefont {Y.}~\bibnamefont {Kato}}, \ and\
  \bibinfo {author} {\bibfnamefont {Y.}~\bibnamefont {Motome}},\ }\href
  {\doibase 10.1103/PhysRevB.97.014408} {\bibfield  {journal} {\bibinfo
  {journal} {Phys. Rev. B}\ }\textbf {\bibinfo {volume} {97}},\ \bibinfo
  {pages} {014408} (\bibinfo {year} {2018})}\BibitemShut {NoStop}%
\bibitem [{\citenamefont {Lefran\ifmmode~\mbox{\c{c}}\else \c{c}\fi{}ois}\
  \emph {et~al.}(2016)\citenamefont {Lefran\ifmmode~\mbox{\c{c}}\else
  \c{c}\fi{}ois}, \citenamefont {Songvilay}, \citenamefont {Robert},
  \citenamefont {Nataf}, \citenamefont {Jordan}, \citenamefont {Chaix},
  \citenamefont {Colin}, \citenamefont {Lejay}, \citenamefont {Hadj-Azzem},
  \citenamefont {Ballou},\ and\ \citenamefont {Simonet}}]{Lefran2016}%
  \BibitemOpen
  \bibfield  {author} {\bibinfo {author} {\bibfnamefont {E.}~\bibnamefont
  {Lefran\ifmmode~\mbox{\c{c}}\else \c{c}\fi{}ois}}, \bibinfo {author}
  {\bibfnamefont {M.}~\bibnamefont {Songvilay}}, \bibinfo {author}
  {\bibfnamefont {J.}~\bibnamefont {Robert}}, \bibinfo {author} {\bibfnamefont
  {G.}~\bibnamefont {Nataf}}, \bibinfo {author} {\bibfnamefont
  {E.}~\bibnamefont {Jordan}}, \bibinfo {author} {\bibfnamefont
  {L.}~\bibnamefont {Chaix}}, \bibinfo {author} {\bibfnamefont {C.~V.}\
  \bibnamefont {Colin}}, \bibinfo {author} {\bibfnamefont {P.}~\bibnamefont
  {Lejay}}, \bibinfo {author} {\bibfnamefont {A.}~\bibnamefont {Hadj-Azzem}},
  \bibinfo {author} {\bibfnamefont {R.}~\bibnamefont {Ballou}}, \ and\ \bibinfo
  {author} {\bibfnamefont {V.}~\bibnamefont {Simonet}},\ }\href {\doibase
  10.1103/PhysRevB.94.214416} {\bibfield  {journal} {\bibinfo  {journal} {Phys.
  Rev. B}\ }\textbf {\bibinfo {volume} {94}},\ \bibinfo {pages} {214416}
  (\bibinfo {year} {2016})}\BibitemShut {NoStop}%
\bibitem [{\citenamefont {Bera}\ \emph {et~al.}(2017)\citenamefont {Bera},
  \citenamefont {Yusuf}, \citenamefont {Kumar},\ and\ \citenamefont
  {Ritter}}]{Bera2017}%
  \BibitemOpen
  \bibfield  {author} {\bibinfo {author} {\bibfnamefont {A.~K.}\ \bibnamefont
  {Bera}}, \bibinfo {author} {\bibfnamefont {S.~M.}\ \bibnamefont {Yusuf}},
  \bibinfo {author} {\bibfnamefont {A.}~\bibnamefont {Kumar}}, \ and\ \bibinfo
  {author} {\bibfnamefont {C.}~\bibnamefont {Ritter}},\ }\href {\doibase
  10.1103/PhysRevB.95.094424} {\bibfield  {journal} {\bibinfo  {journal} {Phys.
  Rev. B}\ }\textbf {\bibinfo {volume} {95}},\ \bibinfo {pages} {094424}
  (\bibinfo {year} {2017})}\BibitemShut {NoStop}%
\bibitem [{\citenamefont {Zhong}\ \emph {et~al.}(2018)\citenamefont {Zhong},
  \citenamefont {Chung}, \citenamefont {Kong}, \citenamefont {Nguyen},
  \citenamefont {Lei},\ and\ \citenamefont {Cava}}]{Zhong_Field-induced2018}%
  \BibitemOpen
  \bibfield  {author} {\bibinfo {author} {\bibfnamefont {R.}~\bibnamefont
  {Zhong}}, \bibinfo {author} {\bibfnamefont {M.}~\bibnamefont {Chung}},
  \bibinfo {author} {\bibfnamefont {T.}~\bibnamefont {Kong}}, \bibinfo {author}
  {\bibfnamefont {L.~T.}\ \bibnamefont {Nguyen}}, \bibinfo {author}
  {\bibfnamefont {S.}~\bibnamefont {Lei}}, \ and\ \bibinfo {author}
  {\bibfnamefont {R.~J.}\ \bibnamefont {Cava}},\ }\href {\doibase
  10.1103/PhysRevB.98.220407} {\bibfield  {journal} {\bibinfo  {journal} {Phys.
  Rev. B}\ }\textbf {\bibinfo {volume} {98}},\ \bibinfo {pages} {220407}
  (\bibinfo {year} {2018})}\BibitemShut {NoStop}%
\bibitem [{\citenamefont {Wildes}\ \emph {et~al.}(2017)\citenamefont {Wildes},
  \citenamefont {Simonet}, \citenamefont {Ressouche}, \citenamefont {Ballou},\
  and\ \citenamefont {McIntyre}}]{wildes2017magnetic}%
  \BibitemOpen
  \bibfield  {author} {\bibinfo {author} {\bibfnamefont {A.}~\bibnamefont
  {Wildes}}, \bibinfo {author} {\bibfnamefont {V.}~\bibnamefont {Simonet}},
  \bibinfo {author} {\bibfnamefont {E.}~\bibnamefont {Ressouche}}, \bibinfo
  {author} {\bibfnamefont {R.}~\bibnamefont {Ballou}}, \ and\ \bibinfo {author}
  {\bibfnamefont {G.}~\bibnamefont {McIntyre}},\ }\href {\doibase
  10.1088/1361-648X/aa8a43} {\bibfield  {journal} {\bibinfo  {journal} {J.
  Phys. Condens. Matter}\ }\textbf {\bibinfo {volume} {29}},\ \bibinfo {pages}
  {455801} (\bibinfo {year} {2017})}\BibitemShut {NoStop}%
\bibitem [{\citenamefont {Yan}\ \emph {et~al.}(2019)\citenamefont {Yan},
  \citenamefont {Okamoto}, \citenamefont {Wu}, \citenamefont {Zheng},
  \citenamefont {Zhou}, \citenamefont {Cao},\ and\ \citenamefont
  {McGuire}}]{Yan_Magnetic2019}%
  \BibitemOpen
  \bibfield  {author} {\bibinfo {author} {\bibfnamefont {J.-Q.}\ \bibnamefont
  {Yan}}, \bibinfo {author} {\bibfnamefont {S.}~\bibnamefont {Okamoto}},
  \bibinfo {author} {\bibfnamefont {Y.}~\bibnamefont {Wu}}, \bibinfo {author}
  {\bibfnamefont {Q.}~\bibnamefont {Zheng}}, \bibinfo {author} {\bibfnamefont
  {H.~D.}\ \bibnamefont {Zhou}}, \bibinfo {author} {\bibfnamefont {H.~B.}\
  \bibnamefont {Cao}}, \ and\ \bibinfo {author} {\bibfnamefont {M.~A.}\
  \bibnamefont {McGuire}},\ }\href {\doibase 10.1103/PhysRevMaterials.3.074405}
  {\bibfield  {journal} {\bibinfo  {journal} {Phys. Rev. Materials}\ }\textbf
  {\bibinfo {volume} {3}},\ \bibinfo {pages} {074405} (\bibinfo {year}
  {2019})}\BibitemShut {NoStop}%
\bibitem [{\citenamefont {Zhong}\ \emph {et~al.}(2020)\citenamefont {Zhong},
  \citenamefont {Gao}, \citenamefont {Ong},\ and\ \citenamefont
  {Cava}}]{zhong2020weak}%
  \BibitemOpen
  \bibfield  {author} {\bibinfo {author} {\bibfnamefont {R.}~\bibnamefont
  {Zhong}}, \bibinfo {author} {\bibfnamefont {T.}~\bibnamefont {Gao}}, \bibinfo
  {author} {\bibfnamefont {N.~P.}\ \bibnamefont {Ong}}, \ and\ \bibinfo
  {author} {\bibfnamefont {R.~J.}\ \bibnamefont {Cava}},\ }\href {\doibase
  10.1126/sciadv.aay6953} {\bibfield  {journal} {\bibinfo  {journal} {Science
  advances}\ }\textbf {\bibinfo {volume} {6}},\ \bibinfo {pages} {eaay6953}
  (\bibinfo {year} {2020})}\BibitemShut {NoStop}%
\bibitem [{\citenamefont {Yuan}\ \emph {et~al.}(2020)\citenamefont {Yuan},
  \citenamefont {Khait}, \citenamefont {Shu}, \citenamefont {Chou},
  \citenamefont {Stone}, \citenamefont {Clancy}, \citenamefont {Paramekanti},\
  and\ \citenamefont {Kim}}]{Yuan2020}%
  \BibitemOpen
  \bibfield  {author} {\bibinfo {author} {\bibfnamefont {B.}~\bibnamefont
  {Yuan}}, \bibinfo {author} {\bibfnamefont {I.}~\bibnamefont {Khait}},
  \bibinfo {author} {\bibfnamefont {G.-J.}\ \bibnamefont {Shu}}, \bibinfo
  {author} {\bibfnamefont {F.~C.}\ \bibnamefont {Chou}}, \bibinfo {author}
  {\bibfnamefont {M.~B.}\ \bibnamefont {Stone}}, \bibinfo {author}
  {\bibfnamefont {J.~P.}\ \bibnamefont {Clancy}}, \bibinfo {author}
  {\bibfnamefont {A.}~\bibnamefont {Paramekanti}}, \ and\ \bibinfo {author}
  {\bibfnamefont {Y.-J.}\ \bibnamefont {Kim}},\ }\href {\doibase
  10.1103/PhysRevX.10.011062} {\bibfield  {journal} {\bibinfo  {journal} {Phys.
  Rev. X}\ }\textbf {\bibinfo {volume} {10}},\ \bibinfo {pages} {011062}
  (\bibinfo {year} {2020})}\BibitemShut {NoStop}%
\bibitem [{\citenamefont {Nasu}\ \emph {et~al.}(2015)\citenamefont {Nasu},
  \citenamefont {Udagawa},\ and\ \citenamefont {Motome}}]{PhysRevB.92.115122}%
  \BibitemOpen
  \bibfield  {author} {\bibinfo {author} {\bibfnamefont {J.}~\bibnamefont
  {Nasu}}, \bibinfo {author} {\bibfnamefont {M.}~\bibnamefont {Udagawa}}, \
  and\ \bibinfo {author} {\bibfnamefont {Y.}~\bibnamefont {Motome}},\ }\href
  {\doibase 10.1103/PhysRevB.92.115122} {\bibfield  {journal} {\bibinfo
  {journal} {Phys. Rev. B}\ }\textbf {\bibinfo {volume} {92}},\ \bibinfo
  {pages} {115122} (\bibinfo {year} {2015})}\BibitemShut {NoStop}%
\bibitem [{\citenamefont {Knolle}\ \emph
  {et~al.}(2014{\natexlab{a}})\citenamefont {Knolle}, \citenamefont
  {Kovrizhin}, \citenamefont {Chalker},\ and\ \citenamefont
  {Moessner}}]{PhysRevLett.112.207203}%
  \BibitemOpen
  \bibfield  {author} {\bibinfo {author} {\bibfnamefont {J.}~\bibnamefont
  {Knolle}}, \bibinfo {author} {\bibfnamefont {D.~L.}\ \bibnamefont
  {Kovrizhin}}, \bibinfo {author} {\bibfnamefont {J.~T.}\ \bibnamefont
  {Chalker}}, \ and\ \bibinfo {author} {\bibfnamefont {R.}~\bibnamefont
  {Moessner}},\ }\href {\doibase 10.1103/PhysRevLett.112.207203} {\bibfield
  {journal} {\bibinfo  {journal} {Phys. Rev. Lett.}\ }\textbf {\bibinfo
  {volume} {112}},\ \bibinfo {pages} {207203} (\bibinfo {year}
  {2014}{\natexlab{a}})}\BibitemShut {NoStop}%
\bibitem [{\citenamefont {Knolle}\ \emph {et~al.}(2015)\citenamefont {Knolle},
  \citenamefont {Kovrizhin}, \citenamefont {Chalker},\ and\ \citenamefont
  {Moessner}}]{PhysRevB.92.115127}%
  \BibitemOpen
  \bibfield  {author} {\bibinfo {author} {\bibfnamefont {J.}~\bibnamefont
  {Knolle}}, \bibinfo {author} {\bibfnamefont {D.~L.}\ \bibnamefont
  {Kovrizhin}}, \bibinfo {author} {\bibfnamefont {J.~T.}\ \bibnamefont
  {Chalker}}, \ and\ \bibinfo {author} {\bibfnamefont {R.}~\bibnamefont
  {Moessner}},\ }\href {\doibase 10.1103/PhysRevB.92.115127} {\bibfield
  {journal} {\bibinfo  {journal} {Phys. Rev. B}\ }\textbf {\bibinfo {volume}
  {92}},\ \bibinfo {pages} {115127} (\bibinfo {year} {2015})}\BibitemShut
  {NoStop}%
\bibitem [{\citenamefont {Knolle}\ \emph
  {et~al.}(2014{\natexlab{b}})\citenamefont {Knolle}, \citenamefont {Chern},
  \citenamefont {Kovrizhin}, \citenamefont {Moessner},\ and\ \citenamefont
  {Perkins}}]{PhysRevLett.113.187201}%
  \BibitemOpen
  \bibfield  {author} {\bibinfo {author} {\bibfnamefont {J.}~\bibnamefont
  {Knolle}}, \bibinfo {author} {\bibfnamefont {G.-W.}\ \bibnamefont {Chern}},
  \bibinfo {author} {\bibfnamefont {D.~L.}\ \bibnamefont {Kovrizhin}}, \bibinfo
  {author} {\bibfnamefont {R.}~\bibnamefont {Moessner}}, \ and\ \bibinfo
  {author} {\bibfnamefont {N.~B.}\ \bibnamefont {Perkins}},\ }\href {\doibase
  10.1103/PhysRevLett.113.187201} {\bibfield  {journal} {\bibinfo  {journal}
  {Phys. Rev. Lett.}\ }\textbf {\bibinfo {volume} {113}},\ \bibinfo {pages}
  {187201} (\bibinfo {year} {2014}{\natexlab{b}})}\BibitemShut {NoStop}%
\bibitem [{\citenamefont {Sandilands}\ \emph {et~al.}(2015)\citenamefont
  {Sandilands}, \citenamefont {Tian}, \citenamefont {Plumb}, \citenamefont
  {Kim},\ and\ \citenamefont {Burch}}]{PhysRevLett.114.147201}%
  \BibitemOpen
  \bibfield  {author} {\bibinfo {author} {\bibfnamefont {L.~J.}\ \bibnamefont
  {Sandilands}}, \bibinfo {author} {\bibfnamefont {Y.}~\bibnamefont {Tian}},
  \bibinfo {author} {\bibfnamefont {K.~W.}\ \bibnamefont {Plumb}}, \bibinfo
  {author} {\bibfnamefont {Y.-J.}\ \bibnamefont {Kim}}, \ and\ \bibinfo
  {author} {\bibfnamefont {K.~S.}\ \bibnamefont {Burch}},\ }\href {\doibase
  10.1103/PhysRevLett.114.147201} {\bibfield  {journal} {\bibinfo  {journal}
  {Phys. Rev. Lett.}\ }\textbf {\bibinfo {volume} {114}},\ \bibinfo {pages}
  {147201} (\bibinfo {year} {2015})}\BibitemShut {NoStop}%
\bibitem [{\citenamefont {Winter}\ \emph
  {et~al.}(2017{\natexlab{b}})\citenamefont {Winter}, \citenamefont {Riedl},
  \citenamefont {Maksimov}, \citenamefont {Chernyshev}, \citenamefont
  {Honecker},\ and\ \citenamefont {Valent{\'\i}}}]{winter2017breakdown}%
  \BibitemOpen
  \bibfield  {author} {\bibinfo {author} {\bibfnamefont {S.~M.}\ \bibnamefont
  {Winter}}, \bibinfo {author} {\bibfnamefont {K.}~\bibnamefont {Riedl}},
  \bibinfo {author} {\bibfnamefont {P.~A.}\ \bibnamefont {Maksimov}}, \bibinfo
  {author} {\bibfnamefont {A.~L.}\ \bibnamefont {Chernyshev}}, \bibinfo
  {author} {\bibfnamefont {A.}~\bibnamefont {Honecker}}, \ and\ \bibinfo
  {author} {\bibfnamefont {R.}~\bibnamefont {Valent{\'\i}}},\ }\href {\doibase
  10.1038/s41467-017-01177-0} {\bibfield  {journal} {\bibinfo  {journal}
  {Nature communications}\ }\textbf {\bibinfo {volume} {8}},\ \bibinfo {pages}
  {1152} (\bibinfo {year} {2017}{\natexlab{b}})}\BibitemShut {NoStop}%
\bibitem [{\citenamefont {Song}\ \emph {et~al.}(2016)\citenamefont {Song},
  \citenamefont {You},\ and\ \citenamefont {Balents}}]{Song2016}%
  \BibitemOpen
  \bibfield  {author} {\bibinfo {author} {\bibfnamefont {X.-Y.}\ \bibnamefont
  {Song}}, \bibinfo {author} {\bibfnamefont {Y.-Z.}\ \bibnamefont {You}}, \
  and\ \bibinfo {author} {\bibfnamefont {L.}~\bibnamefont {Balents}},\ }\href
  {\doibase 10.1103/PhysRevLett.117.037209} {\bibfield  {journal} {\bibinfo
  {journal} {Phys. Rev. Lett.}\ }\textbf {\bibinfo {volume} {117}},\ \bibinfo
  {pages} {037209} (\bibinfo {year} {2016})}\BibitemShut {NoStop}%
\bibitem [{\citenamefont {Nasu}\ \emph {et~al.}(2016)\citenamefont {Nasu},
  \citenamefont {Knolle}, \citenamefont {Kovrizhin}, \citenamefont {Motome},\
  and\ \citenamefont {Moessner}}]{Nasu2016nphys}%
  \BibitemOpen
  \bibfield  {author} {\bibinfo {author} {\bibfnamefont {J.}~\bibnamefont
  {Nasu}}, \bibinfo {author} {\bibfnamefont {J.}~\bibnamefont {Knolle}},
  \bibinfo {author} {\bibfnamefont {D.~L.}\ \bibnamefont {Kovrizhin}}, \bibinfo
  {author} {\bibfnamefont {Y.}~\bibnamefont {Motome}}, \ and\ \bibinfo {author}
  {\bibfnamefont {R.}~\bibnamefont {Moessner}},\ }\href {\doibase
  10.1038/nphys3809} {\bibfield  {journal} {\bibinfo  {journal} {Nat. Phys.}\
  }\textbf {\bibinfo {volume} {12}},\ \bibinfo {pages} {912} (\bibinfo {year}
  {2016})}\BibitemShut {NoStop}%
\bibitem [{\citenamefont {Hal\'asz}\ \emph {et~al.}(2016)\citenamefont
  {Hal\'asz}, \citenamefont {Perkins},\ and\ \citenamefont {van~den
  Brink}}]{Halasz2016}%
  \BibitemOpen
  \bibfield  {author} {\bibinfo {author} {\bibfnamefont {G.~B.}\ \bibnamefont
  {Hal\'asz}}, \bibinfo {author} {\bibfnamefont {N.~B.}\ \bibnamefont
  {Perkins}}, \ and\ \bibinfo {author} {\bibfnamefont {J.}~\bibnamefont
  {van~den Brink}},\ }\href {\doibase 10.1103/PhysRevLett.117.127203}
  {\bibfield  {journal} {\bibinfo  {journal} {Phys. Rev. Lett.}\ }\textbf
  {\bibinfo {volume} {117}},\ \bibinfo {pages} {127203} (\bibinfo {year}
  {2016})}\BibitemShut {NoStop}%
\bibitem [{\citenamefont {Yoshitake}\ \emph {et~al.}(2016)\citenamefont
  {Yoshitake}, \citenamefont {Nasu},\ and\ \citenamefont
  {Motome}}]{yoshitake2016}%
  \BibitemOpen
  \bibfield  {author} {\bibinfo {author} {\bibfnamefont {J.}~\bibnamefont
  {Yoshitake}}, \bibinfo {author} {\bibfnamefont {J.}~\bibnamefont {Nasu}}, \
  and\ \bibinfo {author} {\bibfnamefont {Y.}~\bibnamefont {Motome}},\ }\href
  {\doibase 10.1103/PhysRevLett.117.157203} {\bibfield  {journal} {\bibinfo
  {journal} {Phys. Rev. Lett.}\ }\textbf {\bibinfo {volume} {117}},\ \bibinfo
  {pages} {157203} (\bibinfo {year} {2016})}\BibitemShut {NoStop}%
\bibitem [{\citenamefont {Tanaka}\ \emph {et~al.}(2022)\citenamefont {Tanaka},
  \citenamefont {Mizukami}, \citenamefont {Harasawa}, \citenamefont
  {Hashimoto}, \citenamefont {Hwang}, \citenamefont {Kurita}, \citenamefont
  {Tanaka}, \citenamefont {Fujimoto}, \citenamefont {Matsuda}, \citenamefont
  {Moon} \emph {et~al.}}]{tanaka2022thermodynamic}%
  \BibitemOpen
  \bibfield  {author} {\bibinfo {author} {\bibfnamefont {O.}~\bibnamefont
  {Tanaka}}, \bibinfo {author} {\bibfnamefont {Y.}~\bibnamefont {Mizukami}},
  \bibinfo {author} {\bibfnamefont {R.}~\bibnamefont {Harasawa}}, \bibinfo
  {author} {\bibfnamefont {K.}~\bibnamefont {Hashimoto}}, \bibinfo {author}
  {\bibfnamefont {K.}~\bibnamefont {Hwang}}, \bibinfo {author} {\bibfnamefont
  {N.}~\bibnamefont {Kurita}}, \bibinfo {author} {\bibfnamefont
  {H.}~\bibnamefont {Tanaka}}, \bibinfo {author} {\bibfnamefont
  {S.}~\bibnamefont {Fujimoto}}, \bibinfo {author} {\bibfnamefont
  {Y.}~\bibnamefont {Matsuda}}, \bibinfo {author} {\bibfnamefont {E.-G.}\
  \bibnamefont {Moon}},  \emph {et~al.},\ }\href {\doibase
  10.1038/s41567-021-01488-6} {\bibfield  {journal} {\bibinfo  {journal} {Nat.
  Phys.}\ }\textbf {\bibinfo {volume} {18}},\ \bibinfo {pages} {429} (\bibinfo
  {year} {2022})}\BibitemShut {NoStop}%
\bibitem [{\citenamefont {Nasu}\ \emph {et~al.}(2017)\citenamefont {Nasu},
  \citenamefont {Yoshitake},\ and\ \citenamefont {Motome}}]{Nasu2017}%
  \BibitemOpen
  \bibfield  {author} {\bibinfo {author} {\bibfnamefont {J.}~\bibnamefont
  {Nasu}}, \bibinfo {author} {\bibfnamefont {J.}~\bibnamefont {Yoshitake}}, \
  and\ \bibinfo {author} {\bibfnamefont {Y.}~\bibnamefont {Motome}},\ }\href
  {\doibase 10.1103/PhysRevLett.119.127204} {\bibfield  {journal} {\bibinfo
  {journal} {Phys. Rev. Lett.}\ }\textbf {\bibinfo {volume} {119}},\ \bibinfo
  {pages} {127204} (\bibinfo {year} {2017})}\BibitemShut {NoStop}%
\bibitem [{\citenamefont {Cookmeyer}\ and\ \citenamefont
  {Moore}(2018)}]{Cookmeyer2018}%
  \BibitemOpen
  \bibfield  {author} {\bibinfo {author} {\bibfnamefont {J.}~\bibnamefont
  {Cookmeyer}}\ and\ \bibinfo {author} {\bibfnamefont {J.~E.}\ \bibnamefont
  {Moore}},\ }\href {\doibase 10.1103/PhysRevB.98.060412} {\bibfield  {journal}
  {\bibinfo  {journal} {Phys. Rev. B}\ }\textbf {\bibinfo {volume} {98}},\
  \bibinfo {pages} {060412} (\bibinfo {year} {2018})}\BibitemShut {NoStop}%
\bibitem [{\citenamefont {Kasahara}\ \emph {et~al.}(2018)\citenamefont
  {Kasahara}, \citenamefont {Ohnishi}, \citenamefont {Mizukami}, \citenamefont
  {Tanaka}, \citenamefont {Ma}, \citenamefont {Sugii}, \citenamefont {Kurita},
  \citenamefont {Tanaka}, \citenamefont {Nasu}, \citenamefont {Motome},
  \citenamefont {Shibauchi},\ and\ \citenamefont
  {Matsuda}}]{kasahara2018majorana}%
  \BibitemOpen
  \bibfield  {author} {\bibinfo {author} {\bibfnamefont {Y.}~\bibnamefont
  {Kasahara}}, \bibinfo {author} {\bibfnamefont {T.}~\bibnamefont {Ohnishi}},
  \bibinfo {author} {\bibfnamefont {Y.}~\bibnamefont {Mizukami}}, \bibinfo
  {author} {\bibfnamefont {O.}~\bibnamefont {Tanaka}}, \bibinfo {author}
  {\bibfnamefont {S.}~\bibnamefont {Ma}}, \bibinfo {author} {\bibfnamefont
  {K.}~\bibnamefont {Sugii}}, \bibinfo {author} {\bibfnamefont
  {N.}~\bibnamefont {Kurita}}, \bibinfo {author} {\bibfnamefont
  {H.}~\bibnamefont {Tanaka}}, \bibinfo {author} {\bibfnamefont
  {J.}~\bibnamefont {Nasu}}, \bibinfo {author} {\bibfnamefont {Y.}~\bibnamefont
  {Motome}}, \bibinfo {author} {\bibfnamefont {T.}~\bibnamefont {Shibauchi}}, \
  and\ \bibinfo {author} {\bibfnamefont {Y.}~\bibnamefont {Matsuda}},\ }\href
  {\doibase 10.1038/s41586-018-0274-0} {\bibfield  {journal} {\bibinfo
  {journal} {Nature}\ }\textbf {\bibinfo {volume} {559}},\ \bibinfo {pages}
  {227} (\bibinfo {year} {2018})}\BibitemShut {NoStop}%
\bibitem [{\citenamefont {Hentrich}\ \emph {et~al.}(2019)\citenamefont
  {Hentrich}, \citenamefont {Roslova}, \citenamefont {Isaeva}, \citenamefont
  {Doert}, \citenamefont {Brenig}, \citenamefont {B\"uchner},\ and\
  \citenamefont {Hess}}]{Hentrich2019}%
  \BibitemOpen
  \bibfield  {author} {\bibinfo {author} {\bibfnamefont {R.}~\bibnamefont
  {Hentrich}}, \bibinfo {author} {\bibfnamefont {M.}~\bibnamefont {Roslova}},
  \bibinfo {author} {\bibfnamefont {A.}~\bibnamefont {Isaeva}}, \bibinfo
  {author} {\bibfnamefont {T.}~\bibnamefont {Doert}}, \bibinfo {author}
  {\bibfnamefont {W.}~\bibnamefont {Brenig}}, \bibinfo {author} {\bibfnamefont
  {B.}~\bibnamefont {B\"uchner}}, \ and\ \bibinfo {author} {\bibfnamefont
  {C.}~\bibnamefont {Hess}},\ }\href {\doibase 10.1103/PhysRevB.99.085136}
  {\bibfield  {journal} {\bibinfo  {journal} {Phys. Rev. B}\ }\textbf {\bibinfo
  {volume} {99}},\ \bibinfo {pages} {085136} (\bibinfo {year}
  {2019})}\BibitemShut {NoStop}%
\bibitem [{\citenamefont {Hickey}\ and\ \citenamefont
  {Trebst}(2019)}]{hickey2019emergence}%
  \BibitemOpen
  \bibfield  {author} {\bibinfo {author} {\bibfnamefont {C.}~\bibnamefont
  {Hickey}}\ and\ \bibinfo {author} {\bibfnamefont {S.}~\bibnamefont
  {Trebst}},\ }\href {\doibase 10.1038/s41467-019-08459-9} {\bibfield
  {journal} {\bibinfo  {journal} {Nature communications}\ }\textbf {\bibinfo
  {volume} {10}},\ \bibinfo {pages} {530} (\bibinfo {year} {2019})}\BibitemShut
  {NoStop}%
\bibitem [{\citenamefont {Gordon}\ \emph {et~al.}(2019)\citenamefont {Gordon},
  \citenamefont {Catuneanu}, \citenamefont {S{\o}rensen},\ and\ \citenamefont
  {Kee}}]{Gordon2019}%
  \BibitemOpen
  \bibfield  {author} {\bibinfo {author} {\bibfnamefont {J.~S.}\ \bibnamefont
  {Gordon}}, \bibinfo {author} {\bibfnamefont {A.}~\bibnamefont {Catuneanu}},
  \bibinfo {author} {\bibfnamefont {E.~S.}\ \bibnamefont {S{\o}rensen}}, \ and\
  \bibinfo {author} {\bibfnamefont {H.-Y.}\ \bibnamefont {Kee}},\ }\href
  {\doibase 10.1038/s41467-019-10405-8} {\bibfield  {journal} {\bibinfo
  {journal} {Nat. Commu.}\ }\textbf {\bibinfo {volume} {10}} (\bibinfo {year}
  {2019}),\ 10.1038/s41467-019-10405-8}\BibitemShut {NoStop}%
\bibitem [{\citenamefont {Gao}\ \emph {et~al.}(2019)\citenamefont {Gao},
  \citenamefont {Hickey}, \citenamefont {Xiang}, \citenamefont {Trebst},\ and\
  \citenamefont {Chen}}]{Gao2019_thermal}%
  \BibitemOpen
  \bibfield  {author} {\bibinfo {author} {\bibfnamefont {Y.~H.}\ \bibnamefont
  {Gao}}, \bibinfo {author} {\bibfnamefont {C.}~\bibnamefont {Hickey}},
  \bibinfo {author} {\bibfnamefont {T.}~\bibnamefont {Xiang}}, \bibinfo
  {author} {\bibfnamefont {S.}~\bibnamefont {Trebst}}, \ and\ \bibinfo {author}
  {\bibfnamefont {G.}~\bibnamefont {Chen}},\ }\href {\doibase
  10.1103/PhysRevResearch.1.013014} {\bibfield  {journal} {\bibinfo  {journal}
  {Phys. Rev. Research}\ }\textbf {\bibinfo {volume} {1}},\ \bibinfo {pages}
  {013014} (\bibinfo {year} {2019})}\BibitemShut {NoStop}%
\bibitem [{\citenamefont {Lee}\ \emph {et~al.}(2020)\citenamefont {Lee},
  \citenamefont {Kaneko}, \citenamefont {Chern}, \citenamefont {Okubo},
  \citenamefont {Yamaji}, \citenamefont {Kawashima},\ and\ \citenamefont
  {Kim}}]{Lee2020_Magnetic}%
  \BibitemOpen
  \bibfield  {author} {\bibinfo {author} {\bibfnamefont {H.-Y.}\ \bibnamefont
  {Lee}}, \bibinfo {author} {\bibfnamefont {R.}~\bibnamefont {Kaneko}},
  \bibinfo {author} {\bibfnamefont {L.~E.}\ \bibnamefont {Chern}}, \bibinfo
  {author} {\bibfnamefont {T.}~\bibnamefont {Okubo}}, \bibinfo {author}
  {\bibfnamefont {Y.}~\bibnamefont {Yamaji}}, \bibinfo {author} {\bibfnamefont
  {N.}~\bibnamefont {Kawashima}}, \ and\ \bibinfo {author} {\bibfnamefont
  {Y.~B.}\ \bibnamefont {Kim}},\ }\href {\doibase 10.1038/s41467-020-15320-x}
  {\bibfield  {journal} {\bibinfo  {journal} {Nat. Commun.}\ }\textbf {\bibinfo
  {volume} {11}},\ \bibinfo {pages} {1639} (\bibinfo {year}
  {2020})}\BibitemShut {NoStop}%
\bibitem [{\citenamefont {Yokoi}\ \emph {et~al.}(2021)\citenamefont {Yokoi},
  \citenamefont {Ma}, \citenamefont {Kasahara}, \citenamefont {Kasahara},
  \citenamefont {Shibauchi}, \citenamefont {Kurita}, \citenamefont {Tanaka},
  \citenamefont {Nasu}, \citenamefont {Motome}, \citenamefont {Hickey} \emph
  {et~al.}}]{yokoi2021half}%
  \BibitemOpen
  \bibfield  {author} {\bibinfo {author} {\bibfnamefont {T.}~\bibnamefont
  {Yokoi}}, \bibinfo {author} {\bibfnamefont {S.}~\bibnamefont {Ma}}, \bibinfo
  {author} {\bibfnamefont {Y.}~\bibnamefont {Kasahara}}, \bibinfo {author}
  {\bibfnamefont {S.}~\bibnamefont {Kasahara}}, \bibinfo {author}
  {\bibfnamefont {T.}~\bibnamefont {Shibauchi}}, \bibinfo {author}
  {\bibfnamefont {N.}~\bibnamefont {Kurita}}, \bibinfo {author} {\bibfnamefont
  {H.}~\bibnamefont {Tanaka}}, \bibinfo {author} {\bibfnamefont
  {J.}~\bibnamefont {Nasu}}, \bibinfo {author} {\bibfnamefont {Y.}~\bibnamefont
  {Motome}}, \bibinfo {author} {\bibfnamefont {C.}~\bibnamefont {Hickey}},
  \emph {et~al.},\ }\href {\doibase 10.1126/science.aay5551} {\bibfield
  {journal} {\bibinfo  {journal} {Science}\ }\textbf {\bibinfo {volume}
  {373}},\ \bibinfo {pages} {568} (\bibinfo {year} {2021})}\BibitemShut
  {NoStop}%
\bibitem [{\citenamefont {Chern}\ \emph {et~al.}(2021)\citenamefont {Chern},
  \citenamefont {Zhang},\ and\ \citenamefont {Kim}}]{Chern2021_Sign}%
  \BibitemOpen
  \bibfield  {author} {\bibinfo {author} {\bibfnamefont {L.~E.}\ \bibnamefont
  {Chern}}, \bibinfo {author} {\bibfnamefont {E.~Z.}\ \bibnamefont {Zhang}}, \
  and\ \bibinfo {author} {\bibfnamefont {Y.~B.}\ \bibnamefont {Kim}},\ }\href
  {\doibase 10.1103/PhysRevLett.126.147201} {\bibfield  {journal} {\bibinfo
  {journal} {Phys. Rev. Lett.}\ }\textbf {\bibinfo {volume} {126}},\ \bibinfo
  {pages} {147201} (\bibinfo {year} {2021})}\BibitemShut {NoStop}%
\bibitem [{\citenamefont {Zhang}\ \emph {et~al.}(2021)\citenamefont {Zhang},
  \citenamefont {Chern},\ and\ \citenamefont {Kim}}]{Zhang2021}%
  \BibitemOpen
  \bibfield  {author} {\bibinfo {author} {\bibfnamefont {E.~Z.}\ \bibnamefont
  {Zhang}}, \bibinfo {author} {\bibfnamefont {L.~E.}\ \bibnamefont {Chern}}, \
  and\ \bibinfo {author} {\bibfnamefont {Y.~B.}\ \bibnamefont {Kim}},\ }\href
  {\doibase 10.1103/PhysRevB.103.174402} {\bibfield  {journal} {\bibinfo
  {journal} {Phys. Rev. B}\ }\textbf {\bibinfo {volume} {103}},\ \bibinfo
  {pages} {174402} (\bibinfo {year} {2021})}\BibitemShut {NoStop}%
\bibitem [{\citenamefont {Koyama}\ and\ \citenamefont
  {Nasu}(2021)}]{Koyama2021}%
  \BibitemOpen
  \bibfield  {author} {\bibinfo {author} {\bibfnamefont {S.}~\bibnamefont
  {Koyama}}\ and\ \bibinfo {author} {\bibfnamefont {J.}~\bibnamefont {Nasu}},\
  }\href {\doibase 10.1103/PhysRevB.104.075121} {\bibfield  {journal} {\bibinfo
   {journal} {Phys. Rev. B}\ }\textbf {\bibinfo {volume} {104}},\ \bibinfo
  {pages} {075121} (\bibinfo {year} {2021})}\BibitemShut {NoStop}%
\bibitem [{\citenamefont {Czajka}\ \emph {et~al.}(2021)\citenamefont {Czajka},
  \citenamefont {Gao}, \citenamefont {Hirschberger}, \citenamefont
  {Lampen-Kelley}, \citenamefont {Banerjee}, \citenamefont {Yan}, \citenamefont
  {Mandrus}, \citenamefont {Nagler},\ and\ \citenamefont
  {Ong}}]{czajka2021oscillations}%
  \BibitemOpen
  \bibfield  {author} {\bibinfo {author} {\bibfnamefont {P.}~\bibnamefont
  {Czajka}}, \bibinfo {author} {\bibfnamefont {T.}~\bibnamefont {Gao}},
  \bibinfo {author} {\bibfnamefont {M.}~\bibnamefont {Hirschberger}}, \bibinfo
  {author} {\bibfnamefont {P.}~\bibnamefont {Lampen-Kelley}}, \bibinfo {author}
  {\bibfnamefont {A.}~\bibnamefont {Banerjee}}, \bibinfo {author}
  {\bibfnamefont {J.}~\bibnamefont {Yan}}, \bibinfo {author} {\bibfnamefont
  {D.~G.}\ \bibnamefont {Mandrus}}, \bibinfo {author} {\bibfnamefont {S.~E.}\
  \bibnamefont {Nagler}}, \ and\ \bibinfo {author} {\bibfnamefont
  {N.}~\bibnamefont {Ong}},\ }\href {\doibase 10.1038/s41567-021-01243-x}
  {\bibfield  {journal} {\bibinfo  {journal} {Nature Physics}\ }\textbf
  {\bibinfo {volume} {17}},\ \bibinfo {pages} {915} (\bibinfo {year}
  {2021})}\BibitemShut {NoStop}%
\bibitem [{\citenamefont {Hwang}\ \emph {et~al.}(2022)\citenamefont {Hwang},
  \citenamefont {Go}, \citenamefont {Seong}, \citenamefont {Shibauchi},\ and\
  \citenamefont {Moon}}]{hwang2022identification}%
  \BibitemOpen
  \bibfield  {author} {\bibinfo {author} {\bibfnamefont {K.}~\bibnamefont
  {Hwang}}, \bibinfo {author} {\bibfnamefont {A.}~\bibnamefont {Go}}, \bibinfo
  {author} {\bibfnamefont {J.~H.}\ \bibnamefont {Seong}}, \bibinfo {author}
  {\bibfnamefont {T.}~\bibnamefont {Shibauchi}}, \ and\ \bibinfo {author}
  {\bibfnamefont {E.-G.}\ \bibnamefont {Moon}},\ }\href {\doibase
  10.1038/s41467-021-27943-9} {\bibfield  {journal} {\bibinfo  {journal} {Nat.
  Commun.}\ }\textbf {\bibinfo {volume} {13}},\ \bibinfo {pages} {1} (\bibinfo
  {year} {2022})}\BibitemShut {NoStop}%
\bibitem [{\citenamefont {Feldmeier}\ \emph {et~al.}(2020)\citenamefont
  {Feldmeier}, \citenamefont {Natori}, \citenamefont {Knap},\ and\
  \citenamefont {Knolle}}]{Feldmeier2020}%
  \BibitemOpen
  \bibfield  {author} {\bibinfo {author} {\bibfnamefont {J.}~\bibnamefont
  {Feldmeier}}, \bibinfo {author} {\bibfnamefont {W.}~\bibnamefont {Natori}},
  \bibinfo {author} {\bibfnamefont {M.}~\bibnamefont {Knap}}, \ and\ \bibinfo
  {author} {\bibfnamefont {J.}~\bibnamefont {Knolle}},\ }\href {\doibase
  10.1103/PhysRevB.102.134423} {\bibfield  {journal} {\bibinfo  {journal}
  {Phys. Rev. B}\ }\textbf {\bibinfo {volume} {102}},\ \bibinfo {pages}
  {134423} (\bibinfo {year} {2020})}\BibitemShut {NoStop}%
\bibitem [{\citenamefont {K\"onig}\ \emph {et~al.}(2020)\citenamefont
  {K\"onig}, \citenamefont {Randeria},\ and\ \citenamefont
  {J\"ack}}]{Konig2020}%
  \BibitemOpen
  \bibfield  {author} {\bibinfo {author} {\bibfnamefont {E.~J.}\ \bibnamefont
  {K\"onig}}, \bibinfo {author} {\bibfnamefont {M.~T.}\ \bibnamefont
  {Randeria}}, \ and\ \bibinfo {author} {\bibfnamefont {B.}~\bibnamefont
  {J\"ack}},\ }\href {\doibase 10.1103/PhysRevLett.125.267206} {\bibfield
  {journal} {\bibinfo  {journal} {Phys. Rev. Lett.}\ }\textbf {\bibinfo
  {volume} {125}},\ \bibinfo {pages} {267206} (\bibinfo {year}
  {2020})}\BibitemShut {NoStop}%
\bibitem [{\citenamefont {Pereira}\ and\ \citenamefont
  {Egger}(2020)}]{Pereira2020}%
  \BibitemOpen
  \bibfield  {author} {\bibinfo {author} {\bibfnamefont {R.~G.}\ \bibnamefont
  {Pereira}}\ and\ \bibinfo {author} {\bibfnamefont {R.}~\bibnamefont
  {Egger}},\ }\href {\doibase 10.1103/PhysRevLett.125.227202} {\bibfield
  {journal} {\bibinfo  {journal} {Phys. Rev. Lett.}\ }\textbf {\bibinfo
  {volume} {125}},\ \bibinfo {pages} {227202} (\bibinfo {year}
  {2020})}\BibitemShut {NoStop}%
\bibitem [{\citenamefont {Udagawa}\ \emph {et~al.}(2021)\citenamefont
  {Udagawa}, \citenamefont {Takayoshi},\ and\ \citenamefont
  {Oka}}]{Udagawa2021}%
  \BibitemOpen
  \bibfield  {author} {\bibinfo {author} {\bibfnamefont {M.}~\bibnamefont
  {Udagawa}}, \bibinfo {author} {\bibfnamefont {S.}~\bibnamefont {Takayoshi}},
  \ and\ \bibinfo {author} {\bibfnamefont {T.}~\bibnamefont {Oka}},\ }\href
  {\doibase 10.1103/PhysRevLett.126.127201} {\bibfield  {journal} {\bibinfo
  {journal} {Phys. Rev. Lett.}\ }\textbf {\bibinfo {volume} {126}},\ \bibinfo
  {pages} {127201} (\bibinfo {year} {2021})}\BibitemShut {NoStop}%
\bibitem [{\citenamefont {Klocke}\ \emph {et~al.}(2021)\citenamefont {Klocke},
  \citenamefont {Aasen}, \citenamefont {Mong}, \citenamefont {Demler},\ and\
  \citenamefont {Alicea}}]{Klocke2021}%
  \BibitemOpen
  \bibfield  {author} {\bibinfo {author} {\bibfnamefont {K.}~\bibnamefont
  {Klocke}}, \bibinfo {author} {\bibfnamefont {D.}~\bibnamefont {Aasen}},
  \bibinfo {author} {\bibfnamefont {R.~S.~K.}\ \bibnamefont {Mong}}, \bibinfo
  {author} {\bibfnamefont {E.~A.}\ \bibnamefont {Demler}}, \ and\ \bibinfo
  {author} {\bibfnamefont {J.}~\bibnamefont {Alicea}},\ }\href {\doibase
  10.1103/PhysRevLett.126.177204} {\bibfield  {journal} {\bibinfo  {journal}
  {Phys. Rev. Lett.}\ }\textbf {\bibinfo {volume} {126}},\ \bibinfo {pages}
  {177204} (\bibinfo {year} {2021})}\BibitemShut {NoStop}%
\bibitem [{\citenamefont {Wei}\ \emph {et~al.}(2021)\citenamefont {Wei},
  \citenamefont {Mitrovi\ifmmode~\acute{c}\else \'{c}\fi{}},\ and\
  \citenamefont {Feldman}}]{Wei2021}%
  \BibitemOpen
  \bibfield  {author} {\bibinfo {author} {\bibfnamefont {Z.}~\bibnamefont
  {Wei}}, \bibinfo {author} {\bibfnamefont {V.~F.}\ \bibnamefont
  {Mitrovi\ifmmode~\acute{c}\else \'{c}\fi{}}}, \ and\ \bibinfo {author}
  {\bibfnamefont {D.~E.}\ \bibnamefont {Feldman}},\ }\href {\doibase
  10.1103/PhysRevLett.127.167204} {\bibfield  {journal} {\bibinfo  {journal}
  {Phys. Rev. Lett.}\ }\textbf {\bibinfo {volume} {127}},\ \bibinfo {pages}
  {167204} (\bibinfo {year} {2021})}\BibitemShut {NoStop}%
\bibitem [{\citenamefont {Jang}\ \emph {et~al.}(2021)\citenamefont {Jang},
  \citenamefont {Kato},\ and\ \citenamefont {Motome}}]{Jang2021}%
  \BibitemOpen
  \bibfield  {author} {\bibinfo {author} {\bibfnamefont {S.-H.}\ \bibnamefont
  {Jang}}, \bibinfo {author} {\bibfnamefont {Y.}~\bibnamefont {Kato}}, \ and\
  \bibinfo {author} {\bibfnamefont {Y.}~\bibnamefont {Motome}},\ }\href
  {\doibase 10.1103/PhysRevB.104.085142} {\bibfield  {journal} {\bibinfo
  {journal} {Phys. Rev. B}\ }\textbf {\bibinfo {volume} {104}},\ \bibinfo
  {pages} {085142} (\bibinfo {year} {2021})}\BibitemShut {NoStop}%
\bibitem [{\citenamefont {Liu}\ \emph {et~al.}(shed)\citenamefont {Liu},
  \citenamefont {Slagle}, \citenamefont {Burch},\ and\ \citenamefont
  {Alicea}}]{Yue-Liu2021pre}%
  \BibitemOpen
  \bibfield  {author} {\bibinfo {author} {\bibfnamefont {Y.}~\bibnamefont
  {Liu}}, \bibinfo {author} {\bibfnamefont {K.}~\bibnamefont {Slagle}},
  \bibinfo {author} {\bibfnamefont {K.~S.}\ \bibnamefont {Burch}}, \ and\
  \bibinfo {author} {\bibfnamefont {J.}~\bibnamefont {Alicea}},\ }\href
  {http://arxiv.org/abs/2111.09325} {\bibfield  {journal} {\bibinfo  {journal}
  {preprint}\ ,\ \bibinfo {pages} {arXiv:2111.09325}} (\bibinfo {year}
  {unpublished})}\BibitemShut {NoStop}%
\bibitem [{\citenamefont {Yao}\ and\ \citenamefont {Lee}(2011)}]{Yao_Lee2011}%
  \BibitemOpen
  \bibfield  {author} {\bibinfo {author} {\bibfnamefont {H.}~\bibnamefont
  {Yao}}\ and\ \bibinfo {author} {\bibfnamefont {D.-H.}\ \bibnamefont {Lee}},\
  }\href {\doibase 10.1103/PhysRevLett.107.087205} {\bibfield  {journal}
  {\bibinfo  {journal} {Phys. Rev. Lett.}\ }\textbf {\bibinfo {volume} {107}},\
  \bibinfo {pages} {087205} (\bibinfo {year} {2011})}\BibitemShut {NoStop}%
\bibitem [{\citenamefont {de~Carvalho}\ \emph {et~al.}(2018)\citenamefont
  {de~Carvalho}, \citenamefont {Freire}, \citenamefont {Miranda},\ and\
  \citenamefont {Pereira}}]{Carvalho2018}%
  \BibitemOpen
  \bibfield  {author} {\bibinfo {author} {\bibfnamefont {V.~S.}\ \bibnamefont
  {de~Carvalho}}, \bibinfo {author} {\bibfnamefont {H.}~\bibnamefont {Freire}},
  \bibinfo {author} {\bibfnamefont {E.}~\bibnamefont {Miranda}}, \ and\
  \bibinfo {author} {\bibfnamefont {R.~G.}\ \bibnamefont {Pereira}},\ }\href
  {\doibase 10.1103/PhysRevB.98.155105} {\bibfield  {journal} {\bibinfo
  {journal} {Phys. Rev. B}\ }\textbf {\bibinfo {volume} {98}},\ \bibinfo
  {pages} {155105} (\bibinfo {year} {2018})}\BibitemShut {NoStop}%
\bibitem [{\citenamefont {Aftergood}\ and\ \citenamefont
  {Takei}(2020)}]{Aftergood2020}%
  \BibitemOpen
  \bibfield  {author} {\bibinfo {author} {\bibfnamefont {J.}~\bibnamefont
  {Aftergood}}\ and\ \bibinfo {author} {\bibfnamefont {S.}~\bibnamefont
  {Takei}},\ }\href {\doibase 10.1103/PhysRevResearch.2.033439} {\bibfield
  {journal} {\bibinfo  {journal} {Phys. Rev. Research}\ }\textbf {\bibinfo
  {volume} {2}},\ \bibinfo {pages} {033439} (\bibinfo {year}
  {2020})}\BibitemShut {NoStop}%
\bibitem [{\citenamefont {Minakawa}\ \emph {et~al.}(2020)\citenamefont
  {Minakawa}, \citenamefont {Murakami}, \citenamefont {Koga},\ and\
  \citenamefont {Nasu}}]{Minakawa2020}%
  \BibitemOpen
  \bibfield  {author} {\bibinfo {author} {\bibfnamefont {T.}~\bibnamefont
  {Minakawa}}, \bibinfo {author} {\bibfnamefont {Y.}~\bibnamefont {Murakami}},
  \bibinfo {author} {\bibfnamefont {A.}~\bibnamefont {Koga}}, \ and\ \bibinfo
  {author} {\bibfnamefont {J.}~\bibnamefont {Nasu}},\ }\href {\doibase
  10.1103/PhysRevLett.125.047204} {\bibfield  {journal} {\bibinfo  {journal}
  {Phys. Rev. Lett.}\ }\textbf {\bibinfo {volume} {125}},\ \bibinfo {pages}
  {047204} (\bibinfo {year} {2020})}\BibitemShut {NoStop}%
\bibitem [{\citenamefont {Taguchi}\ \emph {et~al.}(2021)\citenamefont
  {Taguchi}, \citenamefont {Murakami}, \citenamefont {Koga},\ and\
  \citenamefont {Nasu}}]{Taguchi2021}%
  \BibitemOpen
  \bibfield  {author} {\bibinfo {author} {\bibfnamefont {H.}~\bibnamefont
  {Taguchi}}, \bibinfo {author} {\bibfnamefont {Y.}~\bibnamefont {Murakami}},
  \bibinfo {author} {\bibfnamefont {A.}~\bibnamefont {Koga}}, \ and\ \bibinfo
  {author} {\bibfnamefont {J.}~\bibnamefont {Nasu}},\ }\href {\doibase
  10.1103/PhysRevB.104.125139} {\bibfield  {journal} {\bibinfo  {journal}
  {Phys. Rev. B}\ }\textbf {\bibinfo {volume} {104}},\ \bibinfo {pages}
  {125139} (\bibinfo {year} {2021})}\BibitemShut {NoStop}%
\bibitem [{\citenamefont {Koga}\ \emph {et~al.}(2021)\citenamefont {Koga},
  \citenamefont {Murakami},\ and\ \citenamefont {Nasu}}]{Koga_majorana2021}%
  \BibitemOpen
  \bibfield  {author} {\bibinfo {author} {\bibfnamefont {A.}~\bibnamefont
  {Koga}}, \bibinfo {author} {\bibfnamefont {Y.}~\bibnamefont {Murakami}}, \
  and\ \bibinfo {author} {\bibfnamefont {J.}~\bibnamefont {Nasu}},\ }\href
  {\doibase 10.1103/PhysRevB.103.214421} {\bibfield  {journal} {\bibinfo
  {journal} {Phys. Rev. B}\ }\textbf {\bibinfo {volume} {103}},\ \bibinfo
  {pages} {214421} (\bibinfo {year} {2021})}\BibitemShut {NoStop}%
\bibitem [{\citenamefont {Taguchi}\ \emph {et~al.}(2022)\citenamefont
  {Taguchi}, \citenamefont {Murakami},\ and\ \citenamefont
  {Koga}}]{Taguchi_Thermally2022}%
  \BibitemOpen
  \bibfield  {author} {\bibinfo {author} {\bibfnamefont {H.}~\bibnamefont
  {Taguchi}}, \bibinfo {author} {\bibfnamefont {Y.}~\bibnamefont {Murakami}}, \
  and\ \bibinfo {author} {\bibfnamefont {A.}~\bibnamefont {Koga}},\ }\href
  {\doibase 10.1103/PhysRevB.105.125137} {\bibfield  {journal} {\bibinfo
  {journal} {Phys. Rev. B}\ }\textbf {\bibinfo {volume} {105}},\ \bibinfo
  {pages} {125137} (\bibinfo {year} {2022})}\BibitemShut {NoStop}%
\bibitem [{\citenamefont {Takikawa}\ \emph {et~al.}(shed)\citenamefont
  {Takikawa}, \citenamefont {Yamada},\ and\ \citenamefont
  {Fujimoto}}]{Takikawa2021pre}%
  \BibitemOpen
  \bibfield  {author} {\bibinfo {author} {\bibfnamefont {D.}~\bibnamefont
  {Takikawa}}, \bibinfo {author} {\bibfnamefont {M.~G.}\ \bibnamefont
  {Yamada}}, \ and\ \bibinfo {author} {\bibfnamefont {S.}~\bibnamefont
  {Fujimoto}},\ }\href {http://arxiv.org/abs/2104.11115} {\bibfield  {journal}
  {\bibinfo  {journal} {preprint}\ ,\ \bibinfo {pages} {arXiv:2104.11115}}
  (\bibinfo {year} {unpublished})}\BibitemShut {NoStop}%
\bibitem [{\citenamefont {Chen}\ and\ \citenamefont
  {Hu}(2007)}]{PhysRevB.76.193101}%
  \BibitemOpen
  \bibfield  {author} {\bibinfo {author} {\bibfnamefont {H.-D.}\ \bibnamefont
  {Chen}}\ and\ \bibinfo {author} {\bibfnamefont {J.}~\bibnamefont {Hu}},\
  }\href {\doibase 10.1103/PhysRevB.76.193101} {\bibfield  {journal} {\bibinfo
  {journal} {Phys. Rev. B}\ }\textbf {\bibinfo {volume} {76}},\ \bibinfo
  {pages} {193101} (\bibinfo {year} {2007})}\BibitemShut {NoStop}%
\bibitem [{\citenamefont {Feng}\ \emph {et~al.}(2007)\citenamefont {Feng},
  \citenamefont {Zhang},\ and\ \citenamefont {Xiang}}]{PhysRevLett.98.087204}%
  \BibitemOpen
  \bibfield  {author} {\bibinfo {author} {\bibfnamefont {X.-Y.}\ \bibnamefont
  {Feng}}, \bibinfo {author} {\bibfnamefont {G.-M.}\ \bibnamefont {Zhang}}, \
  and\ \bibinfo {author} {\bibfnamefont {T.}~\bibnamefont {Xiang}},\ }\href
  {\doibase 10.1103/PhysRevLett.98.087204} {\bibfield  {journal} {\bibinfo
  {journal} {Phys. Rev. Lett.}\ }\textbf {\bibinfo {volume} {98}},\ \bibinfo
  {pages} {087204} (\bibinfo {year} {2007})}\BibitemShut {NoStop}%
\bibitem [{\citenamefont {Chen}\ and\ \citenamefont
  {Nussinov}(2008)}]{1751-8121-41-7-075001}%
  \BibitemOpen
  \bibfield  {author} {\bibinfo {author} {\bibfnamefont {H.-D.}\ \bibnamefont
  {Chen}}\ and\ \bibinfo {author} {\bibfnamefont {Z.}~\bibnamefont
  {Nussinov}},\ }\href {http://stacks.iop.org/1751-8121/41/i=7/a=075001}
  {\bibfield  {journal} {\bibinfo  {journal} {J. Phys. A}\ }\textbf {\bibinfo
  {volume} {41}},\ \bibinfo {pages} {075001} (\bibinfo {year}
  {2008})}\BibitemShut {NoStop}%
\bibitem [{\citenamefont {Nasu}\ \emph {et~al.}(2014)\citenamefont {Nasu},
  \citenamefont {Udagawa},\ and\ \citenamefont
  {Motome}}]{PhysRevLett.113.197205}%
  \BibitemOpen
  \bibfield  {author} {\bibinfo {author} {\bibfnamefont {J.}~\bibnamefont
  {Nasu}}, \bibinfo {author} {\bibfnamefont {M.}~\bibnamefont {Udagawa}}, \
  and\ \bibinfo {author} {\bibfnamefont {Y.}~\bibnamefont {Motome}},\ }\href
  {\doibase 10.1103/PhysRevLett.113.197205} {\bibfield  {journal} {\bibinfo
  {journal} {Phys. Rev. Lett.}\ }\textbf {\bibinfo {volume} {113}},\ \bibinfo
  {pages} {197205} (\bibinfo {year} {2014})}\BibitemShut {NoStop}%
\bibitem [{\citenamefont {F.}\ and\ \citenamefont
  {M.}(1974)}]{volkov1973collisionless}%
  \BibitemOpen
  \bibfield  {author} {\bibinfo {author} {\bibfnamefont {V.~A.}\ \bibnamefont
  {F.}}\ and\ \bibinfo {author} {\bibfnamefont {K.~S.}\ \bibnamefont {M.}},\
  }\href@noop {} {\bibfield  {journal} {\bibinfo  {journal} {Sov. Phys. JETP}\
  }\textbf {\bibinfo {volume} {38}},\ \bibinfo {pages} {1018} (\bibinfo {year}
  {1974})}\BibitemShut {NoStop}%
\bibitem [{\citenamefont {Tsuji}\ and\ \citenamefont
  {Aoki}(2015)}]{Tsuji_theoryof2015}%
  \BibitemOpen
  \bibfield  {author} {\bibinfo {author} {\bibfnamefont {N.}~\bibnamefont
  {Tsuji}}\ and\ \bibinfo {author} {\bibfnamefont {H.}~\bibnamefont {Aoki}},\
  }\href {\doibase 10.1103/PhysRevB.92.064508} {\bibfield  {journal} {\bibinfo
  {journal} {Phys. Rev. B}\ }\textbf {\bibinfo {volume} {92}},\ \bibinfo
  {pages} {064508} (\bibinfo {year} {2015})}\BibitemShut {NoStop}%
\bibitem [{\citenamefont {Murakami}\ \emph {et~al.}(2020)\citenamefont
  {Murakami}, \citenamefont {Gole\ifmmode~\check{z}\else \v{z}\fi{}},
  \citenamefont {Kaneko}, \citenamefont {Koga}, \citenamefont {Millis},\ and\
  \citenamefont {Werner}}]{Murakami_collective2020}%
  \BibitemOpen
  \bibfield  {author} {\bibinfo {author} {\bibfnamefont {Y.}~\bibnamefont
  {Murakami}}, \bibinfo {author} {\bibfnamefont {D.}~\bibnamefont
  {Gole\ifmmode~\check{z}\else \v{z}\fi{}}}, \bibinfo {author} {\bibfnamefont
  {T.}~\bibnamefont {Kaneko}}, \bibinfo {author} {\bibfnamefont
  {A.}~\bibnamefont {Koga}}, \bibinfo {author} {\bibfnamefont {A.~J.}\
  \bibnamefont {Millis}}, \ and\ \bibinfo {author} {\bibfnamefont
  {P.}~\bibnamefont {Werner}},\ }\href {\doibase 10.1103/PhysRevB.101.195118}
  {\bibfield  {journal} {\bibinfo  {journal} {Phys. Rev. B}\ }\textbf {\bibinfo
  {volume} {101}},\ \bibinfo {pages} {195118} (\bibinfo {year}
  {2020})}\BibitemShut {NoStop}%
\bibitem [{\citenamefont {Willans}\ \emph {et~al.}(2010)\citenamefont
  {Willans}, \citenamefont {Chalker},\ and\ \citenamefont
  {Moessner}}]{Willans2010}%
  \BibitemOpen
  \bibfield  {author} {\bibinfo {author} {\bibfnamefont {A.~J.}\ \bibnamefont
  {Willans}}, \bibinfo {author} {\bibfnamefont {J.~T.}\ \bibnamefont
  {Chalker}}, \ and\ \bibinfo {author} {\bibfnamefont {R.}~\bibnamefont
  {Moessner}},\ }\href {\doibase 10.1103/PhysRevLett.104.237203} {\bibfield
  {journal} {\bibinfo  {journal} {Phys. Rev. Lett.}\ }\textbf {\bibinfo
  {volume} {104}},\ \bibinfo {pages} {237203} (\bibinfo {year}
  {2010})}\BibitemShut {NoStop}%
\bibitem [{\citenamefont {Willans}\ \emph {et~al.}(2011)\citenamefont
  {Willans}, \citenamefont {Chalker},\ and\ \citenamefont
  {Moessner}}]{Willans2011}%
  \BibitemOpen
  \bibfield  {author} {\bibinfo {author} {\bibfnamefont {A.~J.}\ \bibnamefont
  {Willans}}, \bibinfo {author} {\bibfnamefont {J.~T.}\ \bibnamefont
  {Chalker}}, \ and\ \bibinfo {author} {\bibfnamefont {R.}~\bibnamefont
  {Moessner}},\ }\href {\doibase 10.1103/PhysRevB.84.115146} {\bibfield
  {journal} {\bibinfo  {journal} {Phys. Rev. B}\ }\textbf {\bibinfo {volume}
  {84}},\ \bibinfo {pages} {115146} (\bibinfo {year} {2011})}\BibitemShut
  {NoStop}%
\bibitem [{\citenamefont {Beaurepaire}\ \emph {et~al.}(1996)\citenamefont
  {Beaurepaire}, \citenamefont {Merle}, \citenamefont {Daunois},\ and\
  \citenamefont {Bigot}}]{Beaurepaire1996}%
  \BibitemOpen
  \bibfield  {author} {\bibinfo {author} {\bibfnamefont {E.}~\bibnamefont
  {Beaurepaire}}, \bibinfo {author} {\bibfnamefont {J.-C.}\ \bibnamefont
  {Merle}}, \bibinfo {author} {\bibfnamefont {A.}~\bibnamefont {Daunois}}, \
  and\ \bibinfo {author} {\bibfnamefont {J.-Y.}\ \bibnamefont {Bigot}},\ }\href
  {\doibase 10.1103/PhysRevLett.76.4250} {\bibfield  {journal} {\bibinfo
  {journal} {Phys. Rev. Lett.}\ }\textbf {\bibinfo {volume} {76}},\ \bibinfo
  {pages} {4250} (\bibinfo {year} {1996})}\BibitemShut {NoStop}%
\bibitem [{\citenamefont {Hiebert}\ \emph {et~al.}(1997)\citenamefont
  {Hiebert}, \citenamefont {Stankiewicz},\ and\ \citenamefont
  {Freeman}}]{Hiebert1997}%
  \BibitemOpen
  \bibfield  {author} {\bibinfo {author} {\bibfnamefont {W.~K.}\ \bibnamefont
  {Hiebert}}, \bibinfo {author} {\bibfnamefont {A.}~\bibnamefont
  {Stankiewicz}}, \ and\ \bibinfo {author} {\bibfnamefont {M.~R.}\ \bibnamefont
  {Freeman}},\ }\href {\doibase 10.1103/PhysRevLett.79.1134} {\bibfield
  {journal} {\bibinfo  {journal} {Phys. Rev. Lett.}\ }\textbf {\bibinfo
  {volume} {79}},\ \bibinfo {pages} {1134} (\bibinfo {year}
  {1997})}\BibitemShut {NoStop}%
\bibitem [{\citenamefont {Kise}\ \emph {et~al.}(2000)\citenamefont {Kise},
  \citenamefont {Ogasawara}, \citenamefont {Ashida}, \citenamefont {Tomioka},
  \citenamefont {Tokura},\ and\ \citenamefont {Kuwata-Gonokami}}]{Kise2000}%
  \BibitemOpen
  \bibfield  {author} {\bibinfo {author} {\bibfnamefont {T.}~\bibnamefont
  {Kise}}, \bibinfo {author} {\bibfnamefont {T.}~\bibnamefont {Ogasawara}},
  \bibinfo {author} {\bibfnamefont {M.}~\bibnamefont {Ashida}}, \bibinfo
  {author} {\bibfnamefont {Y.}~\bibnamefont {Tomioka}}, \bibinfo {author}
  {\bibfnamefont {Y.}~\bibnamefont {Tokura}}, \ and\ \bibinfo {author}
  {\bibfnamefont {M.}~\bibnamefont {Kuwata-Gonokami}},\ }\href {\doibase
  10.1103/PhysRevLett.85.1986} {\bibfield  {journal} {\bibinfo  {journal}
  {Phys. Rev. Lett.}\ }\textbf {\bibinfo {volume} {85}},\ \bibinfo {pages}
  {1986} (\bibinfo {year} {2000})}\BibitemShut {NoStop}%
\bibitem [{\citenamefont {Guidoni}\ \emph {et~al.}(2002)\citenamefont
  {Guidoni}, \citenamefont {Beaurepaire},\ and\ \citenamefont
  {Bigot}}]{Guidoni2002}%
  \BibitemOpen
  \bibfield  {author} {\bibinfo {author} {\bibfnamefont {L.}~\bibnamefont
  {Guidoni}}, \bibinfo {author} {\bibfnamefont {E.}~\bibnamefont
  {Beaurepaire}}, \ and\ \bibinfo {author} {\bibfnamefont {J.-Y.}\ \bibnamefont
  {Bigot}},\ }\href {\doibase 10.1103/PhysRevLett.89.017401} {\bibfield
  {journal} {\bibinfo  {journal} {Phys. Rev. Lett.}\ }\textbf {\bibinfo
  {volume} {89}},\ \bibinfo {pages} {017401} (\bibinfo {year}
  {2002})}\BibitemShut {NoStop}%
\bibitem [{\citenamefont {Ogasawara}\ \emph {et~al.}(2005)\citenamefont
  {Ogasawara}, \citenamefont {Ohgushi}, \citenamefont {Tomioka}, \citenamefont
  {Takahashi}, \citenamefont {Okamoto}, \citenamefont {Kawasaki},\ and\
  \citenamefont {Tokura}}]{Ogasawara2005}%
  \BibitemOpen
  \bibfield  {author} {\bibinfo {author} {\bibfnamefont {T.}~\bibnamefont
  {Ogasawara}}, \bibinfo {author} {\bibfnamefont {K.}~\bibnamefont {Ohgushi}},
  \bibinfo {author} {\bibfnamefont {Y.}~\bibnamefont {Tomioka}}, \bibinfo
  {author} {\bibfnamefont {K.~S.}\ \bibnamefont {Takahashi}}, \bibinfo {author}
  {\bibfnamefont {H.}~\bibnamefont {Okamoto}}, \bibinfo {author} {\bibfnamefont
  {M.}~\bibnamefont {Kawasaki}}, \ and\ \bibinfo {author} {\bibfnamefont
  {Y.}~\bibnamefont {Tokura}},\ }\href {\doibase 10.1103/PhysRevLett.94.087202}
  {\bibfield  {journal} {\bibinfo  {journal} {Phys. Rev. Lett.}\ }\textbf
  {\bibinfo {volume} {94}},\ \bibinfo {pages} {087202} (\bibinfo {year}
  {2005})}\BibitemShut {NoStop}%
\bibitem [{\citenamefont {Kimel}\ \emph {et~al.}(2005)\citenamefont {Kimel},
  \citenamefont {Kirilyuk}, \citenamefont {Usachev}, \citenamefont {Pisarev},
  \citenamefont {Balbashov},\ and\ \citenamefont
  {Rasing}}]{kimel2005ultrafast}%
  \BibitemOpen
  \bibfield  {author} {\bibinfo {author} {\bibfnamefont {A.}~\bibnamefont
  {Kimel}}, \bibinfo {author} {\bibfnamefont {A.}~\bibnamefont {Kirilyuk}},
  \bibinfo {author} {\bibfnamefont {P.}~\bibnamefont {Usachev}}, \bibinfo
  {author} {\bibfnamefont {R.}~\bibnamefont {Pisarev}}, \bibinfo {author}
  {\bibfnamefont {A.}~\bibnamefont {Balbashov}}, \ and\ \bibinfo {author}
  {\bibfnamefont {T.}~\bibnamefont {Rasing}},\ }\href {\doibase
  10.1038/nature03564} {\bibfield  {journal} {\bibinfo  {journal} {Nature}\
  }\textbf {\bibinfo {volume} {435}},\ \bibinfo {pages} {655} (\bibinfo {year}
  {2005})}\BibitemShut {NoStop}%
\bibitem [{\citenamefont {Cinchetti}\ \emph {et~al.}(2006)\citenamefont
  {Cinchetti}, \citenamefont {S\'anchez~Albaneda}, \citenamefont {Hoffmann},
  \citenamefont {Roth}, \citenamefont {W\"ustenberg}, \citenamefont
  {Krau\ss{}}, \citenamefont {Andreyev}, \citenamefont {Schneider},
  \citenamefont {Bauer},\ and\ \citenamefont {Aeschlimann}}]{Cinchetti2006}%
  \BibitemOpen
  \bibfield  {author} {\bibinfo {author} {\bibfnamefont {M.}~\bibnamefont
  {Cinchetti}}, \bibinfo {author} {\bibfnamefont {M.}~\bibnamefont
  {S\'anchez~Albaneda}}, \bibinfo {author} {\bibfnamefont {D.}~\bibnamefont
  {Hoffmann}}, \bibinfo {author} {\bibfnamefont {T.}~\bibnamefont {Roth}},
  \bibinfo {author} {\bibfnamefont {J.-P.}\ \bibnamefont {W\"ustenberg}},
  \bibinfo {author} {\bibfnamefont {M.}~\bibnamefont {Krau\ss{}}}, \bibinfo
  {author} {\bibfnamefont {O.}~\bibnamefont {Andreyev}}, \bibinfo {author}
  {\bibfnamefont {H.~C.}\ \bibnamefont {Schneider}}, \bibinfo {author}
  {\bibfnamefont {M.}~\bibnamefont {Bauer}}, \ and\ \bibinfo {author}
  {\bibfnamefont {M.}~\bibnamefont {Aeschlimann}},\ }\href {\doibase
  10.1103/PhysRevLett.97.177201} {\bibfield  {journal} {\bibinfo  {journal}
  {Phys. Rev. Lett.}\ }\textbf {\bibinfo {volume} {97}},\ \bibinfo {pages}
  {177201} (\bibinfo {year} {2006})}\BibitemShut {NoStop}%
\bibitem [{\citenamefont {Takahashi}\ \emph {et~al.}(2021)\citenamefont
  {Takahashi}, \citenamefont {Tani}, \citenamefont {Abe}, \citenamefont
  {Yamasaki}, \citenamefont {Suzuki}, \citenamefont {Kan}, \citenamefont
  {Shimakawa},\ and\ \citenamefont {Wadati}}]{Takahashi2021}%
  \BibitemOpen
  \bibfield  {author} {\bibinfo {author} {\bibfnamefont {R.}~\bibnamefont
  {Takahashi}}, \bibinfo {author} {\bibfnamefont {Y.}~\bibnamefont {Tani}},
  \bibinfo {author} {\bibfnamefont {H.}~\bibnamefont {Abe}}, \bibinfo {author}
  {\bibfnamefont {M.}~\bibnamefont {Yamasaki}}, \bibinfo {author}
  {\bibfnamefont {I.}~\bibnamefont {Suzuki}}, \bibinfo {author} {\bibfnamefont
  {D.}~\bibnamefont {Kan}}, \bibinfo {author} {\bibfnamefont {Y.}~\bibnamefont
  {Shimakawa}}, \ and\ \bibinfo {author} {\bibfnamefont {H.}~\bibnamefont
  {Wadati}},\ }\href {\doibase 10.1063/5.0058740} {\bibfield  {journal}
  {\bibinfo  {journal} {Appl. Phys. Lett.}\ }\textbf {\bibinfo {volume}
  {119}},\ \bibinfo {pages} {102404} (\bibinfo {year} {2021})}\BibitemShut
  {NoStop}%
\end{thebibliography}%

\end{document}